\RequirePackage{ifpdf}
\documentclass[a4paper,11pt,notoc]{article}
\pdfoutput=1
\usepackage{jheppub}
\usepackage{amsmath,amsbsy,amssymb}
\usepackage{graphicx}
\usepackage{epsfig}
\usepackage{fixmath}
\usepackage{caption}
\usepackage{subcaption}
\usepackage{slashed}
\usepackage{lineno}
\usepackage{afterpage}

\usepackage{xspace}

\graphicspath{plots/} 

\newcommand\TwoFigBottom{-2}

\newcommand\new{\newcommand}         

\newcommand{\mt      }{\ensuremath{m_{t}}\xspace}
\newcommand{\mtin    }{\ensuremath{m^{\mathrm{in}}_{t}}\xspace}
\newcommand{\mtou    }{\ensuremath{m^{\mathrm{out}}_{t}}\xspace}
\newcommand{\mlb     }{\ensuremath{m_{lb}}\xspace}
\newcommand{\mtwo     }{\ensuremath{m_{T2}}\xspace}
\newcommand{\mll     }{\ensuremath{m_{ll}}\xspace}
\newcommand{\etdr     }{\ensuremath{E_T^{\Delta R}}\xspace}

\newcommand{\nlofull}{\mathrm{NLO_{full}}}
\newcommand{\nlodec}{\mathrm{NLO_{NWA}^{NLOdec}}}
\newcommand{\lodec}{\mathrm{NLO_{NWA}^{LOdec}}}
\newcommand{\nlops}{\mathrm{NLO_{PS}}}
\newcommand{\lops}{\mathrm{LO_{PS}}}
\newcommand{\lofull}{\mathrm{LO_{full}}}
\newcommand{\lolo}{\mathrm{LO_{NWA}^{LOdec}}}
\newcommand{\Chiq      }{\ensuremath{\chi^{2}}\xspace}
\newcommand{\Oi       }[2]{\ensuremath{{#1}_{#2}}\xspace}
\newcommand{\nprod}{n_\mrm{max}^\mrm{prod}}
\newcommand{\ndec}{n_\mrm{max}^\mrm{dec}}
\newcommand{\muqprod}{\mu_Q^{\mrm{prod}}}
\newcommand{\muqdec}{\mu_Q^{\mrm{dec}}}

\def\bi{\begin{itemize}}
\def\ei{\end{itemize}}
\def\be{\begin{equation}}
\def\ee{\end{equation}}
\def\bea{\begin{eqnarray}}
\def\eea{\end{eqnarray}}

\def\gev{\mathrm{\:GeV}}
\def\tev{\mathrm{\:TeV}}
\def\tsc{\textsc}

\def\mrm{\mathrm}

\new{\as}[1]      {{\ifmmode\alpha^{#1}_s
                    \else$\alpha^{#1}_s$\fi}}
\new{\lqcd}       {{\ifmmode\Lambda_{\mathrm{ QCD}}
                    \else $\Lambda_{\mathrm{ QCD}}$\fi}}

\new{\mpi}{Max-Planck-Institut f\"ur Physik, %
  F\"ohringer Ring 6, %
  80805 M\"unchen, %
  Germany}

\title{NLO and off-shell effects  in top quark mass determinations}

\author[a]{Gudrun~Heinrich,}
\author[b]{Andreas~Maier,}
\author[a]{Richard~Nisius,}
\author[c]{Johannes~Schlenk,}
\author[d]{Markus~Schulze,}
\author[a]{Ludovic~Scyboz,}
\author[e]{Jan~Winter\hphantom{,}}
\affiliation[a]{\mpi}
\affiliation[b]{Experimental Physics Department, CERN, CH-1211 Geneva 23, Switzerland}
\affiliation[c]{IPPP, University of Durham, Durham DH1 3LE, UK}
\affiliation[d]{Humboldt-Universit\"at zu Berlin, Institut f\"ur Physik, Newtonstra\ss e 15, 12489 Berlin, Germany}
\affiliation[e]{Department of Physics and Astronomy, Michigan State
  University, East Lansing, MI 48824, USA}

\emailAdd{gudrun@mpp.mpg.de}
\emailAdd{andreas.alexander.maier@cern.ch}
\emailAdd{nisius@mpp.mpg.de}
\emailAdd{johannes.k.schlenk@durham.ac.uk}
\emailAdd{markus.schulze@physik.hu-berlin.de}
\emailAdd{scyboz@mpp.mpg.de}
\emailAdd{jwinter@pa.msu.edu}

\preprint{{\small%
    MPP-2017-197\\
    \hphantom{.}\hfill IPPP/17/69\\
    \hphantom{.}\hfill HU-EP-17/22\\
    \hphantom{.}\hfill MSUHEP-170922}}


\abstract{We study the impact of different theoretical descriptions of
top quark pair production on top quark mass measurements in the
di-lepton channel.
To this aim, the full NLO corrections to $pp\rightarrow W^+W^-b\bar b\rightarrow
(e^+ \nu_e)\,(\mu^- \bar{\nu}_{\mu})\,b\bar b$ production are compared
to calculations in the narrow width approximation, where the production
of a top quark pair is calculated at NLO and combined with three
different descriptions of the top quark decay: leading order,
next-to-leading order and via a parton shower.
The different theory predictions then enter the calibration of
template fit functions, which are
used for a fit to pseudo-data. The offsets in the top quark mass
resulting from the fits based on the various theoretical descriptions
are determined.}

\begin{document}


\maketitle

\section{Introduction}
\label{sec:intro}


The top quark mass is one of the most important parameters in the
Standard Model (SM). As the top quark features the largest Yukawa coupling, 
it is closely linked to Higgs physics. Furthermore, 
the Higgs potential and therefore the vacuum stability
of the SM depends critically on the value of the top quark mass.
Processes involving top quarks allow for important precision tests of
the SM and appear amongst the dominant backgrounds for 
many New Physics searches. They also allow to further constrain the
gluon PDF at large $x$-values~\cite{Czakon:2013tha,Guzzi:2014wia,delDuca:2015gca,Czakon:2016olj}.

The measurement of the top quark mass is  complicated  due to the fact that
the reconstruction of $t\bar{t}$ events from complex hadronic and
leptonic final states is an arduous task.
Measurements of the top quark mass have been performed in various
channels by the Tevatron and LHC collaborations, where the latest
combinations can be found in Refs.~\cite{CDF-1402,Abazov:2017ktz,ATLAS-CONF-2017-071,Khachatryan:2015hba}.
While the most precise result in the di-lepton channel has an uncertainty of
$0.84\gev$~\cite{Aaboud:2016igd}, the most precise combined results for the
top quark mass achieve a precision of
about $0.5\gev$~\cite{Khachatryan:2015hba,ATLAS-CONF-2017-071}.
The precision achieved nowadays is the result of joint efforts in the experimental as well as the
theory community to reduce the systematic uncertainties inherent to top quark mass measurements.
For recent theoretical studies with regards to the definition and
extraction of the top quark mass, see e.g.~\cite{Frixione:2014ala,Beneke:2016cbu,Butenschoen:2016lpz,Kawabata:2016aya,Hoang:2017suc,Hoang:2017btd,Hoang:2017kmk,Bevilacqua:2017ipv,Nason:2017cxd,Corcella:2017rpt,Ravasio:2018lzi,Hoang:2018ned}.

The theoretical description of top quark pair production at hadron
colliders has improved substantially in recent years.
For stable top quarks, NNLO corrections to differential distributions are
known~\cite{Czakon:2015owf,Czakon:2016dgf,Czakon:2017dip} and have recently 
been combined with
NLO electroweak corrections~\cite{Czakon:2017wor}.
The impact of electroweak corrections on distributions related to
$t\bar{t}$ production has been studied in
Refs.~\cite{Hollik:2011ps,Kuhn:2013zoa,Pagani:2016caq} for on-shell top quarks, and
in Ref.~\cite{Denner:2016jyo} for both the on-shell case and with complete
off-shell effects. Electroweak corrections to multi-jet merged
on-shell top quark pair production have been calculated in Ref.~\cite{Gutschow:2018tuk}.
Due to their very high complexity, the NNLO fixed-order calculations have so far only
been combined with top quark decays in the narrow-width approximation
(NWA), which factorises the production and decay processes.
Radiative corrections to top quark decays have been calculated in
Refs.~\cite{Bernreuther:2004jv,Melnikov:2009dn,Campbell:2012uf}, and
since have been extended up to NNLO QCD~\cite{Brucherseifer:2013iv,Gao:2017goi}.
Resummation also has been accomplished up to NNLL, together with other
improvements going beyond fixed
order~\cite{Beneke:2011mq,Cacciari:2011hy,Ferroglia:2013awa,Broggio:2014yca,Kidonakis:2015dla,Pecjak:2016nee}.

However, a description of top quark pair production and decay which
predicts the {\em shapes} of distributions with an accuracy required for
improvements on the current experimental precision needs to go beyond the
narrow-width approximation.
NLO QCD calculations of $W^+W^- b\bar{b}$ production, including leptonic
decays of the $W$ bosons,  have been performed in
Refs.~\cite{Denner:2010jp,Denner:2012yc,Bevilacqua:2010qb,Heinrich:2013qaa}. 
These calculations use the 5-flavour scheme, where
the $b$-quarks are treated as massless partons.
In Ref.~\cite{Heinrich:2013qaa}, particular emphasis has been put on the
impact of the non-factorising contributions on the top quark mass
measurements. 
Recently the calculation of the NLO QCD corrections to $W^+W^- b\bar{b}$ production with
full off-shell effects has also been achieved in the lepton plus jets channel~\cite{Denner:2017kzu}.

The $b$-quark mass effects on observables like the invariant mass of a
lepton-$b$-quark pair  ($m_{lb}$)  are very small. 
However, the use of 
massive $b$-quarks (more precisely, the 4-flavour scheme, 4FNS)  has the
(technically) important feature that it avoids collinear singularities
due to $g\to b\bar{b}$ splittings. This implies that any phase space restrictions on
the $b$-quarks can be made without destroying infrared safety, and
thus allows  to consider 0, 1- and 2-jet bins for 
$pp\to e^+\nu_e\mu^-\bar{\nu}_\mu b\bar{b}$ in one and the same setup, 
which is important for cross sections defined by jet vetos.
In Refs.~\cite{Frederix:2013gra,Cascioli:2013wga}, NLO calculations in the 4FNS have been performed.

The next step in complexity towards a realistic description of the
measured final states consists in combining fixed-order calculations with a parton shower. 
The effect of radiative corrections to both, production and decay, in the factorised
approach matched to a parton shower has been investigated in Ref.~\cite{Campbell:2014kua}
within an extension of the {\tt PowHeg}~\cite{Nason:2004rx,Frixione:2007vw,Alioli:2010xd} framework, called {\tt ttb\_NLO\_dec}
in the {\tt POWHEG-BOX-V2}.
Within the {\tt Sherpa} framework, NLO QCD predictions for top quark
pair production with up to three jets matched to a parton shower are
also available, see Refs.~\cite{Hoeche:2014qda,Hoche:2016elu}.
A new NLO multi-jet merging algorithm relevant to top quark pair
production is also available in {\tt Herwig\,7.1}~\cite{Bellm:2017idv}.

Based on an NLO calculation of $W^+W^- b\bar{b}$ production combined
with the {\tt Powheg} framework, first results of the $W^+W^-
b\bar{b}$ calculation in the 5-flavour scheme matched to a parton
shower have been presented in Ref.~\cite{Garzelli:2014dka}. 
However, it has been noticed later that the matching of NLO matrix elements
involving resonances of coloured particles to parton showers poses
problems which can lead to artefacts in the top quark lineshape~\cite{Jezo:2015aia}.
As a consequence, an improvement of the resonance treatment has been
implemented in {\tt POWHEG-BOX-RES}, called ``resonance aware matching'', 
and combined with NLO matrix elements from OpenLoops~\cite{Cascioli:2011va}, to arrive at 
the most complete description so far~\cite{Jezo:2016ujg}, based on the
framework developed in Ref.~\cite{Jezo:2015aia} and the 4FNS calculation of
Ref.~\cite{Cascioli:2013wga}. An alternative algorithm to treat
radiation from heavy quarks in the {\tt Powheg} NLO+PS framework has been
presented in Ref.~\cite{Buonocore:2017lry}.
An improved resonance treatment in the matching to parton showers for off-shell single top production at NLO
has been worked out in Refs.~\cite{Jezo:2015aia,Frederix:2016rdc}, and similarly for off-shell
$t\bar{t}$ and $t\bar{t}H$ production in $e^+e^-$ collisions in Ref.~\cite{Nejad:2016bci}.

\medskip

In this paper, we investigate the impact of different
approximations on the top quark mass measurement simulating a concrete experimental setup. 
In particular, we follow up on an open question raised in Ref.~\cite{Heinrich:2013qaa}, where 
we performed a study of NLO effects in top quark mass
measurements based on the observable \mlb\ in the framework of a top quark mass measurement as performed by ATLAS
using the template method~\cite{Aad:2015nba,Aaboud:2016igd}. 
Substantial distortions in the \mlb\ distribution are induced by scale variations 
calculated by including  the full NLO corrections to the $W^+W^- b\bar{b}$ final state
(with leptonic $W$-decays). On the other hand, in the factorised
approach, where the $t\bar{t}$ cross section calculated at NLO is
combined with  LO top quark decays in the NWA, the shape distortions
due to the scale variations  are minor. 
As the experimental analysis is based on normalised distributions, the
shape differences induced by scale variations translate in a very
sensitive manner into the theoretical uncertainties on the extraction
of the top quark mass.

The question arises where the shape changes come from, i.e. whether they mainly
come from the non-factorisable contributions contained in the full NLO
corrections to $W^+W^- b\bar{b}$, or from factorisable NLO corrections to the top quark decay. 
And, if the latter is true, what is the effect of a parton shower in
combination with the factorised approach, as it should contain the
leading contributions of the NLO corrections to the top
quark decay.
To answer these questions, we compare the NLO calculation of $W^+W^- b\bar{b}$
production of Ref.~\cite{Heinrich:2013qaa} with the calculation based
on the narrow-width approximation 
where both $t\bar{t}$ production {\it and} decay are calculated at NLO,
as described in Ref.~\cite{Melnikov:2009dn}.
We further quantify the impact of a parton shower in the narrow-width
approximation, combining the NLO matrix elements of top quark pair
production with \tsc{Sherpa}~\cite{Gleisberg:2008ta}.

The structure of this paper is as follows. 
In Section~\ref{sec:calculation}, we describe our different
calculations performed to compare theoretical
descriptions of the complex final state of two charged leptons, two
$b$-jets and missing energy. In Section~\ref{sec:results}, we 
compare these different theoretical descriptions for a number of
observables relevant to top quark mass measurements. 
We then quantify in Section \ref{sec:fit} how the differences in
the theoretical descriptions impact a template fit as utilised
in experimental determinations
of the top quark mass, before we conclude in Section~\ref{sec:conclusions}.


\section{The different stages of the theoretical description}
\label{sec:calculation}

We study the following descriptions of top quark pair production in the di-lepton channel:
\begin{description}
\item[~~~~~$\mathbf{NLO_{full}}$:]
  full NLO corrections to $pp\to W^+W^- b\bar{b}$ with leptonic $W$-decays,
\item[~~~~~$\mathbf{NLO_{NWA}^{NLOdec}}$:]
  NLO $t\bar{t}$ production $\otimes$ NLO decay,
\item[~~~~~$\mathbf{NLO_{NWA}^{LOdec}}$:]
  NLO $t\bar{t}$ production $\otimes$ LO decay,
\item[~~~~~$\mathbf{NLO_{PS}}$:]
  NLO $t\bar{t}$ production+shower $\otimes$ decay via parton showering.
\end{description}
We furthermore use the abbreviation $\mathbf{LO_{full}}$
for $W^+W^- b\bar{b}$ calculated at leading order, the abbreviation
$\mathbf{LO_{NWA}^{LOdec}}$ for LO $t\bar{t}$ production $\otimes$ LO
decay and the abbreviation $\mathbf{LO_{PS}}$ for LO $t\bar{t}$
production $\otimes$ decay via parton showering.
We investigate the effects of different levels in the description
of the top quark decay,
isolating the latter from the effects of the non-resonant and
non-factorisable contributions contained in the $\nlofull$ calculation.
This is done by emulating a concrete experimental analysis used for top
quark mass determinations.
As we match to a parton shower only in combination with
LO top quark decays, we do not need to address the problem of
``resonance-aware matching''~\cite{Jezo:2016ujg,Frederix:2016rdc}.
This allows us to get a clear idea of the effects of the various
approximations used here, which in turn can serve as a basis for
future studies entirely relying on showered results.

\begin{figure}[tbp!]
  \centering
  \begin{subfigure}[b]{0.4\textwidth}
    \centering
\includegraphics[width=1\textwidth]{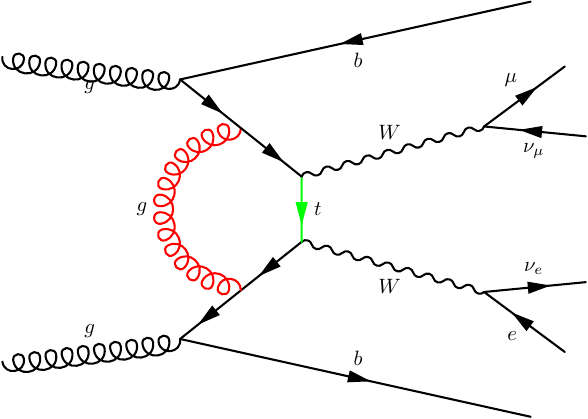}
  \end{subfigure}
  \hskip10mm
  \begin{subfigure}[b]{0.4\textwidth}
    \centering
\includegraphics[width=1\textwidth]{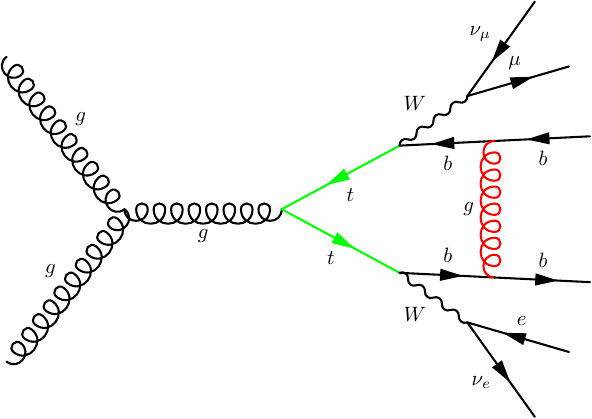}
  \end{subfigure}
  \caption{\label{fig:wwbb_diagrams}
    Examples of one-loop Feynman diagrams contributing to the
    $\nlofull$ calculation, i.e.~the full
    $W^+W^-b\bar{b}$ calculation at NLO: two diagrams are shown
    depicting a non-resonant (left) and a   
    non-factorisable virtual contribution (right).}
\end{figure}

The calculations $\nlofull$ and $\lodec$ have been already described
in detail in Ref.~\cite{Heinrich:2013qaa}.\footnote{%
  In Ref.~\cite{Heinrich:2013qaa}, $\nlofull$ was called
  $W^+W^- b\bar{b}$ and $\lodec$ was called $t\bar{t}$.}
Here we briefly summarise only the main features. We use
\tsc{GoSam}~\cite{Cullen:2011ac,Cullen:2014yla} plus
\tsc{Sherpa}~\cite{Gleisberg:2008ta}, version 2.2.3, where the virtual
corrections generated by \tsc{GoSam} are linked to \tsc{Sherpa} via
the Binoth-Les-Houches-interface~\cite{Binoth:2010xt,Alioli:2013nda}.
This applies not only to the calculations $\nlofull$ and $\lodec$ but
also to the $\nlops$ computation. We note that our full NLO
calculation of the process $pp\rightarrow W^+W^-b\bar b\rightarrow
(e^+ \nu_e)\,(\mu^- \bar{\nu}_{\mu})\,b\bar b$ provides a complete
description of the final state including singly-resonant and
non-resonant top quark contributions. Example diagrams are shown in
Fig.~\ref{fig:wwbb_diagrams}. The computation relies on the
5-flavour scheme, i.e.~the $b$-quark is treated as massless.
To take the top quark decay width into account in a gauge invariant
way, the complex mass scheme~\cite{Denner:2006ic} is used. In our
setup, this entails a replacement of the top quark mass by a
complex number $\mu_t$ evaluated according to
\begin{equation}
\mu_t^2\;=\;m_t^2-i\,m_t\,\Gamma_t~.
\label{eq:cms}
\end{equation}
The $W$-bosons and intermediate $Z$-bosons also have complex masses due to their widths.
Note that we only consider resonant $W$-boson decays.

The results for the $\nlodec$ calculation are obtained
as described in Ref.~\cite{Melnikov:2009dn}.\footnote{%
  The corresponding Monte Carlo generator is publicly available at
  {\tt https://github.com/ TOPAZdevelop/TOPAZ}.}
This framework relies on the factorisation of the matrix elements according to
\begin{eqnarray}\label{eqn:NWA1}
   \mathcal{M}^\mathrm{NWA}_{ij\to t\bar{t}\to b \bar{b} 2l 2\nu}
   \;=\;
   \mathcal{P}_{ij \to t\bar{t}} \otimes \mathcal{D}_{t\to b l^+ \nu}
   \otimes \mathcal{D}_{\bar{t}\to \bar{b} l^- \bar{\nu}}~,
\end{eqnarray}
where $\mathcal{P}_{ij \to t\bar{t}}$ describes the $t\bar{t}$ production process and $\mathcal{D}_{t\to b l \nu}$ the top quark decay dynamics.
Spin correlations are included as indicated by the symbol $\otimes$.
Squaring Eq.~(\ref{eqn:NWA1}) and integrating over the phase space yields the
double-resonant partonic cross section
\begin{eqnarray}\label{eqn:sigmaNWA}
\hat{\sigma}_{ij\to b \bar{b} 2l 2\nu}\;=\;\int \!\! \mathrm{d}\mathrm{PS} \; |\mathcal{M}^\mathrm{NWA}_{ij\to t\bar{t}\to b \bar{b} 2l 2\nu}|^2
+ \mathcal{O}\left(\Gamma_t\big/ m_t\right)
\end{eqnarray}
where off-shell effects are parametrically suppressed by $\Gamma_t \big/ m_t \approx 0.7 \%$.
Expanding Eq.~(\ref{eqn:NWA1}) up to NLO yields
\begin{eqnarray}\label{eqn:NWA2}
   \mathcal{M}^\mathrm{NWA,\; NLO}_{ij\to t\bar{t}\to b \bar{b} 2l 2\nu}
   =&&
   \mathcal{P}_{ij \to t\bar{t}}^\mathrm{LO} \otimes \mathcal{D}^\mathrm{LO}_{t\to b l^+ \nu} \otimes \mathcal{D}^\mathrm{LO}_{\bar{t}\to \bar{b} l^- \bar{\nu}}
  +\mathcal{P}_{ij \to t\bar{t}}^{\delta \mathrm{NLO}} \otimes \mathcal{D}^\mathrm{LO}_{t\to b l^+ \nu} \otimes \mathcal{D}^\mathrm{LO}_{\bar{t}\to \bar{b} l^- \bar{\nu}}
    \nonumber  \\
   +&&
   \mathcal{P}_{ij \to t\bar{t}}^\mathrm{LO} \otimes \left( \mathcal{D}^{\delta \mathrm{NLO}}_{t\to b l^+ \nu} \otimes \mathcal{D}^\mathrm{LO}_{\bar{t}\to \bar{b} l^- \bar{\nu}}
   +\mathcal{D}^\mathrm{LO}_{t\to b l^+ \nu} \otimes \mathcal{D}^{\delta \mathrm{NLO}}_{\bar{t}\to \bar{b} l^- \bar{\nu}} \right)~.
\end{eqnarray}
The NLO corrections to the production process $\mathcal{P}_{ij \to t\bar{t}}^{\delta \mathrm{NLO}}$ involve the virtual and real emission matrix elements
$\mathcal{M}_{gg/q\bar{q} \to t\bar{t}}^{\mathrm{virt}}$, $\mathcal{M}_{gg/q\bar{q} \to t\bar{t}+g}^{\mathrm{real}}$ and
$\mathcal{M}_{qg/\bar{q}g \to t\bar{t}+q/\bar{q}}^{\mathrm{real}}$.
The corresponding NLO decay parts are given by
\begin{eqnarray}\label{eqn:NWAdecay}
\mathcal{D}^\mathrm{virt (real)}_{t\to b l \nu}\;=\;\frac{\mathcal{M}^\mathrm{virt (real)}_{t\to b W (+g)}}{\sqrt{2 m_t \Gamma_t^\mathrm{NLO}}} \otimes
\frac{\mathcal{M}_{W \to l \nu}}{\sqrt{2 M_W  \Gamma_W^\mathrm{NLO}}}~.
\end{eqnarray}
We note that, in contrast to Ref.~\cite{Melnikov:2009dn}, the top quark width $\Gamma_t^\mathrm{NLO}$ in the denominator is not expanded
as $(\Gamma_t^\mathrm{NLO})^{-1/2}=(\Gamma_t^\mathrm{LO})^{-1/2}
\left( 1 - 1/2 \, \alpha_s \, \Gamma_t^{\delta
    \mathrm{NLO}}/\Gamma_t^\mathrm{LO} \right) $, analogously for $(\Gamma_W^\mathrm{NLO})^{-1/2}$.

For our studies relying on $\lodec$ results, we remove all
contributions in the second line of Eq.~(\ref{eqn:NWA2}) and use
$\Gamma_{t,W}^\mathrm{LO}$ instead of $\Gamma_{t,W}^\mathrm{NLO}$. This
treatment guarantees that
$\int \! \mathrm{d}\mathrm{PS} \; |\mathcal{D}_{t\to b l \nu}|^2 =
\mathrm{BR}(t\to bl\nu)$ at LO and NLO, with $\mathrm{BR}(t\to bl\nu)$
denoting the branching ratio for the top quark decay.

Finally, the $\nlops$ computations are based on the
NLO plus parton-shower matching scheme as implemented in
\tsc{Sherpa}~\cite{Hoeche:2011fd}. The original scheme was extended in
Ref.~\cite{Hoeche:2013mua} to incorporate heavy-quark mass effects.
Utilising this scheme, we obtain an NLO+PS accurate description of
$t\bar t$+jets, or, in other words, the NLO description of the
$t\bar t$ production shower. The top quark decays are attached
afterwards such that LO spin correlations are preserved, and each decay
configuration is supplemented by its respective decay shower following
the same procedure as described in Ref.~\cite{Hoche:2014kca}.
For our investigations, we used \tsc{Sherpa}~version~\tsc{2.2.3}.
In the course of this work, it was found that this version treats
radiation emerging from top quark decays in resonant top quark
processes in the same manner as radiation arising from continuum
production processes. This resulted in an omission of the
initial-state spectator mass term that suppresses the ordinary eikonal
radiation of continuum initial-final dipoles.
The problem has been identified and solved by
the implementation of a dedicated dipole-shower algorithm for the
decays, similar to Ref.~\cite{Hamilton:2006ms}. The patch implementing
these changes has been provided by the \tsc{Sherpa} authors and was
used for our results presented below. It will be made available on the
corresponding software download pages, and included in the
\tsc{Sherpa} program from version~\tsc{2.2.5} onwards.


\section{Phenomenological study of observables sensitive to the top
  quark mass}
\label{sec:results}


\subsection{Definition of the observables}

We study the following observables:
\begin{itemize}

\item $m_{lb}$ -- which we define using the invariant mass squared
  \begin{equation}\label{def:mlb}
    \mlb^2\;=\;(p_l+p_b)^2
  \end{equation}
  where $p_l$ denotes the four-momentum of the lepton and $p_b$ the
  four-momentum of the $b$-jet. As there are two top quarks, there are also
 two possible \mlb\ values per event. Since experimentally, it is not possible to reconstruct
 the $b$-quark charge on an event-by-event basis with sufficient accuracy, one also needs a
 criterion to assign a pair of a charged lepton and a $b$-jet
 as the one stemming from the same top quark decay.
Following~\cite{Aaboud:2016igd}, the algorithm applied here is to choose
that $(l^+b\text{-jet},\,l^-b\text{-jet}')$ pairing which
 minimises the sum of the two \mlb\ values per event.
 Finally, the \mlb\ observable used in the analysis is the mean of the two \mlb\
 values per event obtained when applying the above procedure.

\item $m_{T2}$ -- which corresponds to the kinematic variable
  $m_{T2}$~\cite{Lester:1999tx,Barr:2003rg} that, applied to the
  $b\bar{b} 2l 2\nu$ final state, is defined as
  \begin{equation}\label{def:MT2}
    m_{T2}^2\;=\;\min_{\mathbf{p}_T^{\nu_1}+\mathbf{p}_T^{\nu_2}=\mathbf{p}_T^{\mrm{miss}}} \left[ \max \left\{ m_{T}^2 \left( \mathbf{p}_T^{(lb)_1}, \mathbf{p}_T^{\nu_1} \right), m_{T}^2 \left( \mathbf{p}_T^{(lb)_2}, \mathbf{p}_T^{\nu_2} \right) \right\}  \right].
  \end{equation}
Again the pairing of leptons and $b$-jets which minimises
$m_{(lb)_1}+m_{(lb)_2}$ is chosen.
The transverse mass is given by
\begin{equation}
m_T^2\left(\mathbf{p}_T^{(lb)_i}, \mathbf{p}_T^{\nu_i}\right) = m_{(lb)_i}^2 + 2 \left( E_T^{(lb)_i} E_T^{\nu_i} - \mathbf{p}_T^{(lb)_i} \mathbf{p}_T^{\nu_i} \right)\nonumber
\end{equation}
with $E_T = \sqrt{|\mathbf{p}_T|^2+m^2}$, where $m_{\nu_i} = 0$ was used.

\item $E_T^{\Delta R}$ -- which is defined as
  \begin{equation}\label{def:ETdR}
    E_T^{\Delta R}\;=\;\frac{1}{2}\,\left(E_T^{l_1}\Delta R(l_1,b_1)+E_T^{l_2}\Delta R(l_2,b_2)\right)
  \end{equation}
  using the above pairing criterion for leptons and $b$-jets.
  
\end{itemize}

We also consider the following fully leptonic observables, which are
part of the set of observables used for a recent top quark mass
determination from differential leptonic cross sections in Ref.~\cite{ATLAS-CONF-2017-044}:
\begin{itemize}

\item $m_{ll}$ -- as given by the invariant mass squared of the two charged
  leptons, defined as
  \begin{equation}\label{def:mll}
    \mll^2\;=\;(p_{l_1}+p_{l_2})^2~.
  \end{equation}
 
\item $p_{T,\mu}$ -- which is the transverse momentum of the muon.

\item $\eta_{\mu}$ -- which is the rapidity of the muon.

\end{itemize}

\subsection{Input parameters and event requirements}
\label{subsec:input}

We use the  PDF4LHC15\_nlo\_30\_pdfas
sets~\cite{Butterworth:2015oua,Dulat:2015mca,Harland-Lang:2014zoa,Ball:2014uwa}
and a centre-of-mass energy of $\sqrt{s}=13\tev$.
Our default top quark mass is  $m_t=172.5\gev$.
Leading order top quark and $W$ boson widths are used in the LO calculations and the NLO $t\bar{t}\, \otimes$ LO decay calculation, while NLO widths~\citep{Jezabek:1987nf} are used in the remaining NLO calculations.
Widths at NLO appearing in propagators are not expanded in \as{}.
The QCD coupling in the NLO widths is varied according to the chosen
scale. For \as{} evaluated at the central scale $m_t$, the numerical values for the widths are
\begin{equation}
\begin{matrix}
\Gamma_t^\mrm{LO} &=& 1.4806\gev{}&,&\quad \Gamma_t^\mrm{NLO} &=& 1.3535\gev{},\\
\Gamma_W^\mrm{LO} &=& 2.0454\gev{}&,&\quad \Gamma_W^\mrm{NLO} &=& 2.1155\gev{},\\
\Gamma_Z &=& 2.4952\gev{}&.& &&\\
\end{matrix}
\end{equation}
Jets are defined using the anti-$k_T$ algorithm~\cite{Cacciari:2008gp} as implemented in 
{\tt Fastjet}~\cite{Cacciari:2011ma}, with $R=0.4$.
For the electroweak parameters, we employ the following settings:
\begin{equation}\label{eq:ewparam}
  G_{\mu}\;=\;1.16637\cdot10^{-5}\gev{}^{-2},\quad
  M_{W} \;=\;80.385\gev{},\quad
  M_{Z} \;=\;91.1876\gev{}.
\end{equation}

Inspired by Ref.~\cite{Aaboud:2016igd}, and taking into account the
stronger trigger requirements for a $13\tev$ analysis, the following list
of event requirements is used. We require
\begin{itemize}
\item
  exactly two $b$-tagged jets with $p_T^\mrm{jet}>25\gev$ and
  $|\eta^\mrm{jet}|<2.5$. Jets containing a $b\bar{b}$ pair are also
  defined as $b$-jets.
\item
  exactly two oppositely charged leptons which fulfill
  $p_T^{\mu}>28\gev$, $|\eta^{\mu}|<2.5$ for muons and
  $p_T^{e}>28\gev$, $|\eta^{e}|<2.47$ for electrons excluding the
  range $1.37<|\eta^{e}|<1.52$.
  For both types of charged leptons with respect to any jet fulfilling
  the jet requirements, a separation of $\Delta R(l,\mrm{jet})>0.4$ is
  required.
\item
  $p_{T}^{lb}>120\gev$. Using the same lepton $b$-jet assignments as
  for $m_{lb}$, the observable $p_{T}^{lb}$ denotes the mean
  transverse momentum of the two lepton--$b$-quark systems.
\end{itemize}

The $b$-quarks are treated as massless in all fixed-order calculations.
We chose $\mu_R=\mu_F=m_{t}$ as our central scale. The impact of
choosing $H_T/2$ (rather than $m_t$) as the central scale on the top
quark mass determined by our method has been shown to be very
small~\cite{Heinrich:2013qaa}. It furthermore would be difficult to
facilitate an $H_T$ definition for the $\nlops$ approach that matches
the one used in the $\nlofull$ calculation. Even a simplified
$H_T$ definition that involved only the charged lepton and $b$-jet
transverse momenta and neglected the neutrino momenta would be
affected because the parton showering changes the $p_T$ spectrum of
the final state particles. We therefore considered it more consistent
to choose $m_t$ as the central renormalisation and factorisation scale
throughout all calculations. The scale variation bands are obtained by
varying $\mu_R$ and $\mu_F$ simultaneously by a factor of two and one
half with respect to the central scale.
We have also performed 7-point scale variations and found that the 
simultaneous variations always formed the most conservative
uncertainty band in the $\mlb$ and $\mtwo$ distributions, Figs.~\ref{fig:scalevar_mlb} and \ref{fig:scalevar_all_mt2}.

For the parton shower results, we have also investigated the impact of
a dynamic scale, which we call $\mu_{t\bar t}$, 
to compute the matrix elements of the hard
scattering processes producing the top quark pairs.
The scale $\mu_{t\bar t}$ is a ``colour flow inspired'' QCD scale,  introduced
in Ref.~\cite{Hoeche:2013mua}. Using Mandelstam variables $s$, $t$ and
$u$, it is defined as
\begin{align}
  \mu_{t\bar t}^2(q\bar q\to t\bar t)&\;=\;2\,p_q p_t\;=\;m_t^2-t~,\nonumber\\[1mm]
  \mu_{t\bar t}^2(\bar qq\to t\bar t)&\;=\;2\,p_q p_t\;=\;m_t^2-u~,\nonumber\\[1mm]
  \mu_{t\bar t}^2(gg\to t\bar t)&\;=\;\left\{\begin{array}{ccl}
  m_t^2-t&&w_1\varpropto\,\frac{u-m_t^2}{t-m_t^2}+
    \frac{m_t^2}{m_t^2-t}\left\{\frac{4\,t}{t-m_t^2}+\frac{m_t^2}{s}\right\}\\
  &\text{with weight}&\\
  m_t^2-u&&w_2\varpropto\,\frac{t-m_t^2}{u-m_t^2}+
    \frac{m_t^2}{m_t^2-u}\,\left\{\frac{4\,u}{u-m_t^2}+\frac{m_t^2}{s}\right\}~.
  \end{array}\right.
  \label{eq:scale_ttb}
\end{align}
The value for the $gg$ partonic process is chosen randomly
according to the relative size of the two weights $w_1$ and $w_2$.

\begin{table}[tb!]
\begin{center}
\begin{tabular}{|l|l|l|}
\hline&&\\[-3mm]
\multicolumn{1}{|c|}{\textbf{Scheme}} &
\multicolumn{1}{|c|}{\textbf{\boldmath Central scale $\mu_i$}} &
\multicolumn{1}{|c|}{\textbf{\boldmath Variations $\xi_i\,\mu_i$}}\\[1mm]
\hline&&\\[-5pt]
$\mu_F\mu_R\as{\mathrm{PS}}$ &
$\mu_F=\mu_R=\muqprod=\mt$, $\mu_R^\mrm{PS}=p_T^\mrm{emit}$ &
$\xi_R=\xi_F=\xi_R^\mrm{PS}=\{0.5,1.0,2.0\}$\\
&$\mu_F=\mu_R=\muqprod=\mu_{t\bar{t}}$, $\mu_R^\mrm{PS}=p_T^\mrm{emit}$&\\
&&\\[-6pt]
\hline&&\\[-5pt]
$\mu_F\mu_R\mu_Q$ &
$\mu_F=\mu_R=\muqprod=\mt$, $\mu_R^\mrm{PS}=p_T^\mrm{emit}$ &
$\xi_R=\xi_F=\{0.5,1.0,2.0\}$ and\\&&
$\hphantom{\xi_R=\,}\xi_Q=\{\sqrt2,1.0,1/\sqrt2\}$\\
&&\\[-6pt]
\hline&&\\[-5pt]
$\mu_F\mu_R\mu_Q\as{\mathrm{PS}}$ &
$\mu_F=\mu_R=\muqprod=\mt$, $\mu_R^\mrm{PS}=p_T^\mrm{emit}$ &
$\xi_R=\xi_F=\xi_R^\mrm{PS}=\{0.5,1.0,2.0\}$\\&&
and $\xi_Q=\{\sqrt2,1.0,1/\sqrt2\}$\\[-5pt]
&&\\
\hline
\end{tabular}
\end{center}
\caption{\label{tab:scalevars_nlops}%
  Summary of schemes employed to assess the scale variations in the
  $\nlops$ case. In all cases, the decay shower starting scale is kept
  constant, as given by its default $\muqdec=M_W/2$. Note that the
  local $p_T^\mrm{emit}$ values (squared) of each parton branching
  serve as the argument of $\as{}$ in the evaluation of the shower
  kernels. The related variation, $\xi_R^\mrm{PS}\,\mu_R^\mrm{PS}$,
  has been realised through appropriate adjustments of the
  \tsc{Sherpa} parameters {\tt CSS\_IS\_AS\_FAC} and {\tt CSS\_FS\_AS\_FAC}.}
\end{table}

The standard $\mu_R$ and $\mu_F$ variations that we employ for our
fixed-order calculations are not fully appropriate to assess the
theory uncertainties of the $\nlops$ computations, as the
showering depends on further scale and parameter choices. For our
studies, it is interesting to vary $\muqprod$ as well as
$\muqdec$, which are the parameters controlling the overall size of
the resummation domains assigned to the $t\bar t$ production and top
quark decay showers, respectively. 
Within these resummation domains, subsequent shower emissions are
evaluated from the values taken by the ordering variable of the parton
shower.
We therefore also alter the strength
of the parton shower emissions by variations of 
$\mu_R^\mrm{PS}$, the scale entering the evaluation of the strong coupling
$\as{}(\mu_R^\mrm{PS})$ used in the shower kernels. For the
\tsc{Sherpa}~\tsc{CSshower}, the ordering variable is associated with
the local $p_T^\mrm{emit}$ scales of the individual branchings, which
means $\mu_{R,k}^\mrm{PS}\sim p_{T,k}^\mrm{emit}$ for the $k$-th branching.
For the combined variation of several
$\nlops$ parameters, we follow the principle of identifying the
strongest and weakest shower option that one can possibly obtain from
the given individual parameter ranges. This is supposed to lead to a
conservative shower uncertainty estimate.

Our default variation in the $\nlops$ case, denoted by
$\mu_F\mu_R\as{\mathrm{PS}}$, is a
combination of simultaneously varying $\mu_F, \mu_R$ and $\mu_R^\mrm{PS}$ by a factor of
two up and down, with central scale $\mt$.
Alternative ways of uncertainty assessment include the variation of $\muqprod$ and $\muqdec$. 
The \tsc{Sherpa} default is to set  $\muqprod$  equal to
the factorisation scale, while the starting scale of the decay shower
is set to $\muqdec=M_W/2$ and not varied.\footnote{%
  The top quark decays induce a deflection of the colour flow of the
   top quark. The scale of the deflection on average
  corresponds to the mass of the $W$ boson, which therefore serves as
  an appropriate choice for the scale associated with the
  first decay shower branching.}
The different scale variation schemes, which are used by us in the $\nlops$ case are summarised in Table~\ref{tab:scalevars_nlops}.
For each of the schemes shown in Table~\ref{tab:scalevars_nlops}, the uncertainty bands are defined as
the maximum deviation from the central prediction on either side.

\subsection{Numerical predictions}

\subsubsection{Comparison of the different theoretical descriptions}

\begin{table}
\centering
\begingroup
\def\arraystretch{1.3}
\begin{tabular}{|c|c|c|}
\hline
& X=LO $\left[\mathrm{fb}\right]$ & X=NLO $\left[\mathrm{fb}\right]$\\
\hline
$\mathbf{X_{full}}$ & $\left( 739.5\pm   0.3\right)^{ + 31.5\%}_{-22.4\%}$ & $\left( 914\pm   3\right)^{ +  2.1\%}_{ -7.6\%}$\\
$\mathbf{X_{NWA}^{LOdec}}$& $\left( 727.3\pm   0.2\right)^{ + 31.4\%}_{-22.3\%}$ & $\left( 1029\pm   1\right)^{ + 10.4\%}_{-11.5\%}$\\
$\mathbf{X_{NWA}^{NLOdec}}$ & - & $\left( 905\pm   1\right)^{ +  2.3\%}_{ -7.7\%}$\\
$\mathbf{X_{PS}}, \mu=m_t$ & $\left( 637.7\pm   0.9\right)^{ + 29.7\%}_{-21.0\%}$ & $\left( 886\pm   1\right)^{ +  8.5\%}_{ -9.3\%}$\\
$\mathbf{X_{PS}}, \mu=\mu_{t\bar{t}}$ & $\left( 499.7\pm   0.7\right)^{ + 27.6\%}_{-19.3\%}$ & $\left( 805.2\pm   0.9\right)^{ + 12.3\%}_{-10.9\%}$\\
\hline
\end{tabular}
\endgroup
\caption{\label{tab:xs}%
  Fiducial cross sections in various approximations. The first
  uncertainty is the precision of the Monte Carlo phase space integration.
  The scale variation uncertainty obtained by simultaneously varying
  renormalisation and factorisation scales by a factor of two
  (superscript) and one half (subscript) is given in percent.
  For the parton shower results, the  given scale uncertainties are
  obtained by using the variation prescription $\mu_F\mu_R\as{\mathrm{PS}}$
  as detailed in the text.}
\end{table}

In this section, we compare four different NLO descriptions of the $(e^+ \nu_e)\,(\mu^- \bar{\nu}_{\mu})\,b\bar b$ final state
for the  observables described in Section~\ref{sec:calculation}.
Some of the purely leptonic observables have also been used by the ATLAS collaboration
for their recent top quark mass determinations based on $8\tev$ data presented in Ref.~\cite{ATLAS-CONF-2017-044}.
Aiming to quantify the relative differences of the theoretical descriptions,
which should only mildly depend on the centre-of-mass energy,
we show results at the present LHC setting of $13\tev$.
The corresponding fiducial cross sections are summarised in Table~\ref{tab:xs}.
While the level of agreement between the fixed-order full and NWA
calculations is as expected, considerably smaller cross
sections are obtained for the parton shower calculations. Showering leads to a softening
of the final state $b$-jets. In turn, a good fraction of them no longer
satisfy the jet requirements, resulting in an event loss.
The parton shower computation with  $\mu=\mu_{t\bar t}$ leads to an
even smaller fiducial cross section than the computation relying on 
$\mu=m_t$, which is a consequence of the fact that the $\mu_{t\bar t}$
scale is larger, and therefore the value for $\alpha_s$ is smaller. 
In both cases, however, the loss of events due to insufficiently
energetic $b$-jets after parton showering is similar, and amounts to
about 12\%.

\begin{figure}[tbp!]
\centering
\includegraphics[width=0.8\textwidth]{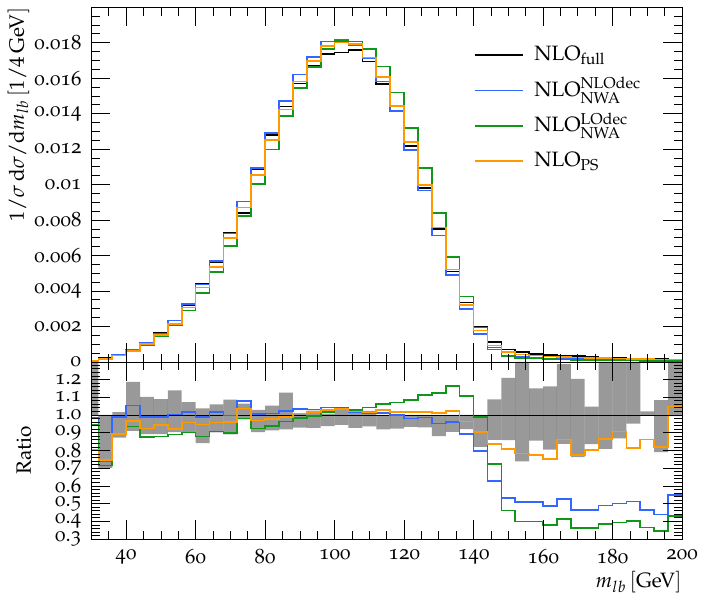}
\caption{\label{fig:scalevar_mlb}%
  Normalised differential cross sections for the invariant mass
  $m_{lb}$ at the $13\tev$ LHC for four different theoretical
  descriptions: $\nlofull$, $\nlodec$, $\lodec$ and $\nlops$. The ratios
  of all descriptions to $\nlofull$ including its scale uncertainty
  band are also shown.}
\end{figure}

In Fig.~\ref{fig:scalevar_mlb}, we present the normalised differential cross sections for $m_{lb}$ based on the four theoretical descriptions,
evaluated at  $\mu_R=\mu_F=m_t$.
In the lower part of the figure, we show their ratio to the $\nlofull$ prediction, including an uncertainty band from
scale variations by a factor of two and one half with respect to the central scale.
We find that 99\% of the total fiducial cross section is accumulated in the range $40$--$150\gev$.
A kinematic edge at $m_{lb}^\mrm{edge}=\sqrt{m_t^2-M_W^2}=152.6\gev$ leads to a sharp drop in the distribution
beyond which it is only populated by non-resonant contributions, additionally clustered radiation and incorrect $b$-lepton pairings.
The significantly larger scale uncertainty for $m_{lb}\ge150\gev$ is due to the fact that NLO is the first non-trivial order populating this region.
This conclusion is further substantiated by the sizeable perturbative correction that we discuss in the following section.
Hence, resummation effects are expected to play a larger role in the vicinity of this kinematic boundary.

We now discuss the impact of off-shell and non-resonant contributions on the $m_{lb}$ distribution.
Their effect is easiest seen by discussing $\nlodec$, displayed in the lower part of Fig.~\ref{fig:scalevar_mlb}.
In the range $30\gev\le\mlb\le130\gev$ this prediction agrees with the full calculation to within a few percent.
The deviations are barely visible within the statistical fluctuations. Around the peak region of the differential cross section for $m_{lb}$, the
NWA calculation overshoots by about 4\%.
This level of agreement is to be expected given the parametric suppression of off-shell effects by $\Gamma_t \big/ m_t$,
which is mildly violated by the applied phase space restrictions.
For $m_{lb}\ge130\gev$, the difference between $\nlodec$ and $\nlofull$ starts to grow
and saturates at about $-50$\% for $\mlb$ values larger than $m_{lb}^\mrm{edge}$.
Again, this is to be expected as the NWA does not apply in this part of the phase space.
In fact, the $\lolo$ prediction (not shown in Fig.~\ref{fig:scalevar_mlb}) vanishes for $m_{lb}\ge m_{lb}^\mrm{edge}$.

\begin{figure}[tbp!]
\centering
\includegraphics[width=0.8\textwidth]{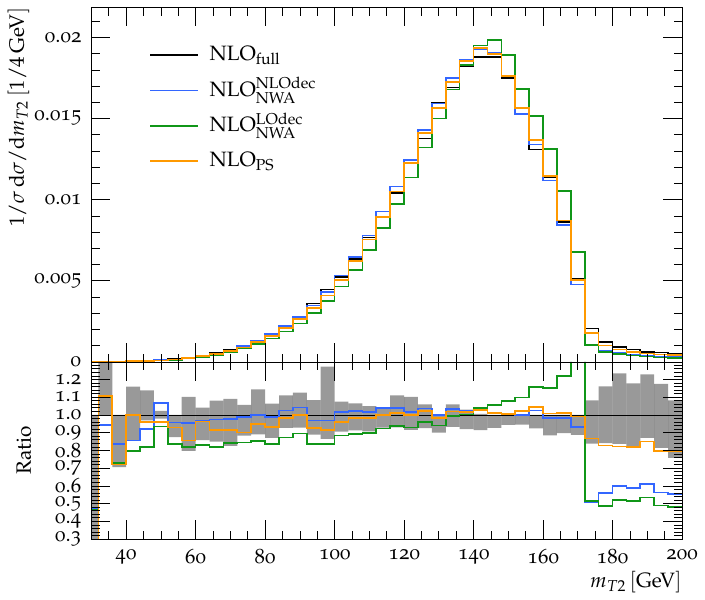}
\caption{\label{fig:scalevar_all_mt2}%
  Normalised differential cross sections for the $m_{T2}$ observable
  at the $13\tev$ LHC, analogous to Fig.~\ref{fig:scalevar_mlb}.
  The ratios of different theoretical descriptions to $\nlofull$
  including its scale uncertainty band are also shown.}
\end{figure}

It is also interesting to study the $\lodec$ prediction to investigate the importance of NLO corrections to the top quark decay.
We find  significant shape differences compared to the full
calculation of the order of about $-10$\% for $\mlb$ around
$50\gev$, rising to about $+20$\% around $\mlb\sim140\gev$.
Therefore, it is crucial in the application of the NWA to account for
a fully consistent NLO treatment of production {\em and} decay. 
For $m_{lb}\ge m_{lb}^\mrm{edge}$, the description completely fails.

Comparing $\nlops$ with $\lodec$, we find that the parton shower treatment of the top quark decay
drives the shape more towards the $\nlofull$ case for $\mlb>m_{lb}^\mrm{edge}$.
For low \mlb values, the parton shower result mostly lies
between the $\lodec$ and $\nlodec$ predictions.

Finally, we discuss the shape differences introduced by the different descriptions in the light of the scale uncertainties.
For clarity of the presentation, we only show the scale band of the
$\nlofull$ reference prediction in the lower part of
Fig.~\ref{fig:scalevar_mlb}.
For the other cases, we refer to Section~\ref{subsec:scaledep}.
We observe that in the bulk of the distribution, shape differences of $\nlodec$ with respect to $\nlofull$ lie inside
the uncertainty bands.
In contrast, both $\lodec$ and $\nlops$ exhibit differences to $\nlofull$ outside their respective
uncertainty bands, $\nlops$ however being much closer to $\nlofull$
than $\lodec$  (see also Fig.~\ref{fig:mlb_scalevar_nwa}).

\begin{figure}[tbp!]
  \centering
  \begin{subfigure}{0.495\textwidth}
    \includegraphics[width=\textwidth]{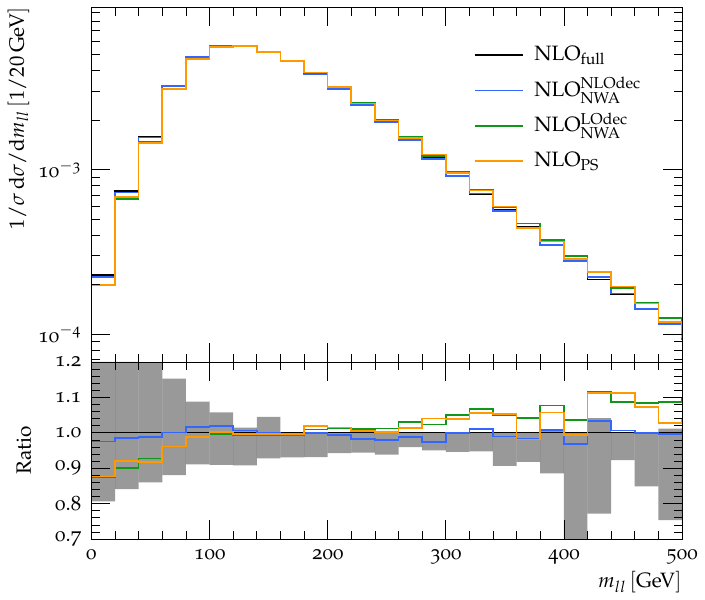}
    \vspace{\TwoFigBottom em}
    \caption{\label{fig:scalevar_all_mll}}
  \end{subfigure}
  \hfill
  \begin{subfigure}{0.495\textwidth}
    \includegraphics[width=\textwidth]{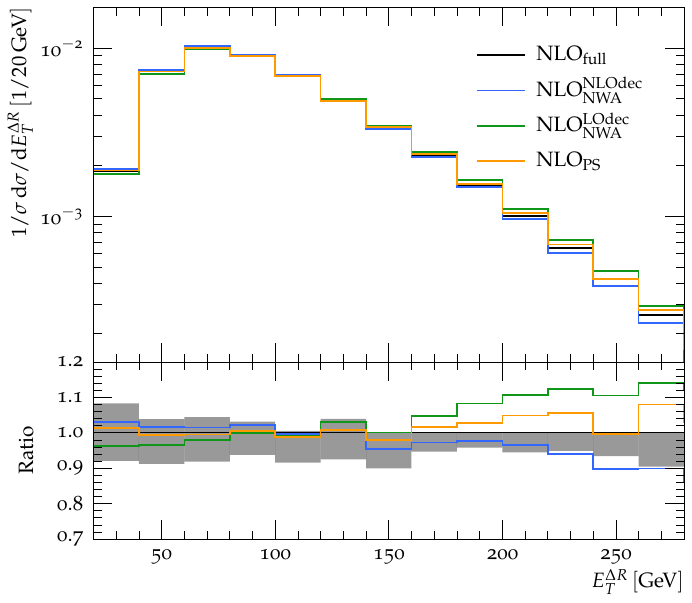}
    \vspace{\TwoFigBottom em}
    \caption{\label{fig:scalevar_all_etdr}}
  \end{subfigure}
  \caption{Normalised differential cross sections for (a)~the
    di-lepton invariant mass, \mll, and (b)~the observable \etdr.
    The ratios of the different theoretical descriptions to $\nlofull$
    including its scale uncertainty band are also shown.}
\end{figure}

In Fig.~\ref{fig:scalevar_all_mt2}, we show the normalised distribution of $m_{T2}$ as defined in Eq.~(\ref{def:MT2}), for the four theoretical descriptions.
By construction, this observable has a sharp kinematic edge at $m_{T2}=m_t$, which is clearly visible and mildly washed out
by off-shell effects, ambiguities related to missing energy and jet recombination.
We find that for the $\nlofull$ prediction, 97\% of the total fiducial cross section is contained below $m_{T2}\le m_t$.
The shapes of the different theoretical descriptions follow patterns very similar to those observed for $m_{lb}$.
In particular, the $\nlodec$ prediction closely follows $\nlofull$ up to the kinematic edge,
with shape differences of a few percent, but in general within the scale uncertainty band.

In Fig.~\ref{fig:scalevar_all_mll}, we show the di-lepton invariant mass $m_{ll}$.
We observe that off-shell effects are small and that all theoretical
descriptions agree at the 10\% level.
This is expected because $\mll$ is an observable which is inclusive in what concerns extra radiation.
The descriptions $\lodec$ and $\nlops$ show a very similar behaviour
and are outside the uncertainty bands of the $\nlofull$ prediction
except for low $\mll$ values.
In Fig.~\ref{fig:scalevar_all_etdr}, we display the \etdr observable defined in Eq.~(\ref{def:ETdR}).
Similar to \mll, also for \etdr, the $\nlofull$ and $\nlodec$
predictions do not exhibit large differences.
However, the shapes of the $\lodec$ and $\nlops$ predictions differ considerably from the  $\nlofull$ prediction.
In contrast to the \mll case, the $\lodec$ and $\nlops$ predictions also differ significantly from each other.

\begin{figure}[tbp!]
\centering
\begin{subfigure}{0.495\textwidth}
\includegraphics[width=\textwidth]{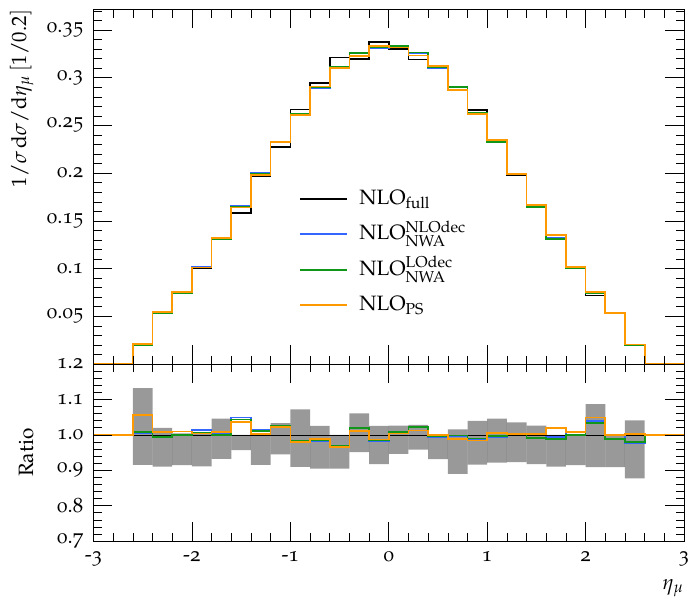}
\vspace{\TwoFigBottom em}
\caption{\label{fig:scalevar_etamu}}
\end{subfigure}
\hfill
\begin{subfigure}{0.495\textwidth}
\includegraphics[width=\textwidth]{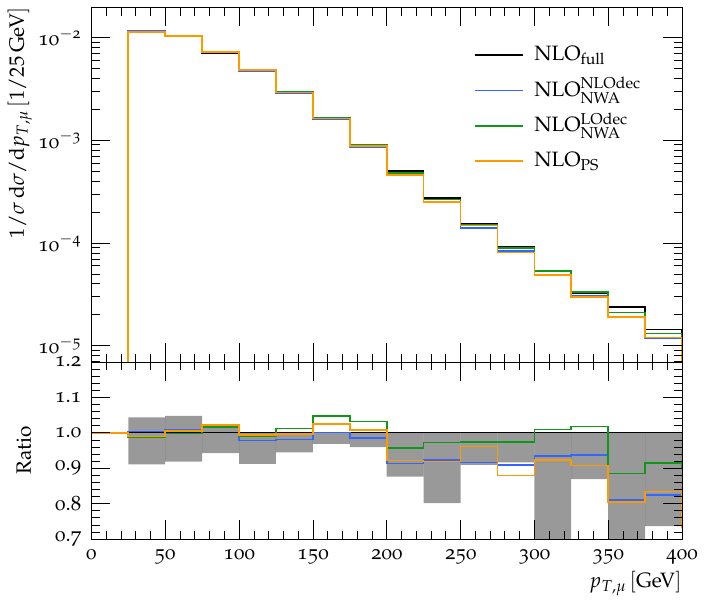}
\vspace{\TwoFigBottom em}
\caption{\label{fig:scalevar_ptmu}}
\end{subfigure}
\caption{Normalised differential cross sections for
  (a)~the rapidity of the muon, $\eta_{\mu}$, and
  (b)~the transverse momentum of the muon, $p_{T,\mu}$.
  The ratios of different theoretical descriptions to $\nlofull$
  including its scale uncertainty band are also shown.}
\end{figure}

In Figs.~\ref{fig:scalevar_etamu} and~\ref{fig:scalevar_ptmu}, we show the muon rapidity $\eta_{\mu}$ and the muon transverse momentum $p_{T,\mu}$, respectively.
Our four theoretical predictions for the $(e^+ \nu_e)\,(\mu^- \bar{\nu}_{\mu})\,b\bar{b}$ final state show a rather different behaviour in these two distributions.
While the whole rapidity spectrum in Fig.~\ref{fig:scalevar_etamu} is properly modelled by all predictions,
the transverse momentum spectrum in Fig.~\ref{fig:scalevar_ptmu} is somewhat softer in the tail for the $\nlodec$ and $\nlops$ calculations with respect to $\nlofull$.
A possible interpretation is that non-resonant contributions in $\nlofull$ contain $W$-bosons stemming from a
hard collision rather than the top quark decay. Therefore they can carry higher energies which lead to a harder transverse momentum spectrum of the muon.

\subsubsection{Scale dependence at LO and NLO}
\label{subsec:scaledep}

In this section, we will only consider the observables $\mlb$, $\mtwo$, $\mll$ and $\etdr$, as they are promising with respect to
at least one of the requirements of being observables with small systematics and/or high sensitivity to the top quark mass.

\begin{figure}[tbp!]
\centering
\begin{subfigure}{0.495\textwidth}
\includegraphics[width=\textwidth]{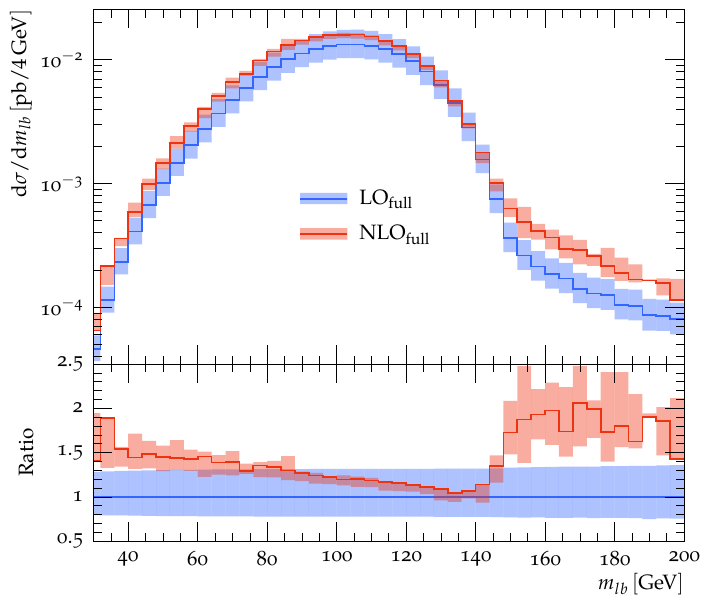}
\vspace{\TwoFigBottom em}
\caption{\label{fig:mlb_scalevar}}
\end{subfigure}
\hfill
\begin{subfigure}{0.495\textwidth}
\includegraphics[width=\textwidth]{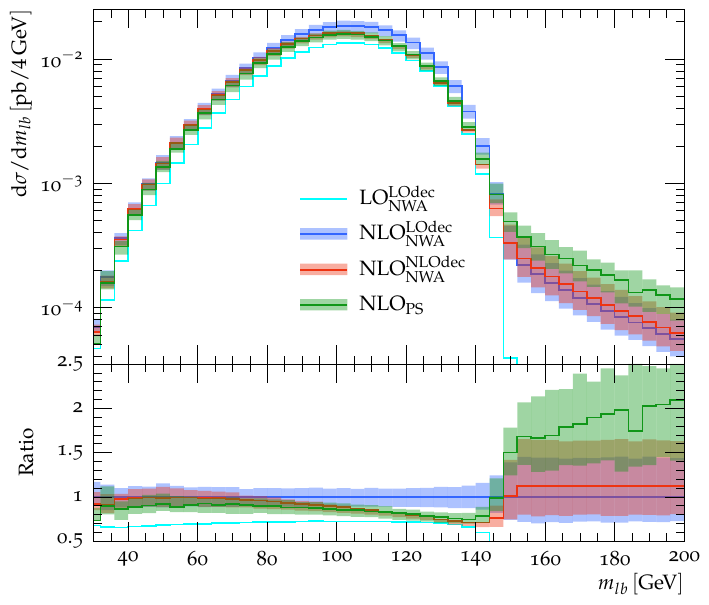}
\vspace{\TwoFigBottom em}
\caption{\label{fig:mlb_scalevar_nwa}}
\end{subfigure}
\caption{Results including scale variation bands for $m_{lb}$, for (a)~the $\lofull$ and $\nlofull$ calculations, (b)~the calculations based on the NWA.
The ratios with respect to (a) $\lofull$ and (b) $\lodec$ are also shown.\label{fig:mlb_scalevar_all}}
\end{figure}

\begin{figure}[tbp!]
\centering
\begin{subfigure}{0.495\textwidth}
\includegraphics[width=\textwidth]{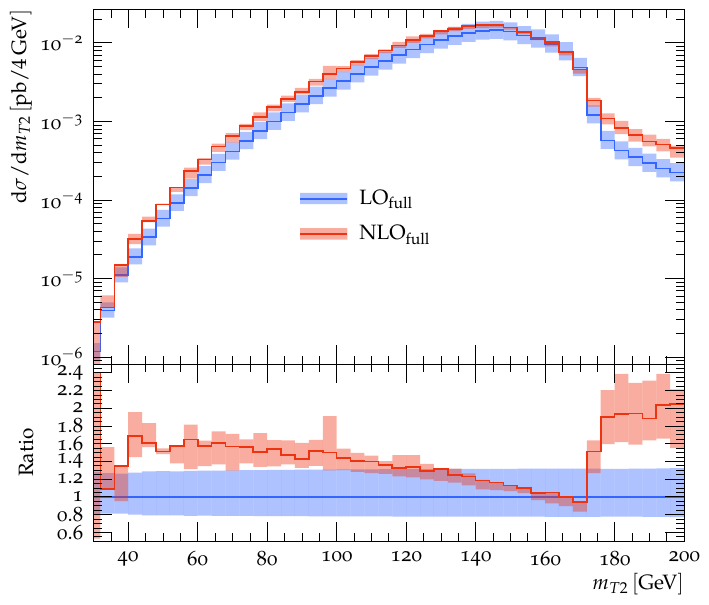}
\vspace{\TwoFigBottom em}
\caption{}
\label{fig:mt2_scalevar}
\end{subfigure}
\hfill
\begin{subfigure}{0.495\textwidth}
\includegraphics[width=\textwidth]{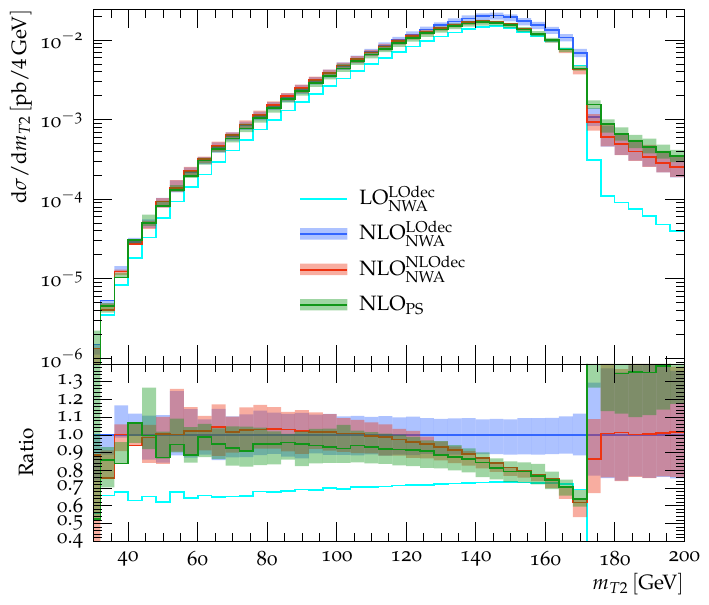}
\vspace{\TwoFigBottom em}
\caption{}
\label{fig:mt2_scalevar_nwa}
\end{subfigure}
\caption{Results including scale variation bands for $m_{T2}$, for (a)~the $\lofull$ and $\nlofull$ calculations, and (b)~the calculations based on the NWA.
The ratios are defined as in Fig.~\ref{fig:mlb_scalevar_all}.}
\end{figure}

\begin{figure}[tbp!]
\centering
\begin{subfigure}{0.495\textwidth}
\includegraphics[width=\textwidth]{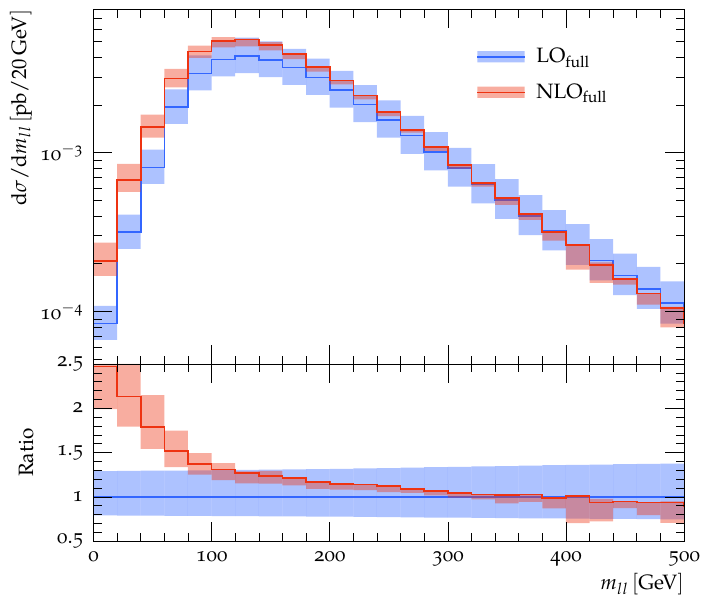}
\vspace{\TwoFigBottom em}
\caption{}
\label{fig:mll_scalevar}
\end{subfigure}
\hfill
\begin{subfigure}{0.495\textwidth}
\includegraphics[width=\textwidth]{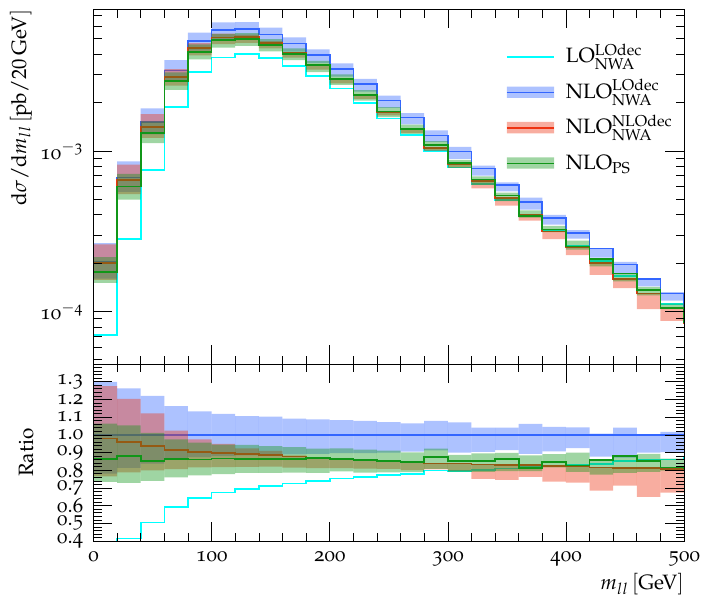}
\vspace{\TwoFigBottom em}
\caption{}
\label{fig:mll_scalevar_nwa}
\end{subfigure}
\caption{Results including scale variation bands for  $m_{ll}$, for (a)~the $\lofull$ and $\nlofull$ calculations, and (b)~the calculations based on the NWA.
The ratios are defined as in Fig.~\ref{fig:mlb_scalevar_all}.}
\end{figure}

\begin{figure}[tbp!]
\centering
\begin{subfigure}{0.495\textwidth}
\includegraphics[width=\textwidth]{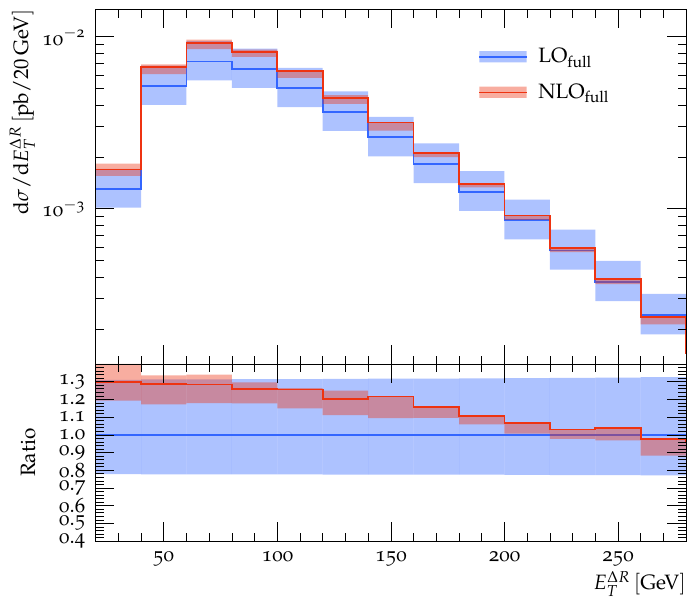}
\vspace{\TwoFigBottom em}
\caption{}
\label{fig:ETdR_scalevar}
\end{subfigure}
\hfill
\begin{subfigure}{0.495\textwidth}
\includegraphics[width=\textwidth]{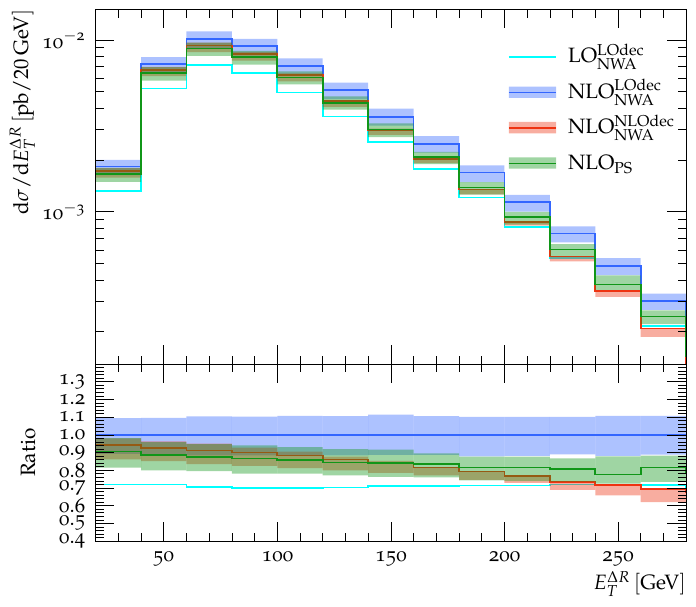}
\vspace{\TwoFigBottom em}
\caption{}
\label{fig:ETdR_scalevar_nwa}
\end{subfigure}
\caption{Results including scale variation bands for \etdr for (a)~the $\lofull$ and $\nlofull$ calculations, and (b)~the calculations based on the NWA.
The ratios are defined as in Fig.~\ref{fig:mlb_scalevar_all}.}
\end{figure}

For $\nlofull$, we compare LO and NLO predictions on the left-hand
side, while in the figures on the right-hand side, we compare
calculations based on the NWA, including scale variations\footnote{For better visibility, 
 the wide and uniform scale variation band for the $\lolo$ result is not shown in Figs.~\ref{fig:mlb_scalevar_nwa} to~\ref{fig:ETdR_scalevar_nwa}.}.
We observe that the NLO corrections in the $\nlofull$ case lead to significant shape differences compared to $\lofull$,
see Figs.~\ref{fig:mlb_scalevar} to~\ref{fig:ETdR_scalevar}.
While this is to be expected in the tails of the distributions, it is
remarkable that the shape difference also affects the central and in particular the regions with low values of the observables.
Given that the differences between the LO and NLO theory predictions in the full $W^+W^-b\bar{b}$ calculation are still sizeable in the bulk of the distributions, large
differences in the top quark mass extracted from templates based on these predictions can be expected.
The shape differences at low values of $\mlb$ and $\mtwo$ are  less pronounced in the calculations based on the NWA (with NLO in the $t\bar{t}$ production),
as can be seen from Figs.~\ref{fig:mlb_scalevar_nwa}  and~\ref{fig:mt2_scalevar_nwa}.
However, there are also significant shape differences in the bulk of the distribution.
In addition, for the $\mlb$ distribution, Fig.~\ref{fig:mlb_scalevar_nwa}, the peak is lower in the $\nlodec$ and the $\nlops$ case
 compared to the $\lodec$ case, which can be easily understood considering the fact that  more radiation, i.e.~a harder distribution in the tail,
softens the peak region.

For the observable $\mll$, the shape differences introduced by the $\nlofull$ calculation at low $\mll$ values are
particularly pronounced in Fig.~\ref{fig:mll_scalevar}.
The calculations based on the NWA in Fig.~\ref{fig:mll_scalevar_nwa},
$\nlodec$ and $\lodec$, cease to overlap at relatively low $\mll$
values ($\mll\sim160\gev$), while
$\nlops$ mostly lies between $\nlodec$ and $\lodec$ in the region
beyond $\mll>200\gev$.
As shown in Table~\ref{tab:xs}, the total
cross section predicted by $\nlops$ is considerably smaller.
This is due to the fact that after the shower, the $b$-jets are softer and therefore a larger fraction of events does not pass the requirement of
two $b$-jets above $p_{T,\mrm{min}}^\mrm{jet}=25\gev$.
Even though the observable $\mll$ does not involve jets, the jet requirements affect this observable,
since we use the data set produced with the same requirements as for the other observables.
A similar pattern is seen in the observable \etdr (Figs.~\ref{fig:ETdR_scalevar} and~\ref{fig:ETdR_scalevar_nwa}).

The scale variation bands in the $\nlofull$ case and the $\nlodec$ case are rather asymmetric:
the central scale leads to the largest differential cross section compared to up- and downwards variations over a large kinematic range of the corresponding observable.
This effect is particularly pronounced for the $\mll$ and \etdr distributions.

\subsubsection{Distributions for several top quark masses}

\begin{figure}[tbp!]
\centering
\begin{subfigure}{0.495\textwidth}
\includegraphics[width=\textwidth]{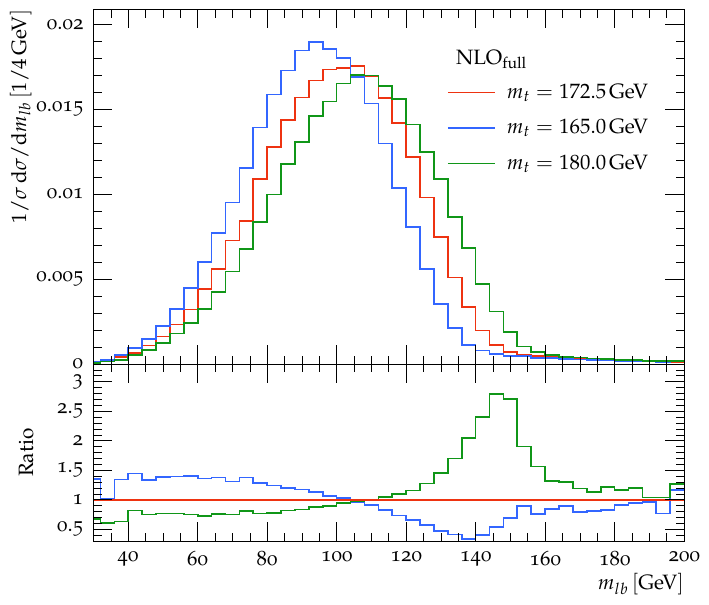}
\vspace{\TwoFigBottom em}
\caption{\label{fig:massvar_fullNLO_mlb}}
\end{subfigure}
\hfill
\begin{subfigure}{0.495\textwidth}
\includegraphics[width=\textwidth]{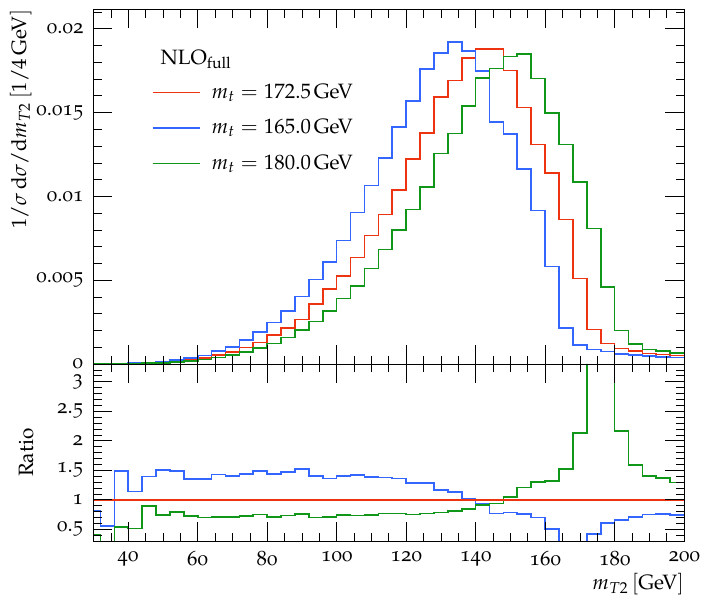}
\vspace{\TwoFigBottom em}
\caption{\label{fig:massvar_fullNLO_mt2}}
\end{subfigure}
\caption{Effect of top quark mass variations on the normalised differential cross sections for $m_{lb}$ and $m_{T2}$.
We also show the ratios to the prediction obtained with $m_t=172.5\gev$.
All results are obtained with the $\nlofull$ description for the $13\tev$ LHC.}
\end{figure}

\begin{figure}[tbp!]
\centering
\begin{subfigure}{0.495\textwidth}
\includegraphics[width=\textwidth]{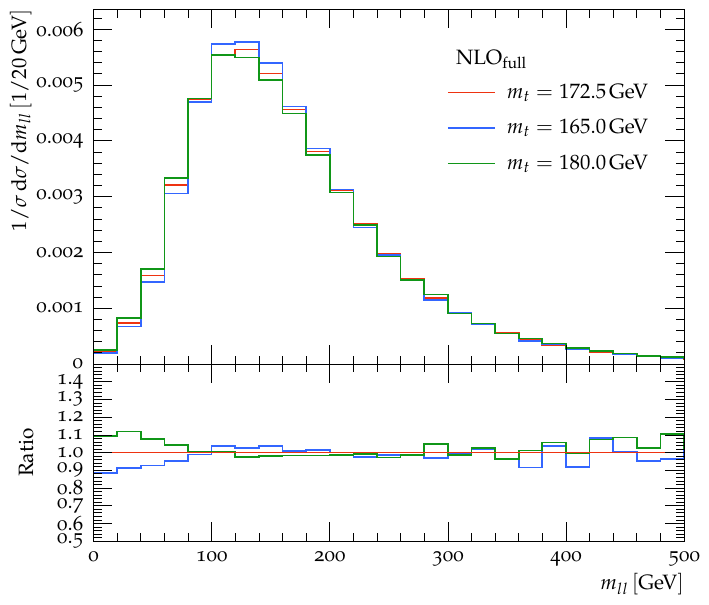}
\vspace{\TwoFigBottom em}
\caption{}
\label{fig:massvar_fullNLO_mll}
\end{subfigure}
\hfill
\begin{subfigure}{0.495\textwidth}
\includegraphics[width=\textwidth]{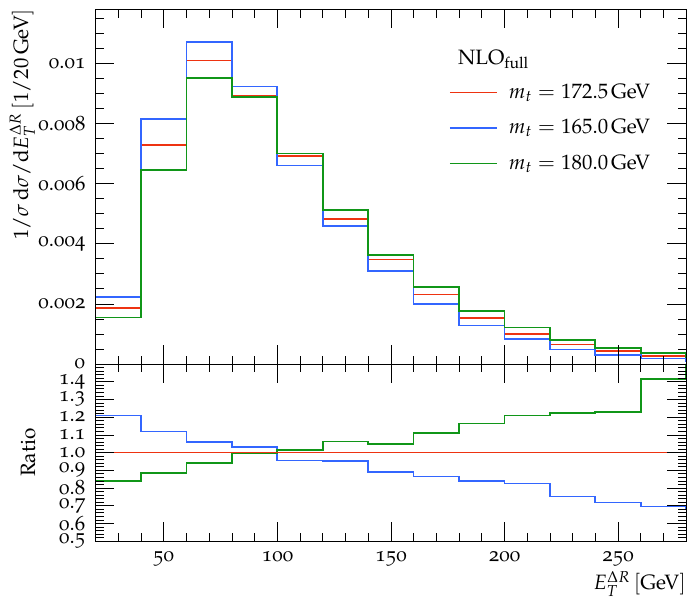}
\vspace{\TwoFigBottom em}
\caption{}
\label{fig:massvar_fullNLO_etdr}
\end{subfigure}
\caption{Effect of top quark mass variation on the normalised differential cross section for $m_{ll}$ and \etdr.
We also show the ratios to the prediction obtained with $m_t=172.5\gev$.
The results are obtained with the $\nlofull$ description for the $13\tev$ LHC.}
\end{figure}

In this section, we investigate the sensitivity of the four observables $\mlb$, $\mtwo$, $\mll$ and $\etdr$ to variations of the top quark mass.
We exploit distributions based on the $\nlofull$ calculation using the three values,
$m_t=165$, $172.5$, $180\gev$, for the top quark mass.

We observe a strong sensitivity of the $\mlb$ and $\mtwo$ distributions to the top quark mass
with ratios up to about three in the given range.
A lower top quark mass naturally leads to a softer spectrum while a higher top quark mass leads to a harder spectrum in these two observables.
The sensitivity of $\mll$ is shown in Fig.~\ref{fig:massvar_fullNLO_mll} and turns out to be very small.
Unfortunately, being a purely leptonic observable,
the low sensitivity counterbalances its expected~\cite{Frixione:2014ala} better experimental systematics.
Compared to the $\mll$ distribution, the \etdr distribution in Fig.~\ref{fig:massvar_fullNLO_etdr} shows a somewhat larger sensitivity to $m_t$, albeit much smaller than what is observed for $\mlb$ and $\mtwo$.


\section{Measurement of the top quark mass based on pseudo-data}
\label{sec:fit}

The top quark mass measurements in the di-lepton channel presented in Refs.~\cite{Aad:2015nba,AMaierPhD:2015,Aaboud:2016igd}
 use the template method.
 In this method, simulated distributions are constructed for different
 input values of the top quark mass, \mtin.
 The distributions (templates) per \mtin\ are then individually fitted to a
 suitable function. Using templates at different \mtin, it is verified that all
 parameters of the function linearly depend on $\mt=\mtin$. Consequently, this
 linearity is imposed in a combined fit to all templates.
 This fit fixes the theory prediction (i.e.~the parametrisation of the theory
 hypothesis) by determining all parameters of the function, except for \mt\ and
 the absolute normalisation.
 The former is to be determined from the data and represents the fit result, while
 the latter is left as a free parameter.
 We therefore follow the experimental procedure to neglect the absolute normalisation in the fit
 to avoid a dependence on the involved experimental determination of the total luminosity
 and detector efficiency.
 This choice makes the results of this study independent of the total cross section of
 the respective calculations, leaving shape changes of the differential distributions
 as the measure for \mt.
 Using those parameter values, a likelihood fit of this function to data is
 performed to obtain the value for \mt\ that best describes the data,
 namely \mtou, together with its statistical uncertainty.

 In experimental analyses, these templates are constructed at the detector
 level, i.e.~mimicking real data.
 Here, an analogous procedure is employed to
 assess the impact of different theory descriptions on
 the template method used to determine the top quark mass.
 In our analysis, the pseudo-data mimicking experimental data (i.e.~the data
 model) in each figure are always generated from those predictions, which are believed
 to be closer to real data, i.e.~those that are considered to give the ``better'' result.
 We simulate a data luminosity of $50/\mrm{fb}$.

 The sensitivity to the theoretical assumptions and their uncertainties is
 assessed by fits to one thousand pseudo-data sets created by random sampling from
 the underlying theory prediction.
 The layout of Figure~\ref{fig:LOvsNLO_NWA_decayLO} is representative for an
 entire set of figures presented in the following. For three different values
 of \mtin, each of these figures shows the observed difference of
 \mtou, the mass measured by the procedure, and \mtin, the mass used
 to generate the pseudo-data.
 The red/blue points correspond to the mean difference observed for all pseudo-data
 sets that are produced  as stated in the second line of the figure
 legends,  and analysed with the template fit functions (the theory
 hypothesis), denoted by ``calibration''  in the legend for
 the red/blue points.
 The uncertainty per point is statistical only and corresponds to
 the expected experimental uncertainty for the assumed data luminosity.
 The points are displaced on the horizontal axis to ensure better
 visibility in the case of overlapping bands.
The horizontal lines stem from a fit of the three points to a
 constant, displaying the {\em average} offset.
  The values given are the (individual) offsets together with their
 statistical uncertainties.
 The bands indicate the effect of the scale variations on the measured \mt.
 They are obtained by replacing the central-scale pseudo-data by those
 derived from
 the associated samples, which were calculated using the varied scales.

 The ranges of the fits have been chosen on a plateau of good fit performance and high mass sensitivity.
The ranges of choice are
\begin{align}
&40\gev\leq \mlb\leq 160\gev\;,\label{eq:fitranges}\\
&80\gev\leq \mtwo\leq 180\gev\;.\nonumber
\end{align}
Note that for the $\nlops$ calculations employing the $\mu_{t\bar t}$
scale, we used a fit range of $50\gev\leq \mlb\leq 150\gev$.

As the range around the kinematic edge is a particularly \mt-sensitive
region, the question arises how much our results depend on the chosen
fit range. Therefore we produced another set of fits where we 
restricted the fit range to $\mlb<140\gev$, and found that the results
are sufficiently stable under this change of the fit range. 
The results of both fit ranges are reported below.

\boldmath
\subsection{Fit results for $m_{lb}$}
\unboldmath

\begin{figure}[tbp!]
  \centering
  \begin{subfigure}{0.495\textwidth}
    \includegraphics[width=\textwidth]{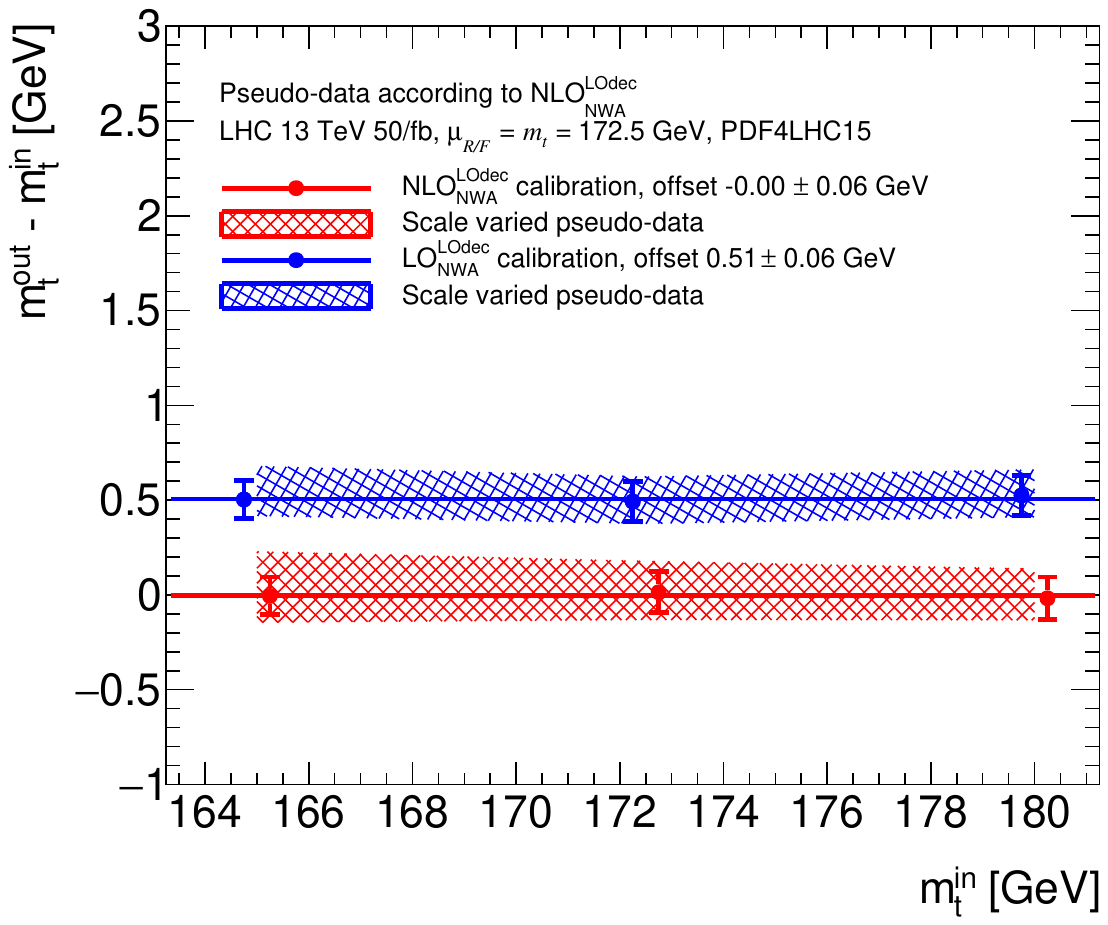}
    \vspace{\TwoFigBottom em}
    \caption{\label{fig:LOvsNLO_NWA_decayLO}}
  \end{subfigure}
  \hfill
  \begin{subfigure}{0.495\textwidth}
    \includegraphics[width=\textwidth]{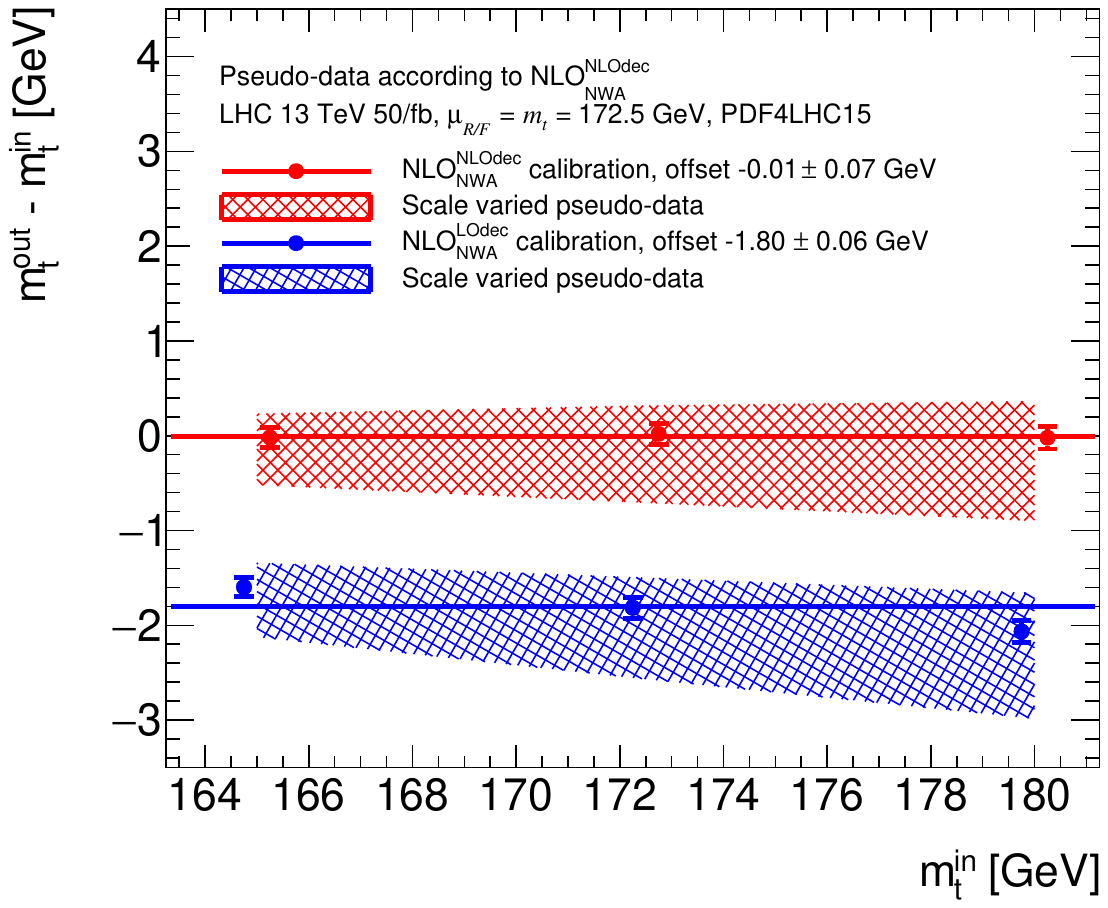}
    \vspace{\TwoFigBottom em}
    \caption{\label{fig:NLO_NWA_decayLOvsNLO}}
  \end{subfigure}
  \caption{\label{fig:NWA}%
    Results of the top quark mass determination using the observable $m_{lb}$
    and (a) pseudo-data generated according to the factorised
    approach with $\lodec$, showing the effect of changing
    the perturbative order in the production process only, and (b)
    pseudo-data obtained from the factorised approach with
    $\nlodec$, showing the effect of changing the perturbative order
    in the decay process only.}
\end{figure}

Figure~\ref{fig:LOvsNLO_NWA_decayLO} shows results of a fit where the pseudo-data have been generated using the factorised approach with $\lodec$.
The fit has been performed once with $\lolo$ as the theory model (blue) and once with $\lodec$ (red).
The vanishing offset (i.e.~it is compatible with zero) for the red lines (here and in all the following figures) proves that the method is closed, i.e.~it finds the input value when the pseudo-data and the calibration coincide.
The offset between the blue and red lines in
Fig.~\ref{fig:LOvsNLO_NWA_decayLO} shows the effect of changing the
perturbative order of the {\em production} process in the theory
model. The offset of  $0.51\pm 0.06\,\gev$
demonstrates that these corrections have an impact
on the mass determination at the level of the present experimental uncertainties. As the fits are based on
normalised differential cross sections, the bands are sensitive to
shape differences induced by the scale variations, rather than to their
overall magnitudes.

Figure~\ref{fig:NLO_NWA_decayLOvsNLO} shows results of a fit where the
pseudo-data have been generated using the factorised approach based on
the $\nlodec$, i.e.~the NWA at NLO, while the theory models differ in
the {\em decay} order only. We observe that the effect of an
${\cal O}(\as{})$ change in the perturbative order of the decay is
more significant than changing the order in the production process.
The offset stemming from the former amounts to $-1.80\pm0.06\gev$, while
switching from LO to NLO in the description of the production process
yields an offset of $0.51\pm0.06\gev$ (cf.~Fig.~\ref{fig:LOvsNLO_NWA_decayLO}).
In addition, the size of the uncertainty bands increases because the NLO corrections to the decay lead to non-uniform scale variation bands.

\begin{figure}[tbp!]
  \centering
  \begin{subfigure}{0.495\textwidth}
    \includegraphics[width=\textwidth]{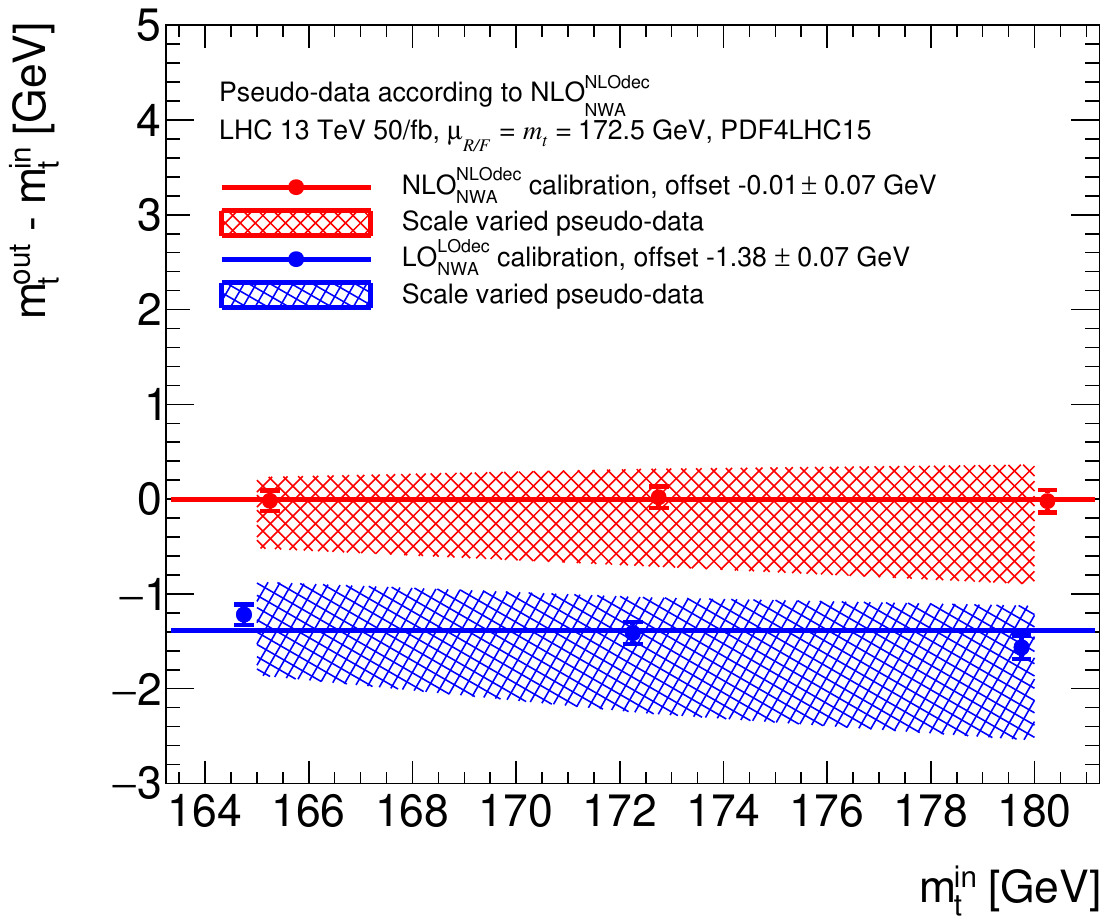}
    \vspace{\TwoFigBottom em}
    \caption{\label{fig:NLO_NWA_decayNLO}}
  \end{subfigure}
  \hfill
  \begin{subfigure}{0.495\textwidth}
    \includegraphics[width=\textwidth]{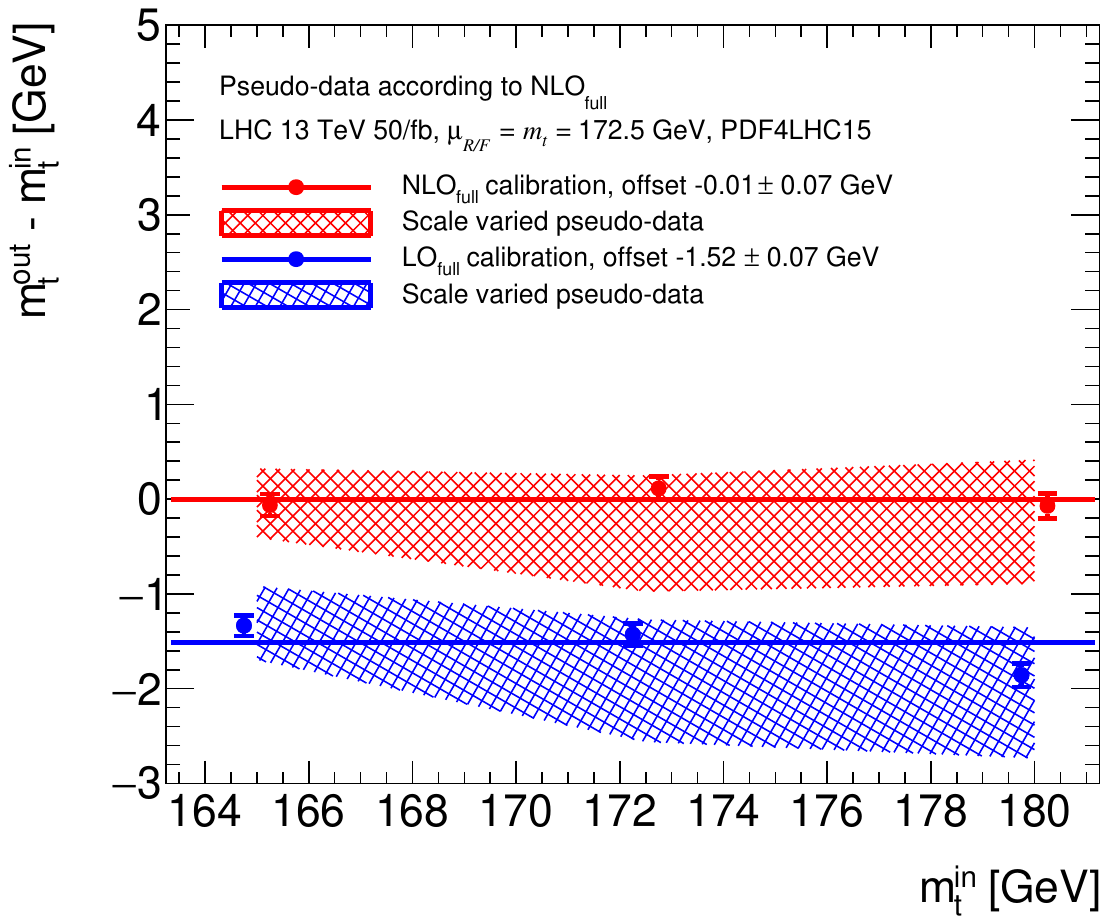}
    \vspace{\TwoFigBottom em}
    \caption{\label{fig:fullNLO}}
  \end{subfigure}
  \caption{\label{fig:NWAboth}%
    Results of the top quark mass determination using the observable $m_{lb}$
    and (a) pseudo-data generated according to the factorised
    approach with $\nlodec$, showing the effect of changing
    the perturbative order in both the production and decay process,
    and (b) pseudo-data derived from $\nlofull$ distributions. In both
    cases, the focus is on the comparison of LO versus NLO calibrations.}
\end{figure}

Figure~\ref{fig:NLO_NWA_decayNLO} shows the effect of changing the
perturbative order in both the production and decay process.
Comparing Figs.~\ref{fig:NWA} and~\ref{fig:NLO_NWA_decayNLO}, we observe that,
within the statistical uncertainties,  the offset in Fig.~\ref{fig:NLO_NWA_decayNLO} coincides with
the sum of the offsets in Figs.~\ref{fig:LOvsNLO_NWA_decayLO} and
\ref{fig:NLO_NWA_decayLOvsNLO}, as is expected for the factorised approach.
Figure~\ref{fig:fullNLO} shows results of a fit where the pseudo-data
have been generated using the $\nlofull$ calculation, and the
calibrations are based on the $\nlofull$ and $\lofull$
descriptions. While the uncertainty bands are comparable to the
factorised case that uses pseudo-data based on $\nlodec$
(Fig.~\ref{fig:NLO_NWA_decayNLO}), the offset increases from
$-1.38\pm0.07\gev$ to $-1.52\pm0.07\gev$.
While this increase in the offset is not conclusive when taking the
statistical uncertainty into account, it still is an indication of the
trend that the inclusion of a richer set of corrections leads to larger
offsets.

\begin{figure}[tbp!]
\centering
\begin{subfigure}{0.495\textwidth}
\includegraphics[width=\textwidth]{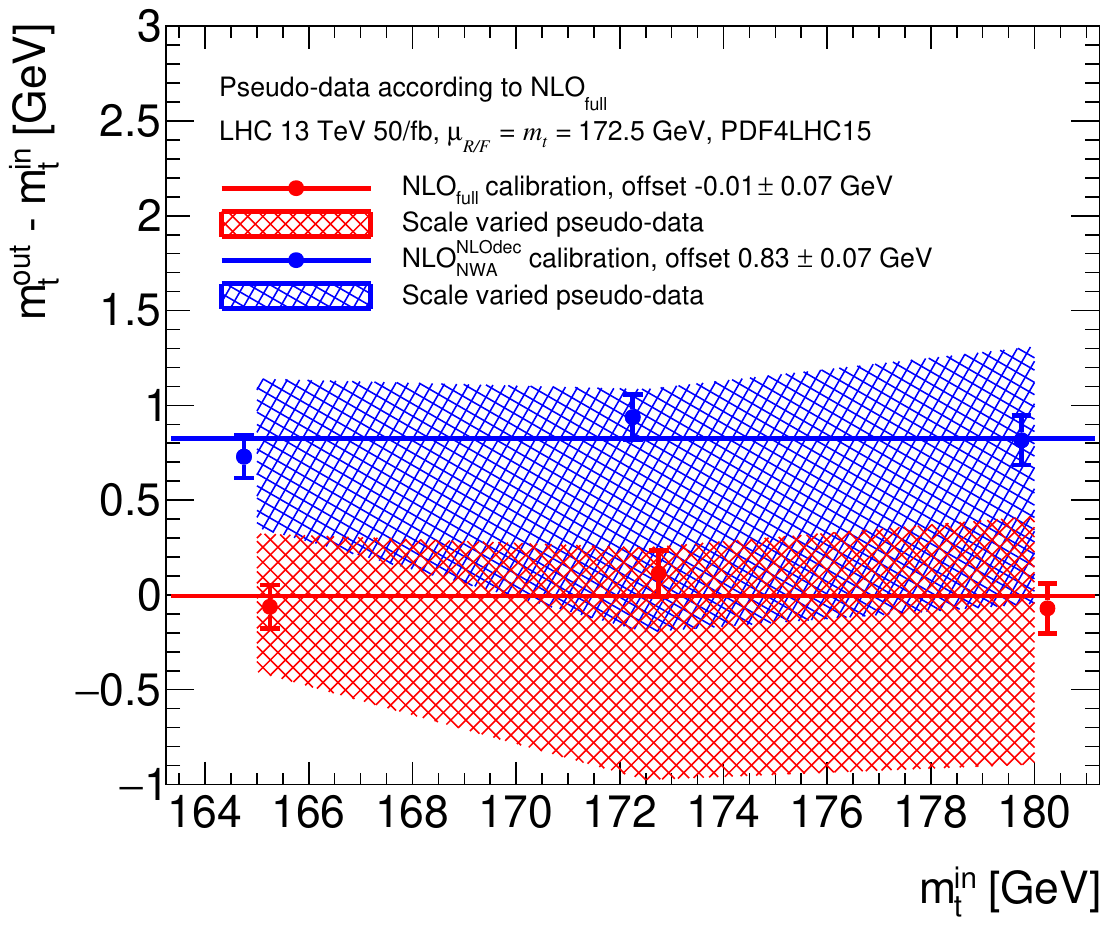}
\vspace{\TwoFigBottom em}
\caption{\label{fig:NLO_WWbb_vs_NWA_decayNLO}}
\end{subfigure}
\hfill
\begin{subfigure}{0.495\textwidth}
\includegraphics[width=\textwidth]{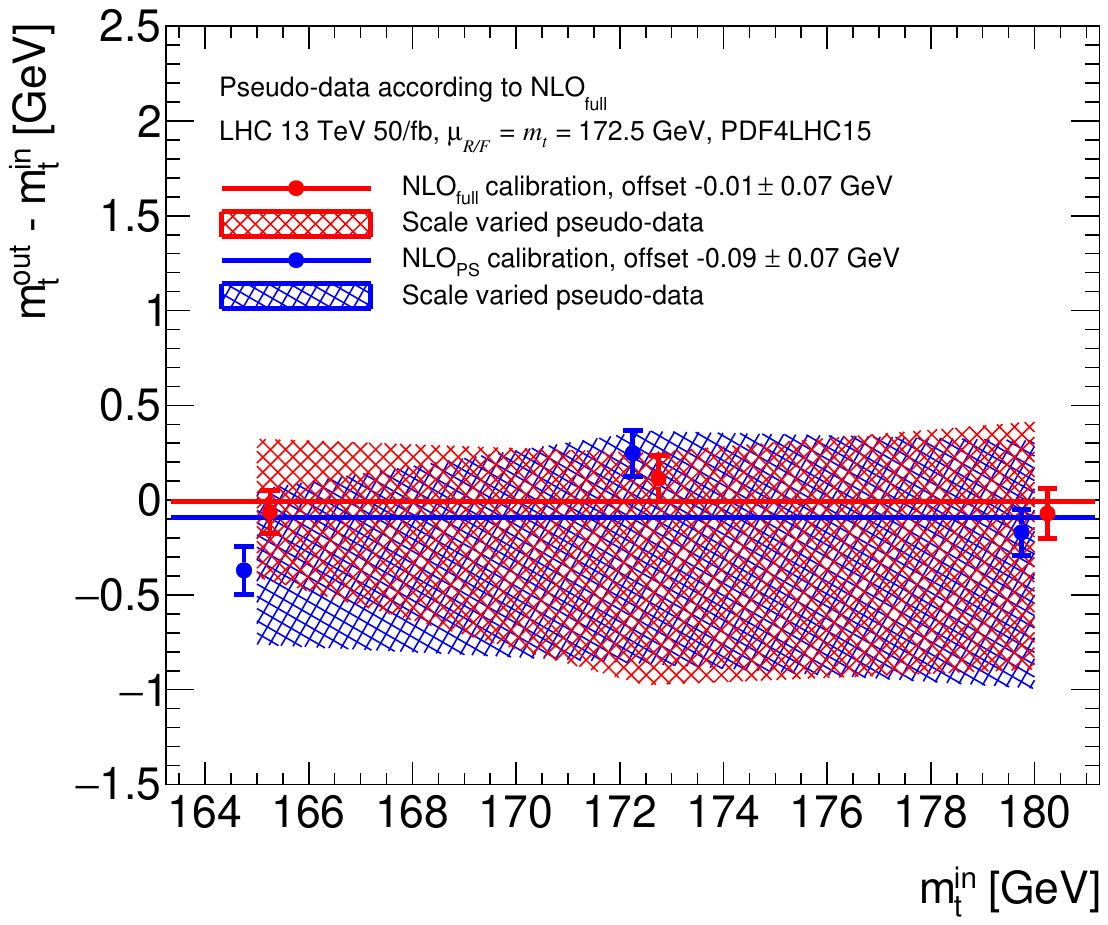}
\vspace{\TwoFigBottom em}
\caption{\label{fig:WWbb_vs_PS_NLO}}
\end{subfigure}
\caption{\label{fig:NLO_WWbb_vs}%
  Top quark mass determination results for the observable $m_{lb}$
  comparing pseudo-data generated according to the $\nlofull$
  predictions with (a) the $\nlodec$ calibration and (b) the $\nlops$
  calibration.}
\end{figure}

In Fig.~\ref{fig:NLO_WWbb_vs_NWA_decayNLO}, we again use pseudo-data
generated according to $\nlofull$,
this time comparing the fit based on the full NLO calibration to the one
obtained with the $\nlodec$ calibration representing the factorised NLO
approach.
We see that the offset of $0.83\pm0.07\gev$ is smaller in magnitude
than in Fig.~\ref{fig:fullNLO}, and goes in the opposite direction.
This indicates that the non-factorisable contributions are suppressed
in the fit range, since the NWA, with the corrections to the decay 
included, is a better approximation than $\lofull$ only.

In Fig.~\ref{fig:WWbb_vs_PS_NLO}, we replace the $\nlodec$ calibration 
by the one from the $\nlops$ prediction.
We observe an offset of $-0.09\pm0.07\gev$, which is
surprisingly small compared to that given in
Fig.~\ref{fig:NLO_WWbb_vs_NWA_decayNLO}. It is expected
that the two NWA-based descriptions, both including 
the leading radiation in the decay, lead to quite
similar results. However, the $\nlops$ simulation differs from the
$\nlodec$ calculation in a number of points.
While $\nlops$ falls short of describing the top quark
decay beyond the soft limit owing to the absence of decay matrix-element
corrections, the parton shower approach generates a very different,
more complete QCD radiation pattern
as a result of including resummation effects in the production
as well as the decay of the top quarks. This means that the two stages
of $t\bar t$\/ production and decay are not  factorised in
exactly the same way as in the $\nlodec$ calculation. These
differences explain why the offset in
Fig.~\ref{fig:NLO_WWbb_vs_NWA_decayNLO} is different from the
one in Fig.~\ref{fig:WWbb_vs_PS_NLO}.
In fact, as can be seen from Figs.~\ref{fig:scalevar_mlb} as well
as~\ref{fig:mlb_scalevar_nwa}, the emission pattern and resummation
effects of the $\nlops$ case are relevant at lower $\mlb$ values and
in particular around (and above) the kinematic edge, and lead to a
shape of the \mlb distribution, which differs from the fixed-order
$\nlodec$ case. Especially for the $\mlb\sim140\gev$ region, we notice
that the agreement between $\nlops$ and $\nlofull$ is better than
between $\nlops$ and $\nlodec$. This is an
indication that in this region, resummation effects are more important
than the inclusion of the radiative correction in the decay.
The nearly vanishing mass offset shown in Fig.~\ref{fig:WWbb_vs_PS_NLO} occurs
due to the fact that the shapes of $\nlops$ and $\nlofull$ do
not differ significantly in most of the fit range, despite their different theoretical content.

\begin{figure}[tbp!]
\centering
\begin{subfigure}{0.495\textwidth}
\includegraphics[width=\textwidth]{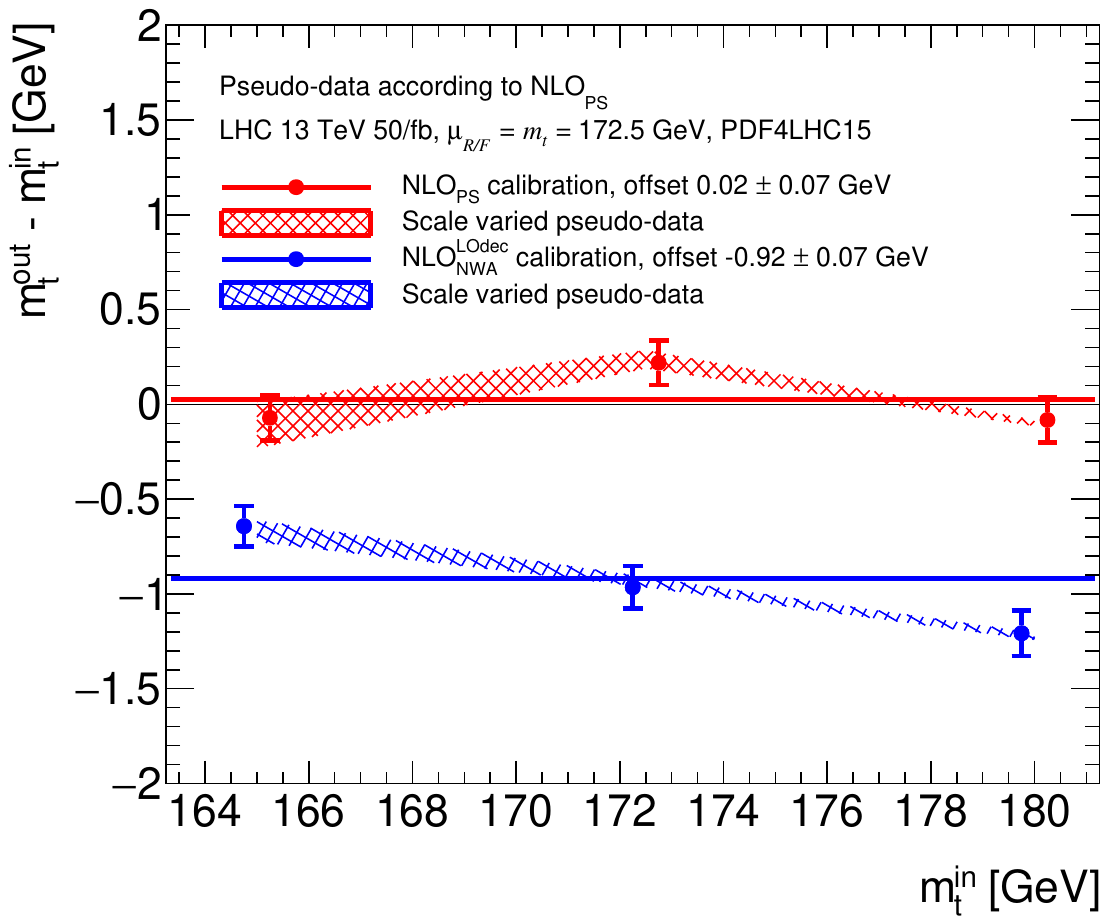}
\vspace{\TwoFigBottom em}
\caption{\label{fig:PS_vs_ttbar_NLO}}
\end{subfigure}
\hfill
\begin{subfigure}{0.495\textwidth}
\includegraphics[width=\textwidth]{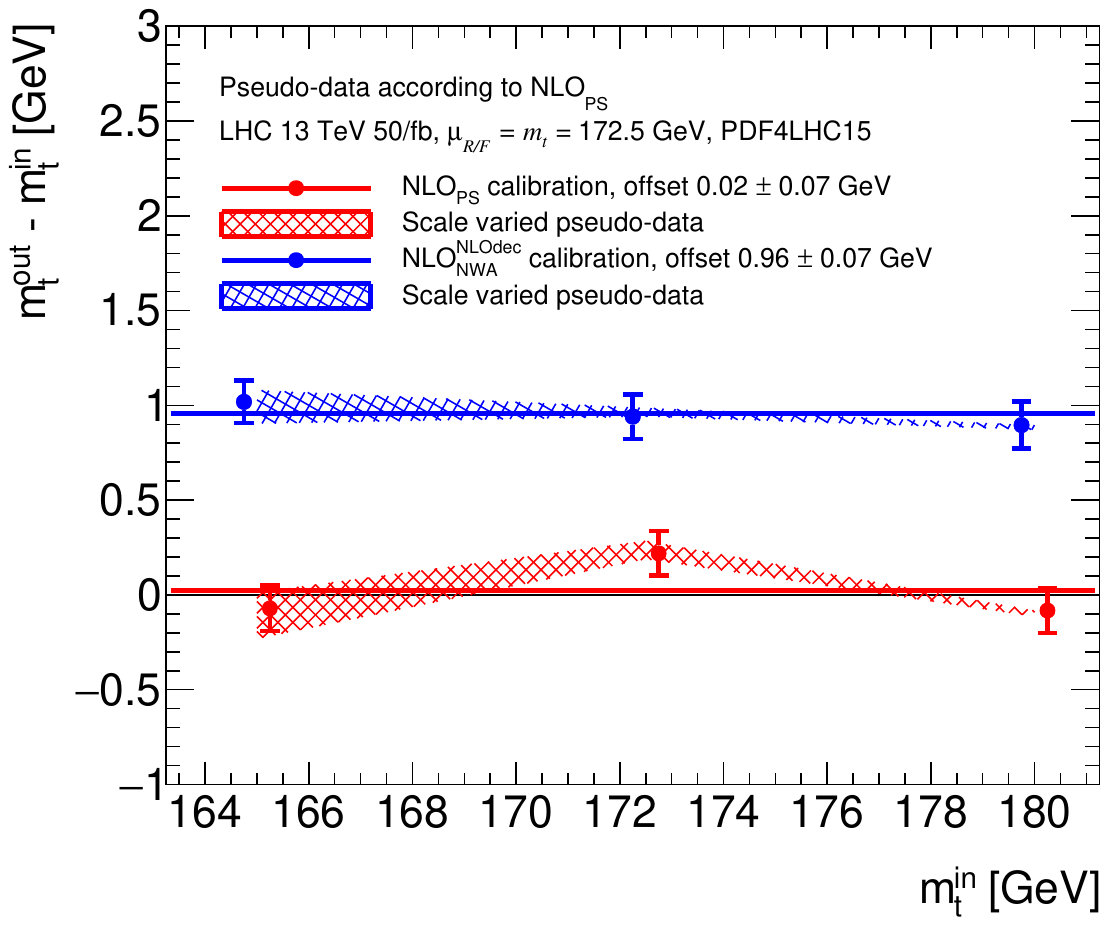}
\vspace{\TwoFigBottom em}
\caption{\label{fig:PS_vs_ttbar_NLO_nlodc}}
\end{subfigure}
\caption{\label{fig:NLO_PS_vs}%
  Top quark mass determination results for the observable $m_{lb}$
  comparing pseudo-data generated according to the $\nlops$
  predictions with  (a) the $\lodec$ calibration and (b) the $\nlodec$ calibration.}
\end{figure}

In Fig.~\ref{fig:NLO_PS_vs}, 
we use pseudo-data generated according to the $\nlops$ prediction
using the scale setting $\mu_F=\mu_R=m_t$. The related
scale variations have been obtained by employing the
$\mu_F\mu_R\as{\mathrm{PS}}$ scheme as described at the end of
Section~\ref{subsec:input}. 
By comparing to Fig.~\ref{fig:LOvsNLO_NWA_decayLO}, 
we observe that the uncertainty bands of $\nlops$ are smaller than the
ones for $\lodec$. However, for the theory models relying on NLO decays,
as shown in Fig.~\ref{fig:NLO_NWA_decayLOvsNLO} for $\nlodec$
and in Fig.~\ref{fig:fullNLO} for $\nlofull$, the bands are much wider.
Hence, we expect that adding a parton shower to the $\nlofull$
calculation, the bands would persist or be only slightly reduced,
analogous to the LO decay situation discussed above.

Unlike the case presented in Fig.~\ref{fig:WWbb_vs_PS_NLO}, the direct
comparison between results from the NWAs and $\nlops$ produces
non-vanishing mass shifts.
If we analyse the $\nlops$ pseudo-data using the
fixed-order $\lodec$ calibration, we find a mass offset of
$-0.92\pm0.07\gev$ as shown in Fig.~\ref{fig:PS_vs_ttbar_NLO}. This
indicates that the parton shower emissions (in both stages),
supplementing the NLO accurate $t\bar t$ production, have a
considerable impact on the results.
In addition, a significant dependence of the $\lodec$ calibration offset
on the top quark mass is observed,
i.e.~the blue points are inconsistent with the constant fit.
This implies that the $\lodec$ \mlb\ distribution has a stronger
dependence on the top quark mass than the one generated by $\nlops$.
A similar trend has been seen in Fig.~\ref{fig:NLO_NWA_decayLOvsNLO}, 
where $\lodec$ is compared to $\nlodec$.
Turning to Fig.~\ref{fig:PS_vs_ttbar_NLO_nlodc}, we show the case
where the $\nlops$ pseudo-data have been confronted with the
improved fixed-order model $\nlodec$. For this case, we would expect a
pseudo-data-theory agreement which is better than the one seen in
Fig.~\ref{fig:PS_vs_ttbar_NLO}, since
both the $\nlops$ and the $\nlodec$ description contain the major
contributions to describe the extra emission in the top quark decays.
However, the offset of
$0.96\pm0.07\gev$ is similar in size (while opposite in direction)
compared to the LO decay case shown in Fig.~\ref{fig:PS_vs_ttbar_NLO}.
This is consistent with the offset differences shown in Table~\ref{tab:offset}, 
for example subtracting the offset given in Fig.~\ref{fig:NLO_NWA_decayLOvsNLO}
 from the one in Fig.~\ref{fig:PS_vs_ttbar_NLO}, 
or alternatively the one in Fig.~\ref{fig:WWbb_vs_PS_NLO} from the one
in Fig.~\ref{fig:NLO_WWbb_vs_NWA_decayNLO}.

Further investigations are needed to understand the source of the 
mass shift observed in  Fig.~\ref{fig:PS_vs_ttbar_NLO_nlodc}. Based on
the current findings, we cannot conclude whether it
originates from ($i$)~the inclusion of resummation effects, or
($ii$)~genuine differences in incorporating the 
fixed-order QCD corrections to the production\footnote{The NLO
  treatment of production times decay is implemented differently in
  $\nlops$ and $\nlodec$. The parton shower calculation uses a
  multiplicative approach, whereas the fixed-order calculation is
  expanded in $\alpha_s$ up to ${\cal O}(\alpha_s^3)$, therefore
  leading to differences which are formally of next-to-next-to leading
  order. }  {\em and} decay of the top quark pairs, or both.
The different radiation
patterns generated by $\nlops$ and $\nlodec$ do not allow for a strict,
same-level comparison between the two approaches, but reducing the
amount of radiation produced by $\nlops$ is expected to bring them closer to
each other, and to diminish the role of resummation effects.

There is no unique way of limiting the scope of the resummation.
To control the generation of a reduced branching pattern, we use an approach where each showering process
can be terminated after a (given) fixed number of emissions, denoted
by $n_\mrm{max}$. For our study, we rely on the fully factorised
version of combining the subshowers, i.e.~we separately restrict the
number of emissions to no more than $n$ ($\nprod=\ndec=n$) in each
subshower (the primary one evolving the $t\bar t$ production and the
secondary one evolving the decays). 
The combination of one-emission production and decay showers
($\nprod=\ndec=1$) can then be used to emulate the $\nlodec$
calculation, which enables us to approximately separate effects ($i$)
from ($ii$).
In addition, comparing the restricted and full $\nlops$ prescriptions
will provide us with a qualitative
estimate of the impact of the full resummation.
Starting from $\nprod=1$ and $\ndec=1$, we can successively restore
the full shower by incrementing the number of emissions.

\begin{figure}[t!]
\centering
\begin{subfigure}{0.495\textwidth}
\includegraphics[width=\textwidth]{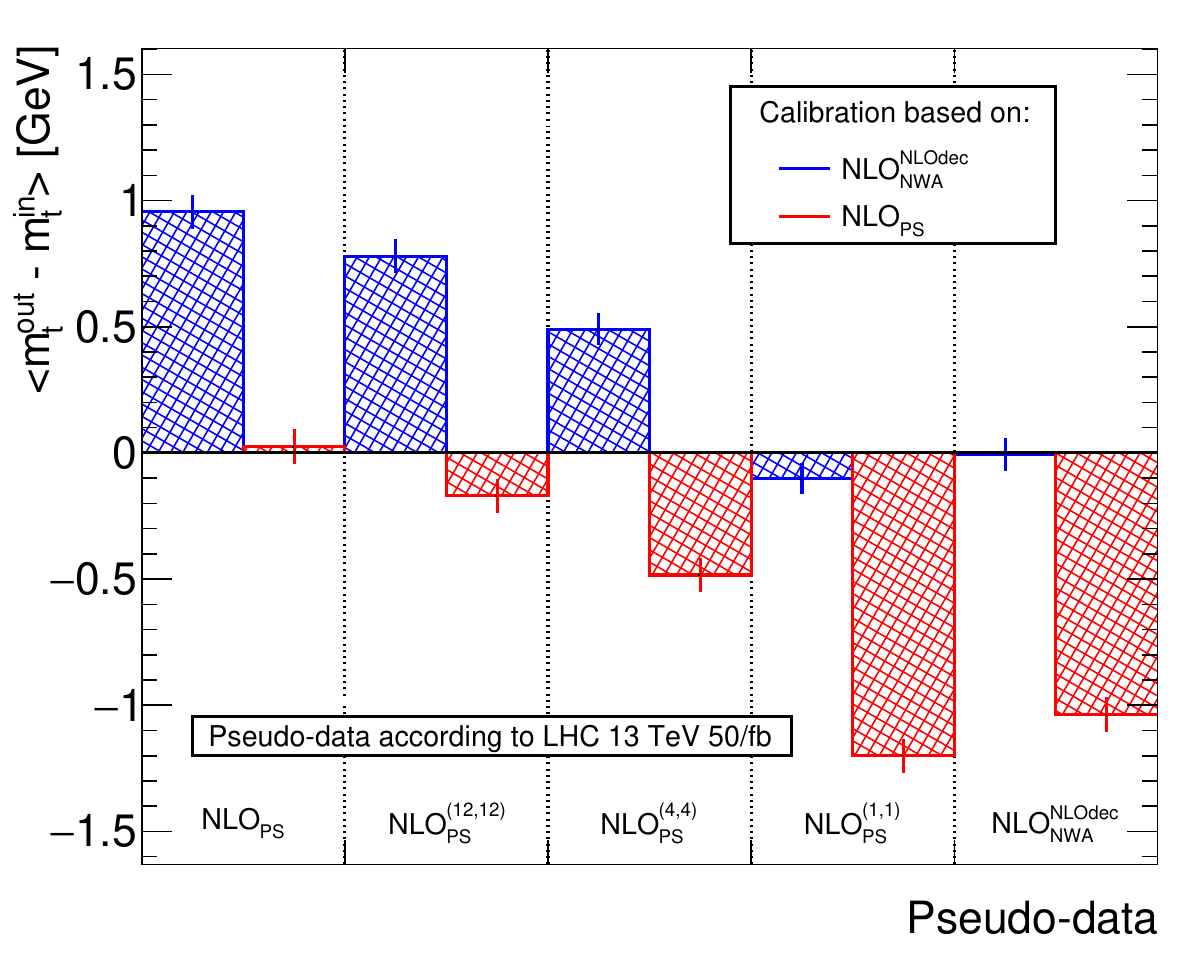}
\vspace{\TwoFigBottom em}
\caption{\label{fig:oneemit_showers_var}}
\end{subfigure}
\hfill
\begin{subfigure}{0.495\textwidth}
\includegraphics[width=\textwidth]{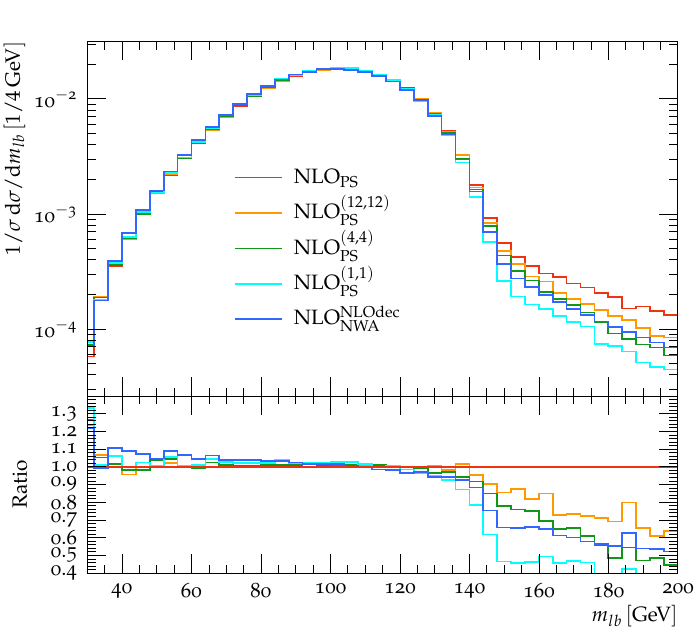}
\vspace{\TwoFigBottom em}
\caption{\label{fig:oneemit_showers_var_dist}}
\end{subfigure}
\caption{\label{fig:oneemit_showers}%
  Results of the restricted-shower study for the $m_{lb}$ observable
  using $\mrm{NLO}_\mrm{PS}^{(\nprod,\ndec)}$
  parton showers that terminate after a certain maximal number of
  emissions in both the production and decay showers.
  In~(a) mass offsets are shown for a number of pseudo-data sets using
  the $\nlodec$ and the $\nlops$ calibrations in the shape analysis.
  The sets of pseudo-data are generated according to the 
  $\nlodec$ description, the default $\nlops$ as well as three $\nlops$
  showers that differ in $\nprod=\ndec=n_\mrm{max}$. The corresponding $m_{lb}$
  distributions for the case $m_t=172.5\gev$ are given in~(b).}
\end{figure}
\begin{figure}[t!]
\centering
\begin{subfigure}{0.495\textwidth}
\includegraphics[width=\textwidth]{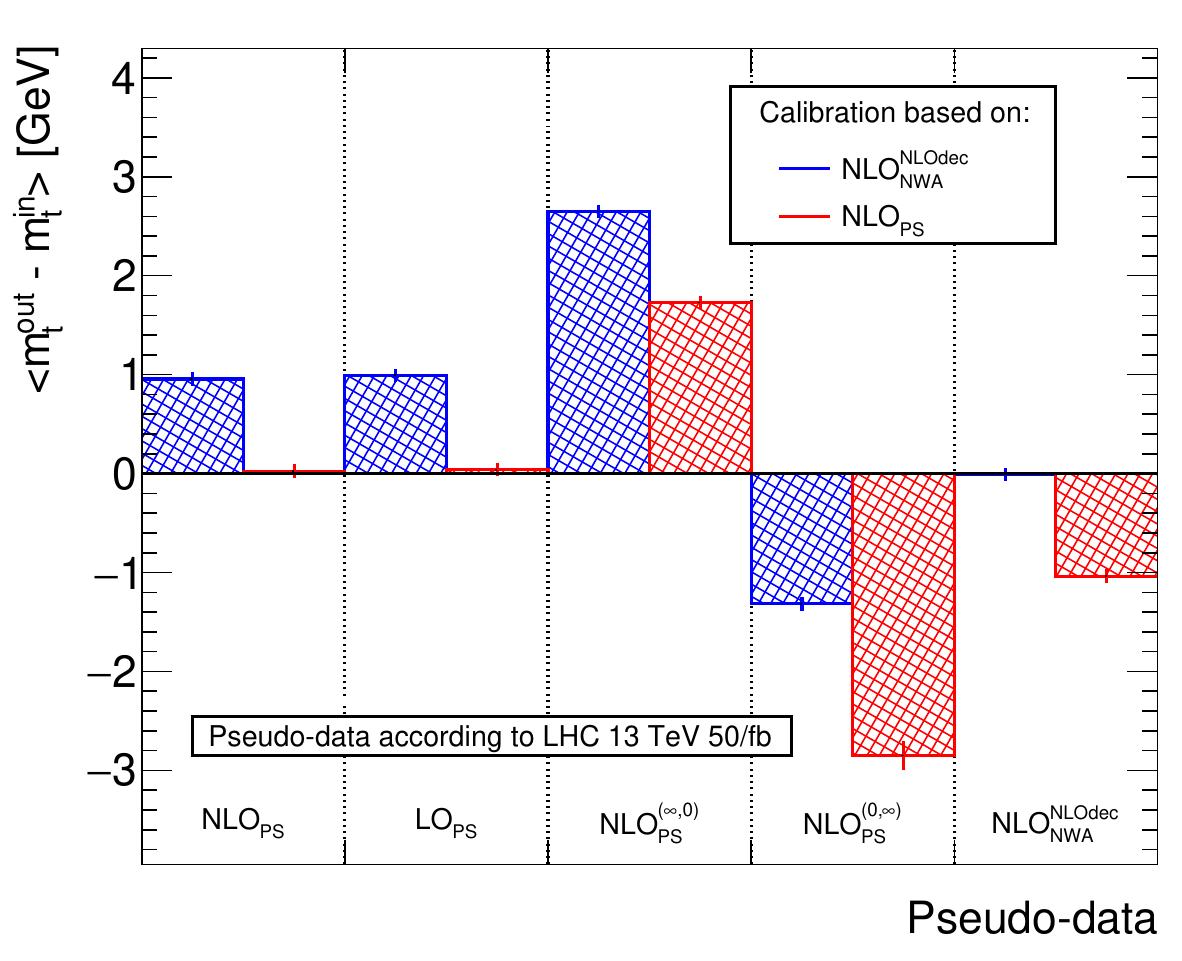}
\vspace{\TwoFigBottom em}
\caption{\label{fig:lo-prod-dec_showers_var}}
\end{subfigure}
\hfill
\begin{subfigure}{0.495\textwidth}
\includegraphics[width=\textwidth]{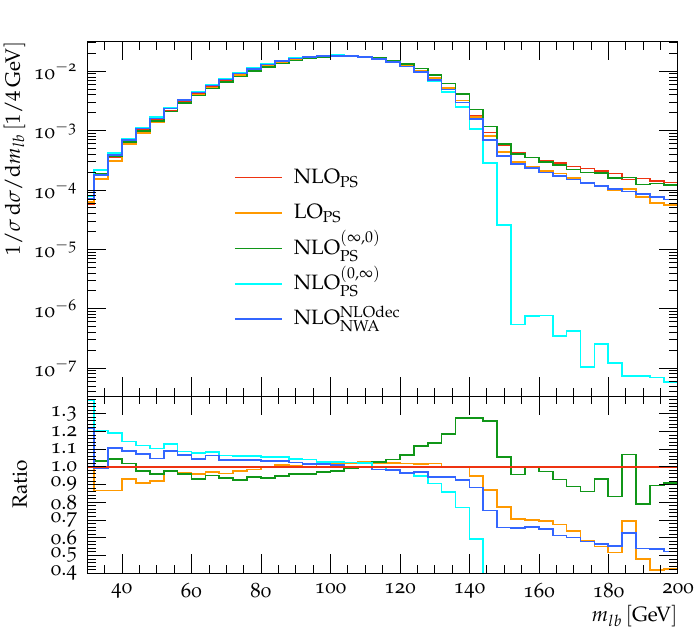}
\vspace{\TwoFigBottom em}
\caption{\label{fig:lo-prod-dec_showers_var_dist}}
\end{subfigure}
\caption{\label{fig:lo-prod-dec_showers}%
   Results of the restricted-shower study for the $m_{lb}$ observable
   using pure production, pure decay and pure LO parton showers only.
   In~(a) mass offsets are shown for a number of pseudo-data sets using    
   the $\nlodec$ and the $\nlops$ calibrations in the shape analysis.
   The sets of pseudo-data are generated according to the $\nlodec$
   description, the default $\nlops$ and $\lops$ showers as well as
   $\nlops$ showers whose evolution is restricted to the production or
   decay stage only. The corresponding $m_{lb}$ distributions for the
   case $m_t=172.5\gev$ are given in~(b).}
\vspace*{-4mm}
\end{figure}

\afterpage{\clearpage}

Figure~\ref{fig:oneemit_showers} summarises the results of the
restricted-shower studies.
 Figure~\ref{fig:oneemit_showers_var} shows the offsets and their
 statistical uncertainties for sets of pseudo-data analysed with two
 calibrations, namely $\nlops$ and $\nlodec$, while the figure to the
 right, Fig.~\ref{fig:oneemit_showers_var_dist}, depicts the
 corresponding \mlb\ distributions.
 The leftmost bin in Fig.~\ref{fig:oneemit_showers_var} corresponds to the
 mass shifts displayed in Fig.~\ref{fig:PS_vs_ttbar_NLO_nlodc}. The blue bar depicts the
 offset of the $\nlops$ pseudo-data, analysed with the $\nlodec$ calibration.
 The almost vanishing red bar shows the closure for the $\nlops$ pseudo-data and
 calibration.
 Moving to the right, the parton shower is more and more restricted,
 allowing for at
 most $12$, $4$ and $1$ emissions in each subshower.
 This results in a smooth transition from the offset of
 $0.96\pm0.07\gev$ to almost
 zero (with an indication of a small overshoot to negative offsets).
 The mass shifts becoming fairly small for more restricted showering
 indicate that most of the differences between the $\nlodec$ and
 $\nlops$ predictions emerge from resummation effects.
 Finally, the rightmost bin is for the $\nlodec$ pseudo-data themselves.

 The calculated offsets, obtained from fits to the \mlb\ distributions
 like the ones in Fig.~\ref{fig:oneemit_showers_var_dist},
 receive contributions from regions with large differential cross
 sections and small differences between restricted shower and
 calibration ($\nlodec$ and $\nlops$) results, as well as from regions
 with small differential cross sections and relatively large differences.
 The interplay of these effects can lead to situations such as the one
 observed here, where the mass offsets obtained from
 $\mrm{NLO}_\mrm{PS}^{(1,1)}$ pseudo-data are closer to the ones
 obtained by using $\nlodec$ pseudo-data, despite the fact that the
 $\mrm{NLO}_\mrm{PS}^{(4,4)}$ curve is closer to $\nlodec$ for \mlb\
 values around the kinematic edge and beyond.

We complete the parton shower studies by presenting offsets and
\mlb distributions for parton shower descriptions where we separately
switch off ($a$)~the NLO corrections to the $t\bar t$ production i.e.~use $\lops$,
($b$)~the emissions in the decay showers, denoted by~$\mrm{NLO}_\mrm{PS}^{(\infty,0)}$,
and ($c$)~the emissions in the production shower, denoted by $\mrm{NLO}_\mrm{PS}^{(0,\infty)}$.
For the corresponding results in Fig.~\ref{fig:lo-prod-dec_showers},
the same calibrations as in Fig.~\ref{fig:oneemit_showers} are used.
We find that the offsets for the $\lops$ and $\nlops$ predictions
agree very well, although the shape of the $\lops$ \mlb\ distribution
in Fig.~\ref{fig:lo-prod-dec_showers_var_dist} substantially deviates
from the $\nlops$ one outside the range $70\gev<\mlb<140\gev$.
This means we observe similar compensating effects in the fit as
discussed for Fig.~\ref{fig:oneemit_showers}.
The small difference in the offsets indicates that the NLO treatment
of the production process included by the $\nlops$ prescription has a
minor impact on the fit. The nearly vanishing offset between
the $\lops$ pseudo-data and $\nlops$ calibration is likely to be a consequence of
the same resummation corrections being applied in both showers.

The $\mrm{NLO}_\mrm{PS}^{(\infty,0)}$ prediction in Fig.~\ref{fig:lo-prod-dec_showers}
can be considered as the shower correction to $t\bar t$ production at NLO ($\lodec$), 
while the $\mrm{NLO}_\mrm{PS}^{(0,\infty)}$ prediction represents the shower approximation to the
radiative corrections in the top quark decays.
The use of the related pseudo-data increases the absolute mass offsets
for both the $\nlops$ and the $\nlodec$ calibration, illustrating that
the production shower predominantly evolves through initial-state
radiation (resulting in larger fitted \mt) while the decay showers are
mostly driven by final-state radiation (yielding smaller \mt).
This is induced by the corresponding \mlb\ distributions in
Fig.~\ref{fig:lo-prod-dec_showers_var_dist}, where we observe that the
$\mrm{NLO}_\mrm{PS}^{(\infty,0)}$ prediction is enhanced for larger \mlb
values, in particular around the kinematic edge of the distribution,
while the $\mrm{NLO}_\mrm{PS}^{(0,\infty)}$ prediction turns out to be
softer than the others, showing a very sharp kinematic edge.
For the $\nlodec$ calibration, the sum of the mass
offsets for production shower pseudo-data, amounting to
$2.65\pm0.07\gev$, and decay showers pseudo-data, amounting to
$-1.32\pm0.07\gev$, is close to the mass shift of $0.96\pm0.07\gev$
obtained for $\nlops$ pseudo-data.
This means that the generation-level factorisation (dissection) of the emission
patterns for production and decays almost completely carries over to
the analysis level.

\begin{figure}[tbp!]
  \centering
  \begin{subfigure}{0.495\textwidth}
    \includegraphics[width=\textwidth]{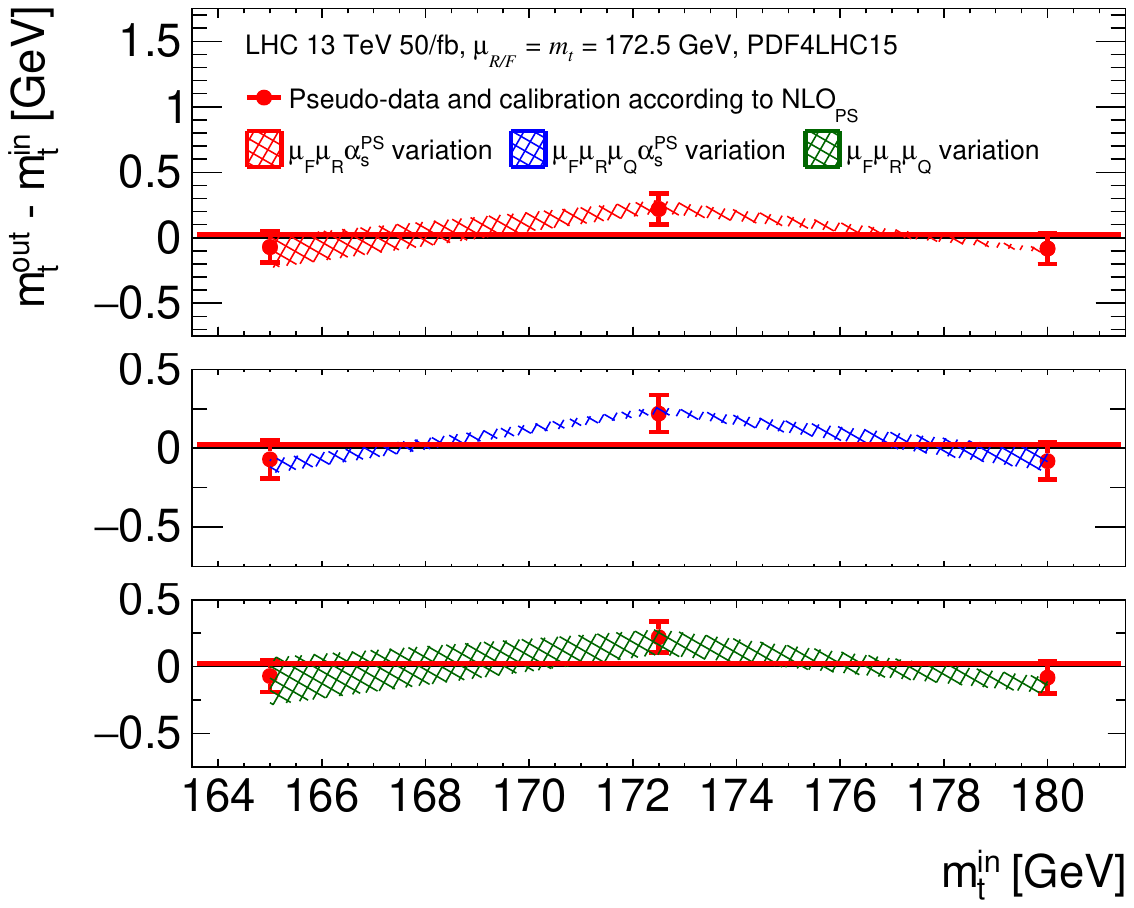}
    \caption{\label{fig:PS_scale_var}}
  \end{subfigure}
  \hfill
  \begin{subfigure}{0.495\textwidth}
    \includegraphics[width=\textwidth]{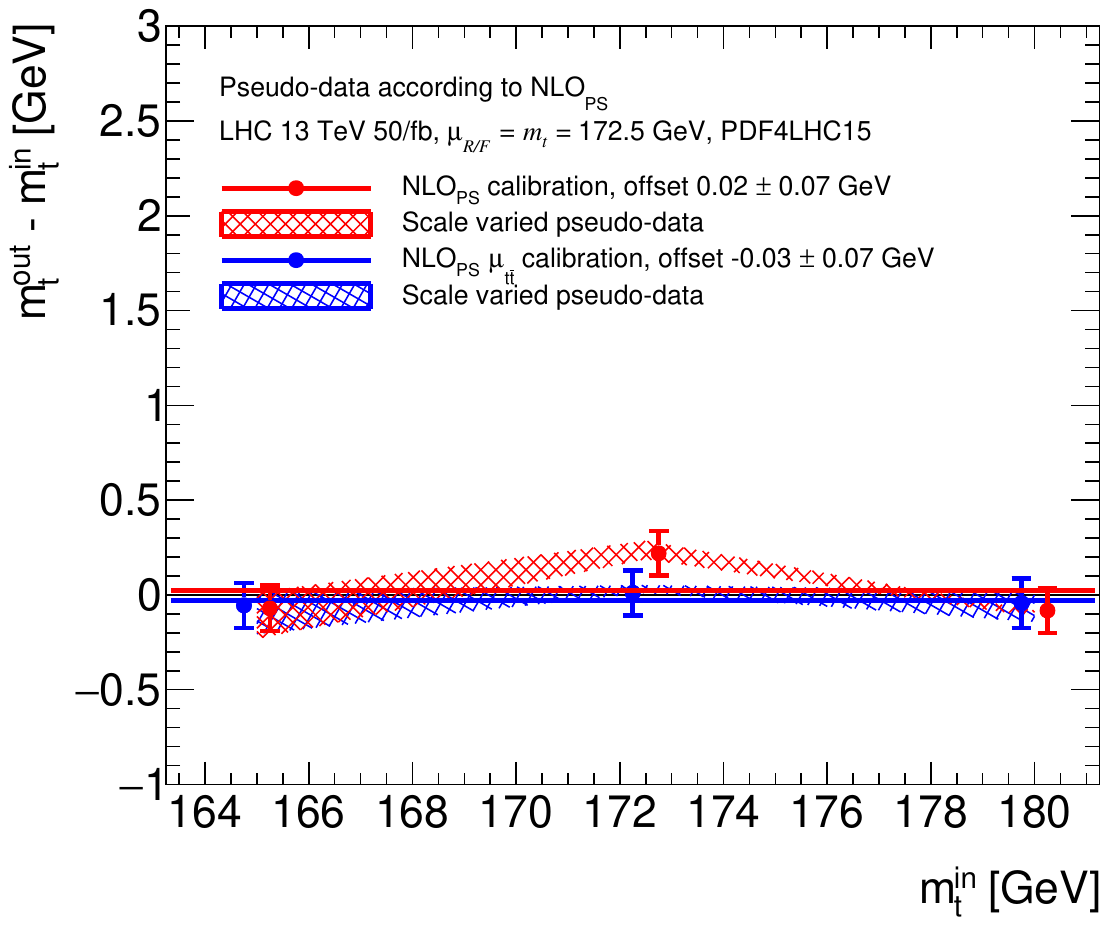}
    \caption{\label{fig:PS_scale_NLO}}
  \end{subfigure}
  \caption{\label{fig:PS_scale}%
    Results for different schemes determining the parton shower scale
    dependence using the $m_{lb}$ observable and pseudo-data as well
    as calibrations derived from $\nlops$ predictions.
    In~(a), the uncertainty bands are shown for the different ways of
    evaluating the shower scale dependence (cf.~Section~\ref{subsec:input}).
    The offsets and uncertainty bands for different central scale
    choices used in the computation of the hard process are shown in~(b).}
\end{figure}

The \mlb distributions of the restricted and full showering show clear
differences. To quantify the significance of these differences, the
parton shower scale uncertainties are assessed.
For the decay showers, we performed a decay shower starting scale variation by
using factors of $0.5$ and $2.0$ applied to the central scale $\muqdec$. Despite this wide range for varying the resummation scales, we find negligible differences in the shapes of the \mlb\ distributions.
Therefore, all variations of the shower description employed here are always based on the fixed value $\muqdec=M_W/2$.
We use the different schemes described in Section~\ref{subsec:input}
to obtain the scale-variation induced theory uncertainties of the $\nlops$
prescription presented in Fig.~\ref{fig:PS_scale_var}.
While the combined variation, $\mu_F\mu_R\mu_Q\as{\mathrm{PS}}$, leads to the smallest uncertainty band, the band based on the $\mu_F\mu_R\mu_Q$ parameter variation is marginally larger.
Most notably, these differences are much smaller than those occurring between
the various theory descriptions discussed above.

Finally, for the $\nlops$ calculations, we compare in Fig.~\ref{fig:PS_scale_NLO}  
the results for the two central-scale choices $\mu_R=\mu_F=m_t$ and $\mu_R=\mu_F=\mu_{t\bar t}$ as defined in Eq.~(\ref{eq:scale_ttb}).
Although the predicted total cross sections listed in the last two rows of Table~\ref{tab:xs} depend on this choice, the two predictions lead to consistent measured top quark masses, i.e.~the associated offsets agree within their uncertainties.

As can be inferred from Fig.~\ref{fig:massvar_fullNLO_mlb}, the sensitivity of
the \mlb\ observable to the top quark mass, and consequently the achievable
statistical uncertainty on \mt\ in data, depends on the fit range used.
In this context, the range $140$--$160\gev$ is a particularly \mt-sensitive
region, which however also features sizeable differences in the theoretical
descriptions, for example as shown in Fig.~\ref{fig:scalevar_mlb}.
Consequently, the resulting offsets listed in Table~\ref{tab:offset}
depend on the chosen fit range.
As an example, restricting the fit to $\mlb<140\gev$ results in
absolute differences in the offsets between full range and reduced range of $min=0.05\gev$ and $max=0.36\gev$, 
where $min$ corresponds to $\nlops$ pseudo-data versus $\nlodec$ calibration, 
and $max$ corresponds to $\nlodec$ pseudo-data versus $\lolo$ calibration.
In general, larger differences are observed either for larger absolute
offsets or for cases with large uncertainty bands.
As a result, within the given uncertainties the general pattern does not depend on the fit range.
An experimental analysis should be optimised for the smallest total
uncertainty, including the variation of the relative importance of statistical
and systematic uncertainties, while changing the fit range. Therefore we
consider the results shown in Table~\ref{tab:offset}, based on the fit ranges
given in Eq.~(\ref{eq:fitranges}), as our nominal values.

\begin{table}[tb!]
\small
\begin{center}
\begin{tabular}{|l|l|c|c|c|c|c|}
\cline{3-6}
\multicolumn{2}{c}{}         & \multicolumn{2}{|c|}{Offset [GeV]} &
\multicolumn{2}{|c|}{Figure} & \multicolumn{1}{c}{}\\\hline
 Pseudo-data &  Calibration & \mlb & \mtwo  &\mlb & \mtwo & \Chiq \\\hline
\hline
&&&&&&\\
 $\lodec$  & $\mathrm{LO_{NWA}^{LOdec}}$ & $+0.51 \pm 0.06$ &$+0.48 \pm 0.04$ &
 \ref{fig:LOvsNLO_NWA_decayLO} & \ref{fig:LOvsNLO_NWA_decayLO_mt2}   & 0.17\\
 $\nlodec$ &                 $\lodec$ & $-1.80 \pm 0.06$ &$-1.67 \pm 0.04$ &
 \ref{fig:NLO_NWA_decayLOvsNLO} & \ref{fig:NLO_NWA_decayLOvsNLO_mt2} & 3.25\\
 $\nlodec$ & $\mathrm{LO_{NWA}^{LOdec}}$ & $-1.38 \pm 0.07$ &$-1.24 \pm 0.05$ &
 \ref{fig:NLO_NWA_decayNLO} & \ref{fig:NLO_NWA_decayNLO_mt2}         & 2.65\\
$\nlofull$ &                $\lofull$ & $-1.52 \pm 0.07$ &$-1.62 \pm 0.05$ &
 \ref{fig:fullNLO} & \ref{fig:fullNLO_mt2}                           & 1.35\\
$\nlofull$ &                $\nlodec$ & $+0.83 \pm 0.07$ &$+0.60 \pm 0.06$ &
 \ref{fig:NLO_WWbb_vs_NWA_decayNLO} & \ref{fig:NLO_WWbb_vs_NWA_decayNLO_mt2}
                                                                     & 6.22\\
$\nlofull$ &                 $\nlops$ & $-0.09 \pm 0.07$ &$-0.07 \pm 0.06$ &
 \ref{fig:WWbb_vs_PS_NLO} & \ref{fig:WWbb_vs_PS_NLO_mt2}             & 0.05\\
  $\nlops$ &                 $\lodec$ & $-0.92 \pm 0.07$ &$-1.17 \pm 0.05$ &
\ref{fig:PS_vs_ttbar_NLO} & \ref{fig:PS_vs_ttbar_NLO_mt2}            & 8.45 \\
  $\nlops$ &                $\nlodec$ & $+0.96 \pm 0.07$ &$+0.68 \pm 0.05$ &
\ref{fig:PS_vs_ttbar_NLO_nlodc} & \ref{fig:PS_vs_ttbar_NLO_nlodc_mt2}& 10.59\\
  $\nlops$ &   $\nlops~(\mu_{t\bar{t}})$ & $-0.03 \pm 0.07$ &$+0.02 \pm 0.05$ &
 \ref{fig:PS_scale_NLO} & \ref{fig:PS_scale_NLO_mt2}                 & 0.34\\
&&&&&&\\
\hline
\end{tabular}
\end{center}
\caption{Summary of the offsets observed when analysing pseudo-data listed in
  the first column with template fit functions calibrated based on various
  theoretical predictions as given in the second column.
 The observed offsets for the two observables \mlb\ and \mtwo\ are reported in
 the second pair of columns, where the corresponding figures are listed in the
 next pair of columns.
 Finally, the \Chiq\ for the differences in the offsets for the two observables
 are displayed in the rightmost column, see text for further details.
\label{tab:offset}}
\end{table}

\boldmath
\subsection{Fit results for $\mtwo$}
\unboldmath

\begin{figure}[tbp!]
\centering
\begin{subfigure}{0.495\textwidth}
\includegraphics[width=\textwidth]{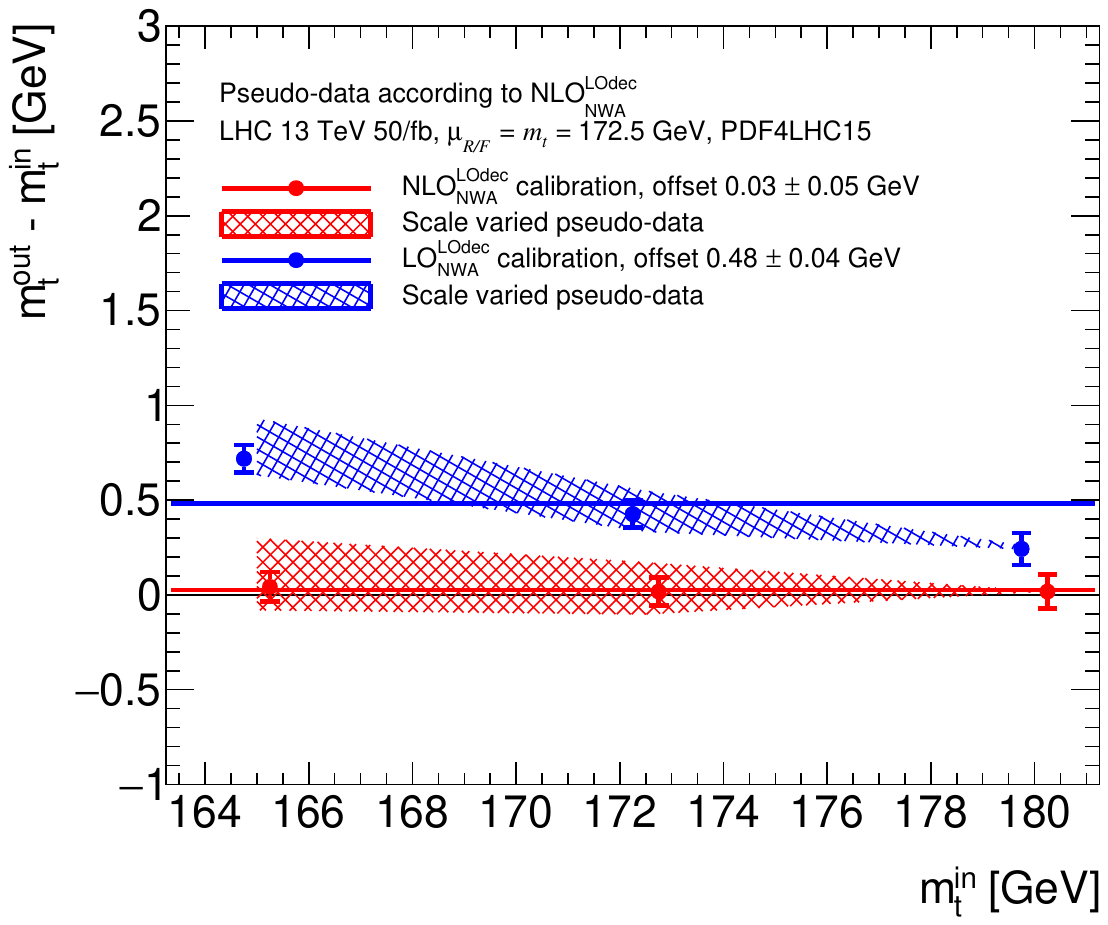}
\vspace{\TwoFigBottom em}
\caption{\label{fig:LOvsNLO_NWA_decayLO_mt2}}
\end{subfigure}
\hfill
\begin{subfigure}{0.495\textwidth}
\includegraphics[width=\textwidth]{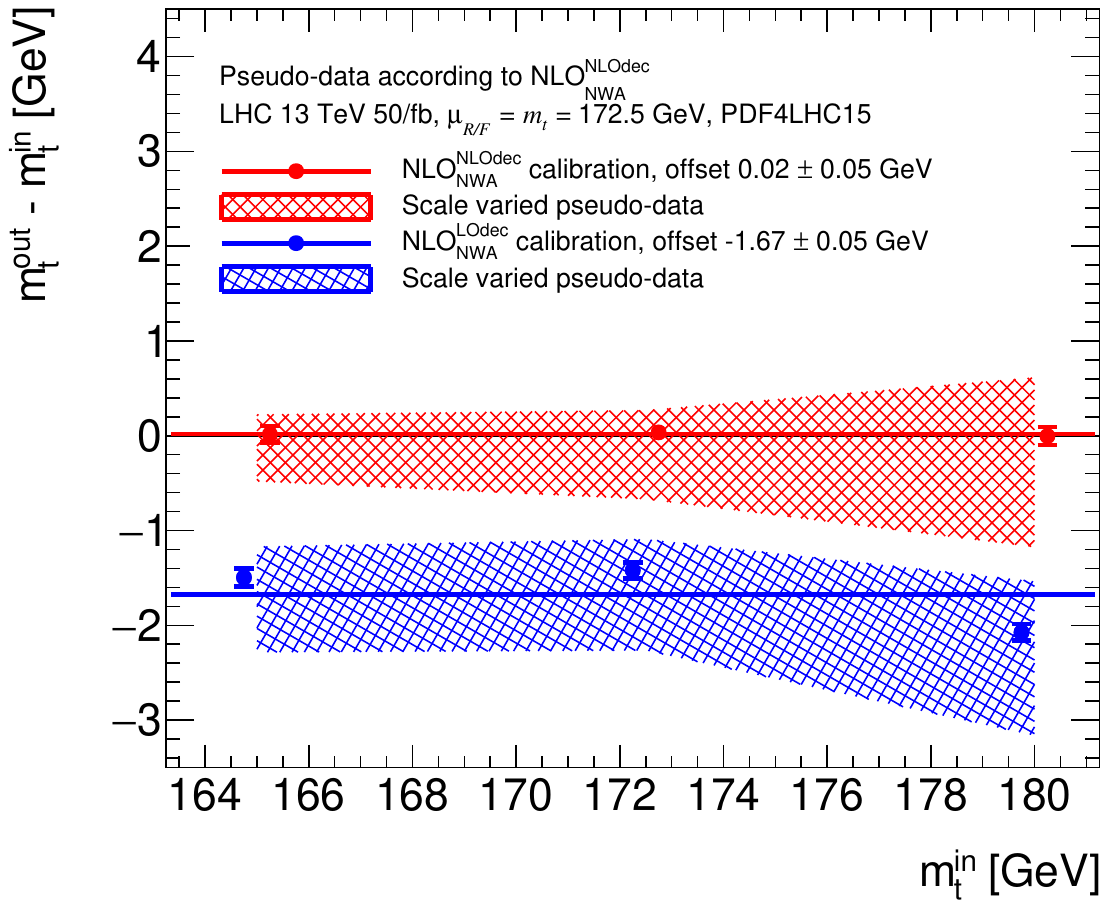}
\vspace{\TwoFigBottom em}
\caption{\label{fig:NLO_NWA_decayLOvsNLO_mt2}}
\end{subfigure}
\caption{\label{fig:NWA_mt2}%
  Same as Fig.~\ref{fig:NWA} but for the observable $m_{T2}$.}
\end{figure}

\begin{figure}[tbp!]
\centering
\begin{subfigure}{0.495\textwidth}
\includegraphics[width=\textwidth]{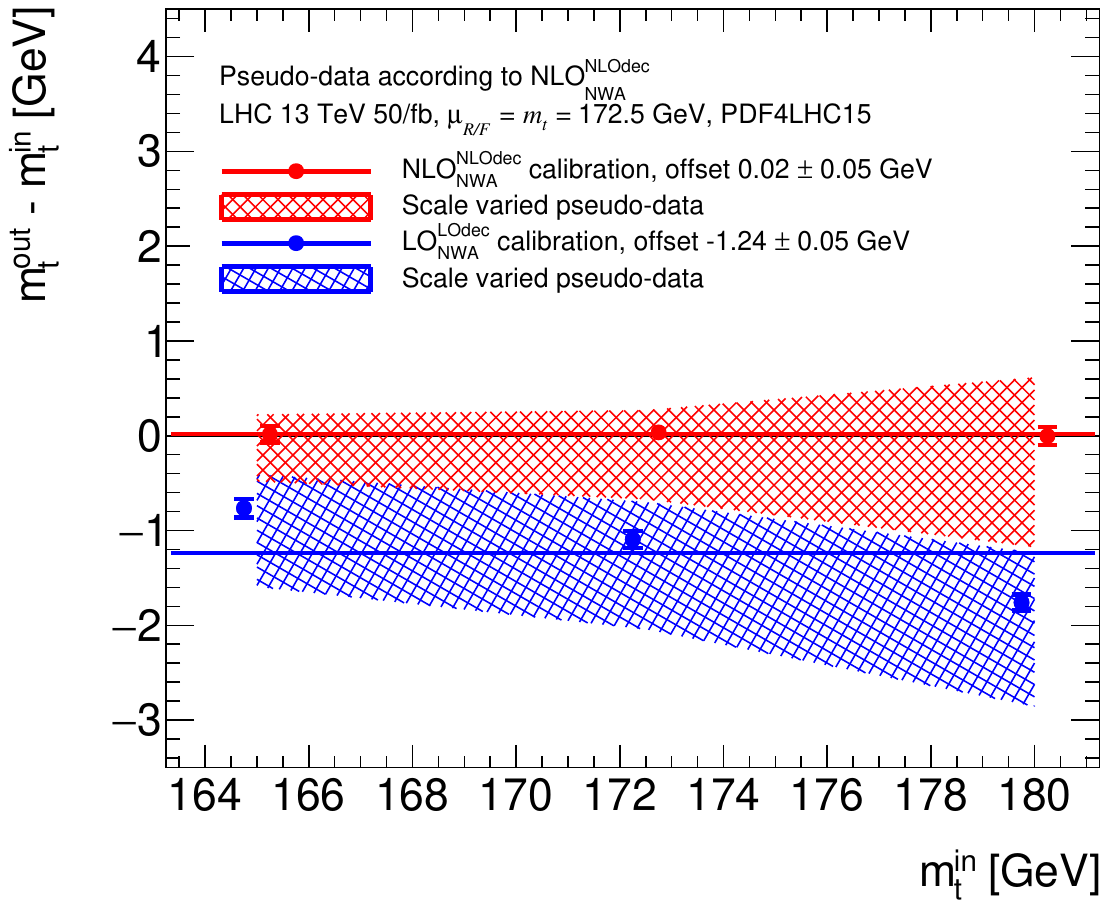}
\vspace{\TwoFigBottom em}
\caption{\label{fig:NLO_NWA_decayNLO_mt2}}
\end{subfigure}
\hfill
\begin{subfigure}{0.495\textwidth}
\includegraphics[width=\textwidth]{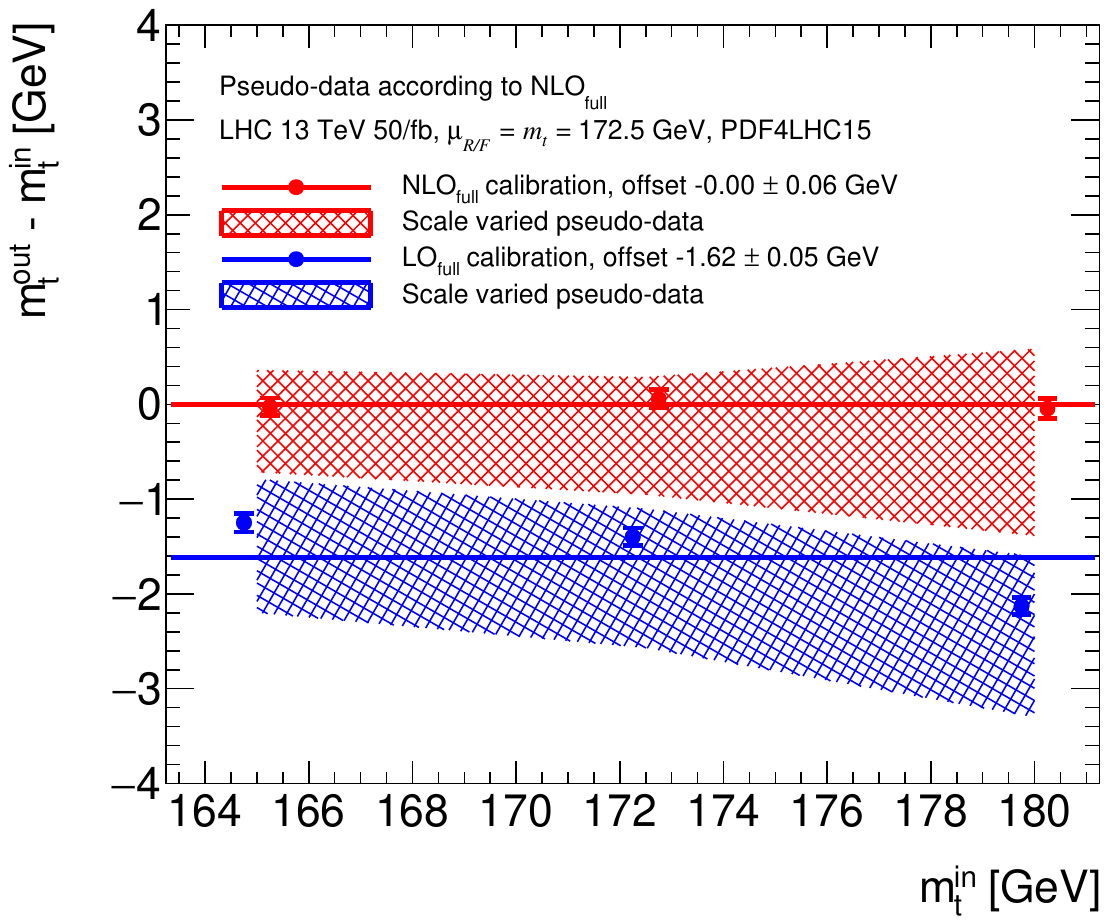}
\vspace{\TwoFigBottom em}
\caption{\label{fig:fullNLO_mt2}}
\end{subfigure}
\caption{\label{fig:NWAboth_mt2}%
  Same as Fig.~\ref{fig:NWAboth} but for the observable $m_{T2}$.}
\end{figure}

\begin{figure}[tbp!]
\centering
\begin{subfigure}{0.495\textwidth}
\includegraphics[width=\textwidth]{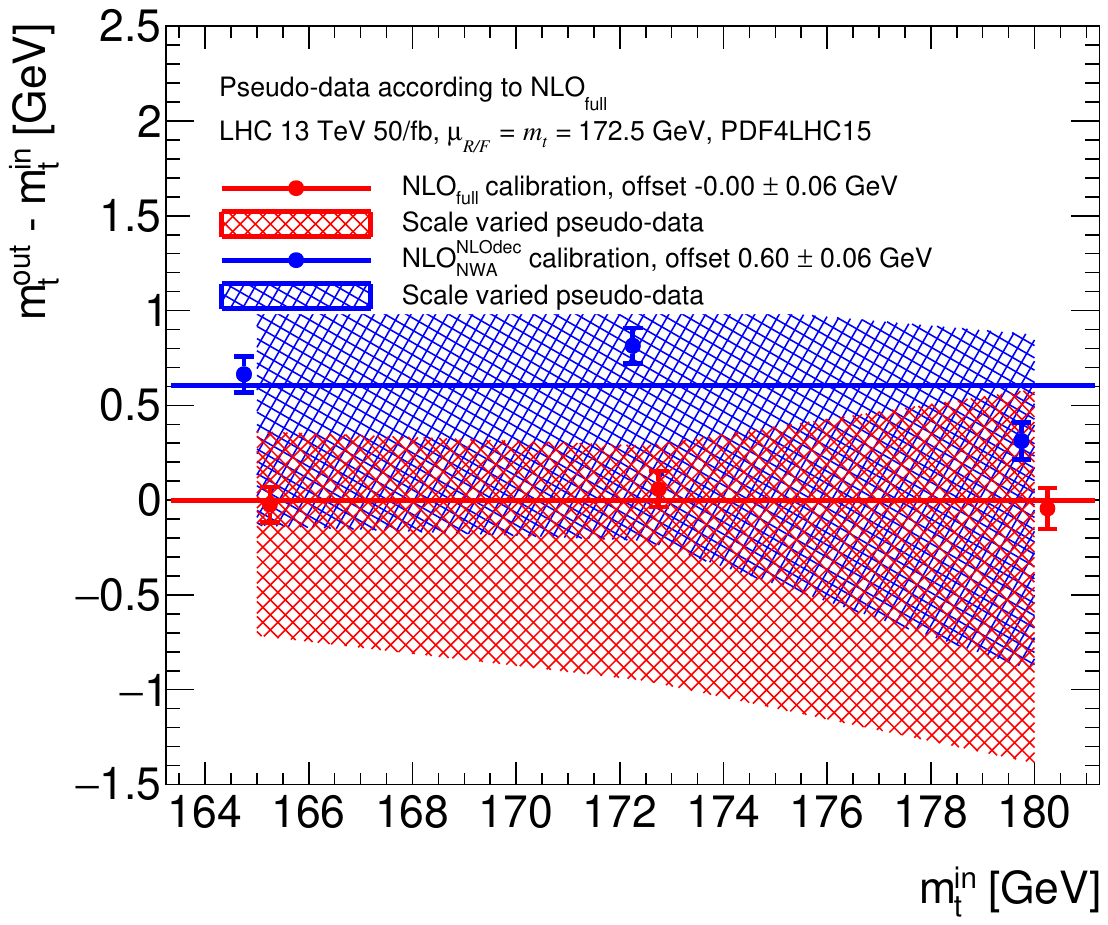}
\vspace{\TwoFigBottom em}
\caption{\label{fig:NLO_WWbb_vs_NWA_decayNLO_mt2}}
\end{subfigure}
\hfill
\begin{subfigure}{0.495\textwidth}
\includegraphics[width=\textwidth]{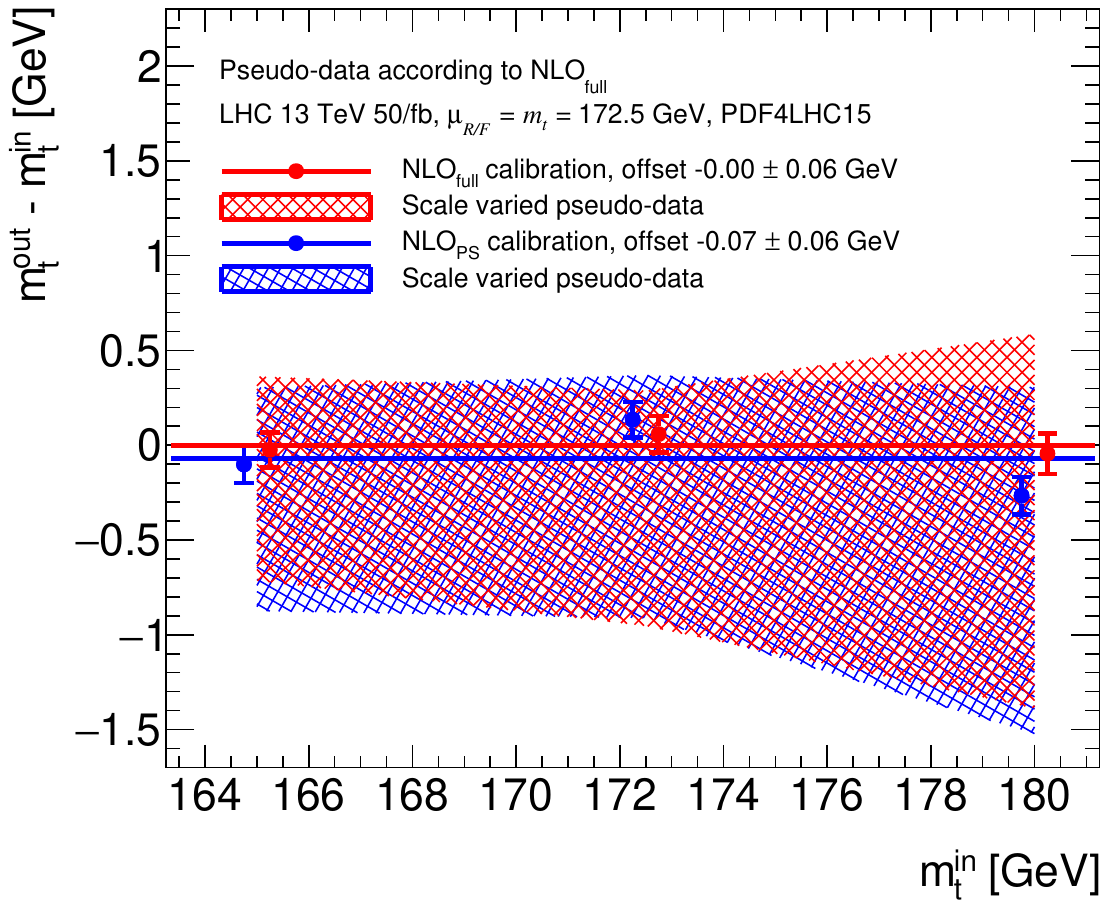}
\vspace{\TwoFigBottom em}
\caption{\label{fig:WWbb_vs_PS_NLO_mt2}}
\end{subfigure}
\caption{\label{fig:NLO_WWbb_vs_mt2}%
  Same as Fig.~\ref{fig:NLO_WWbb_vs} but for the observable $m_{T2}$.}
\end{figure}

\begin{figure}[tbp!]
\centering
\begin{subfigure}{0.495\textwidth}
\includegraphics[width=\textwidth]{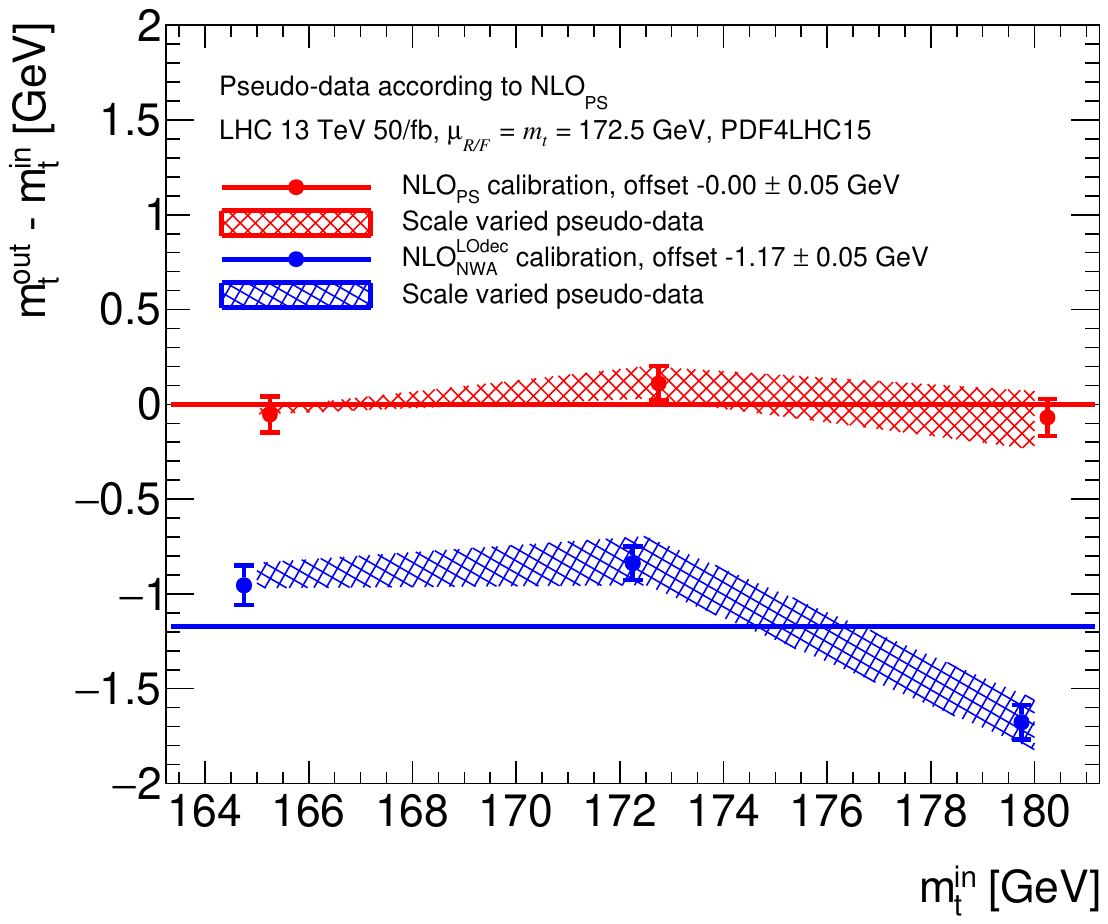}
\vspace{\TwoFigBottom em}
\caption{\label{fig:PS_vs_ttbar_NLO_mt2}}
\end{subfigure}
\hfill
\begin{subfigure}{0.495\textwidth}
\includegraphics[width=\textwidth]{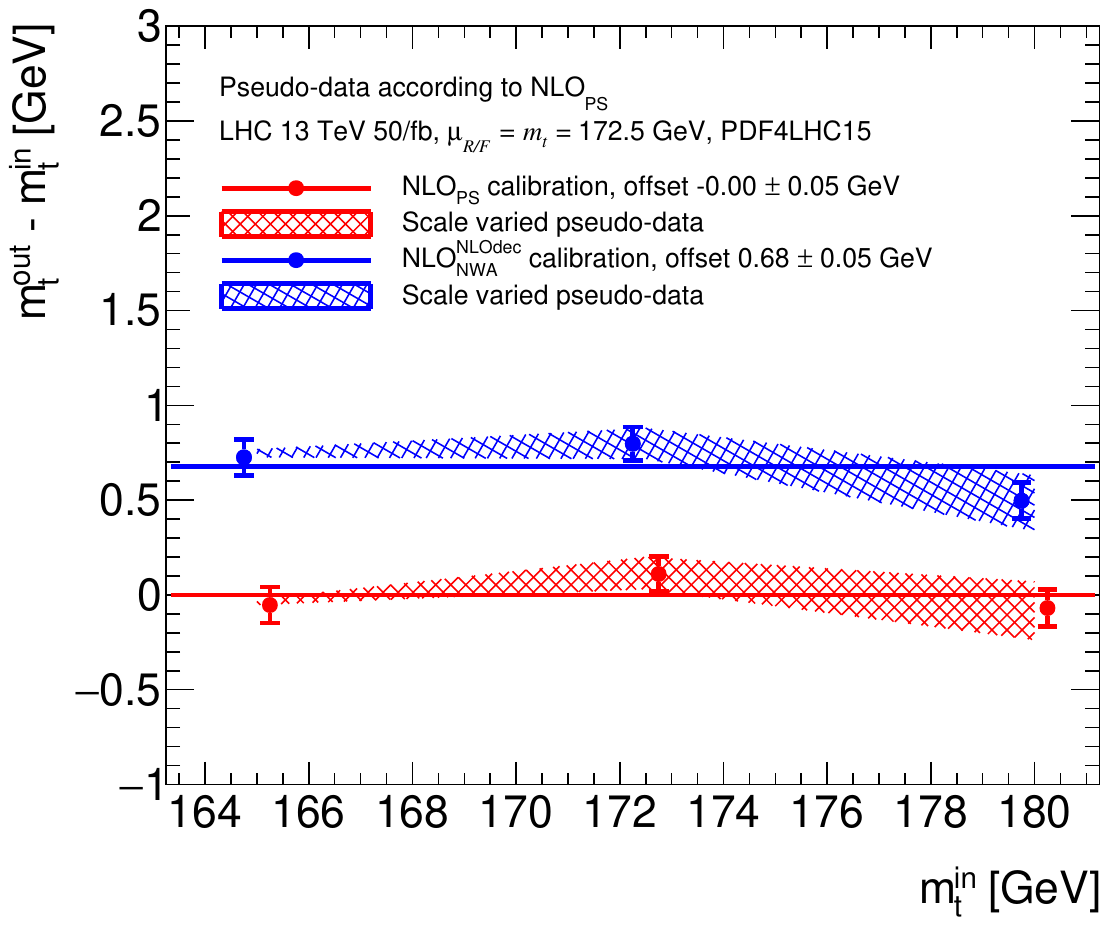}
\vspace{\TwoFigBottom em}
\caption{\label{fig:PS_vs_ttbar_NLO_nlodc_mt2}}
\end{subfigure}
\caption{\label{fig:NLO_PS_vs_mt2}%
  Same as Fig.~\ref{fig:NLO_PS_vs} but for the observable $m_{T2}$.}
\end{figure}

\begin{figure}[tbp!]
\centering
\begin{subfigure}{0.495\textwidth}
\includegraphics[width=\textwidth]{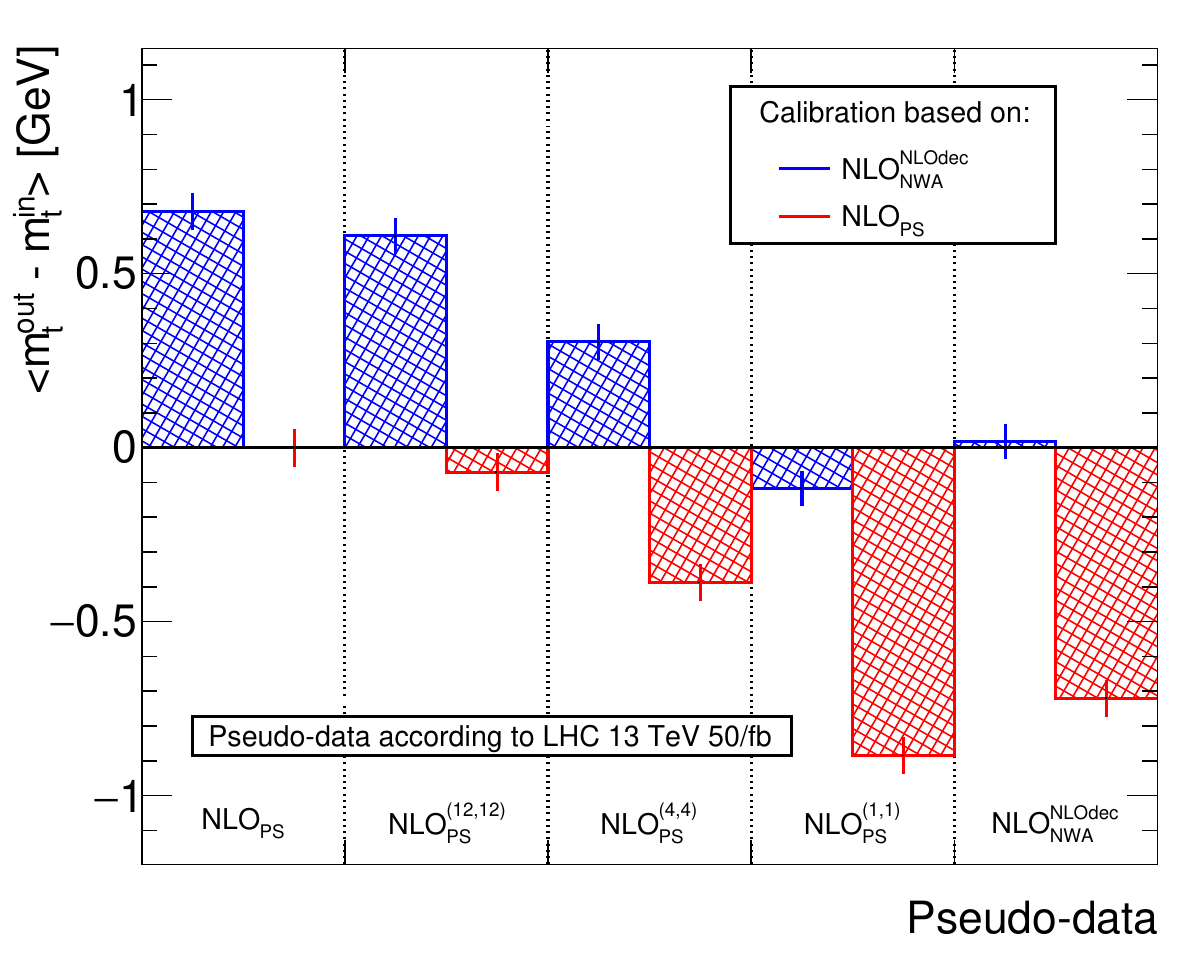}
\vspace{\TwoFigBottom em}
\caption{\label{fig:oneemit_showers_var_mt2}}
\end{subfigure}
\hfill
\begin{subfigure}{0.495\textwidth}
\includegraphics[width=\textwidth]{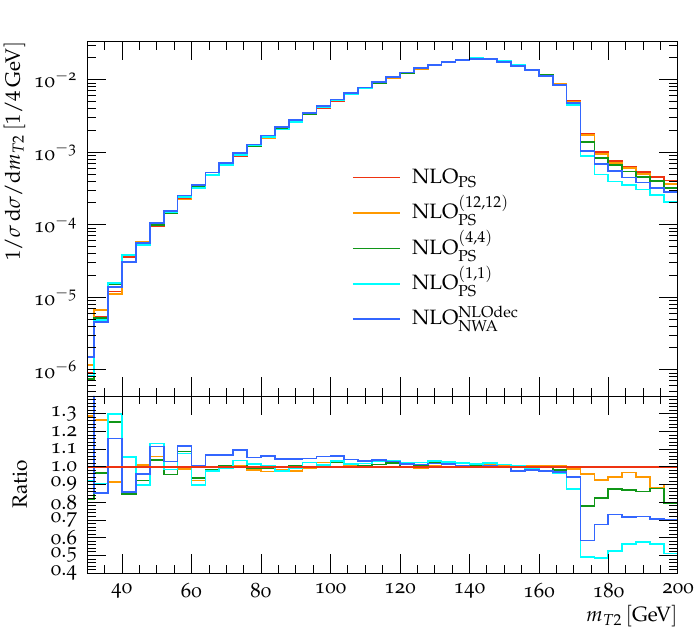}
\vspace{\TwoFigBottom em}
\caption{\label{fig:oneemit_showers_var_mt2_dist}}
\end{subfigure}
\caption{\label{fig:oneemit_showers_mt2}%
  Same as Fig.~\ref{fig:oneemit_showers} but for the observable $m_{T2}$.}
\end{figure}

\begin{figure}[tbp!]
\centering
\begin{subfigure}{0.495\textwidth}
\includegraphics[width=\textwidth]{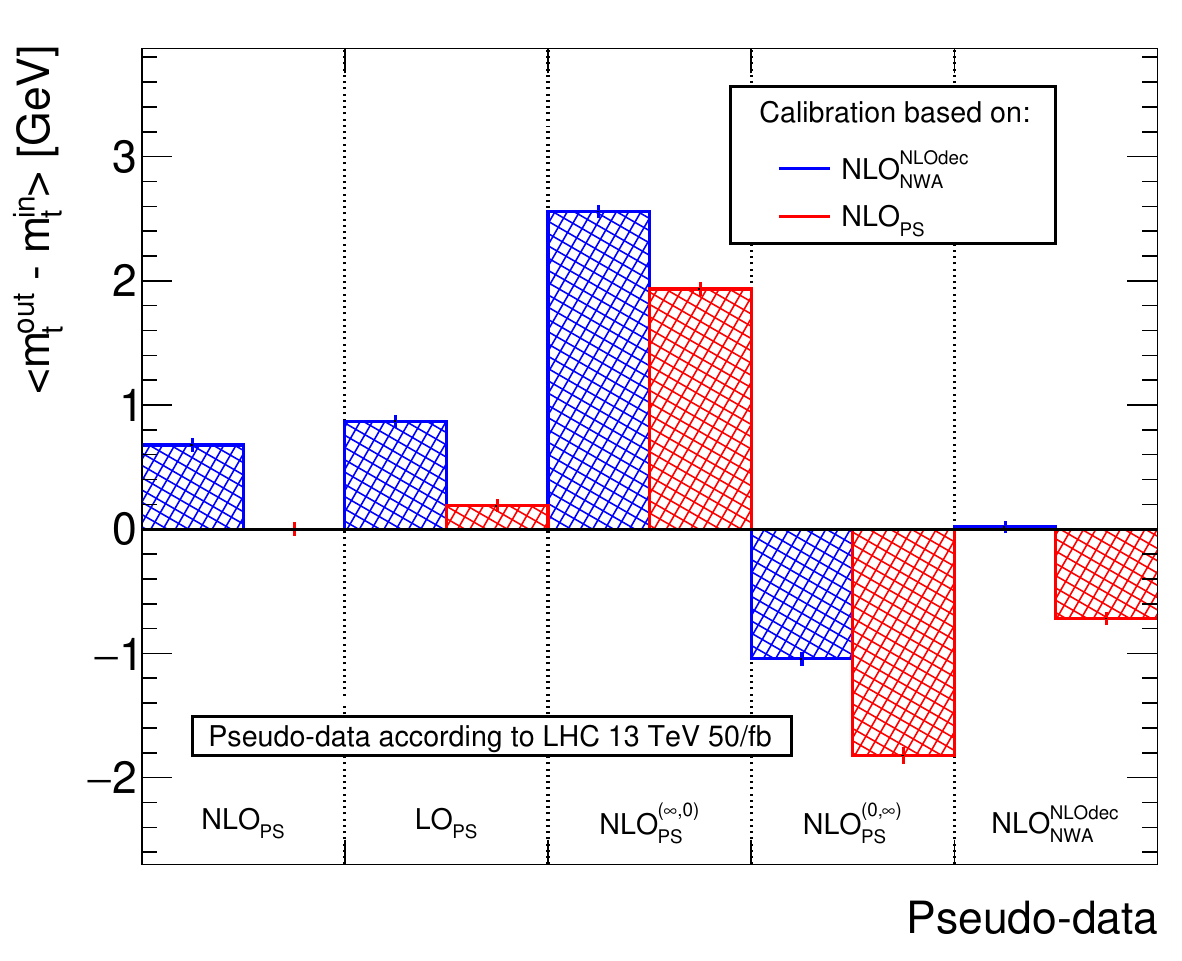}
\vspace{\TwoFigBottom em}
\caption{\label{fig:lo-prod-dec_showers_var_mt2}}
\end{subfigure}
\hfill
\begin{subfigure}{0.495\textwidth}
\includegraphics[width=\textwidth]{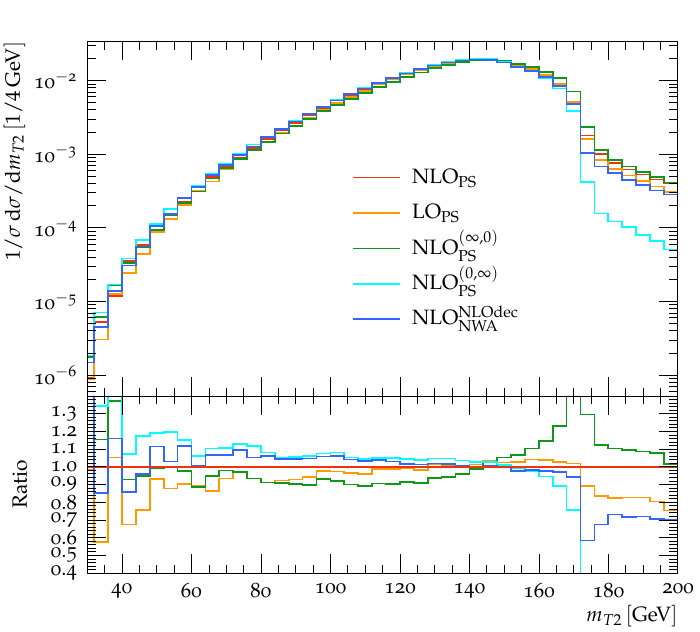}
\vspace{\TwoFigBottom em}
\caption{\label{fig:lo-prod-dec_showers_var_mt2_dist}}
\end{subfigure}
\caption{\label{fig:lo-prod-dec_showers_mt2}%
  Same as Fig.~\ref{fig:lo-prod-dec_showers} but for the observable $m_{T2}$.}
\end{figure}

\begin{figure}[tbp!]
\centering
\begin{subfigure}{0.495\textwidth}
\includegraphics[width=\textwidth]{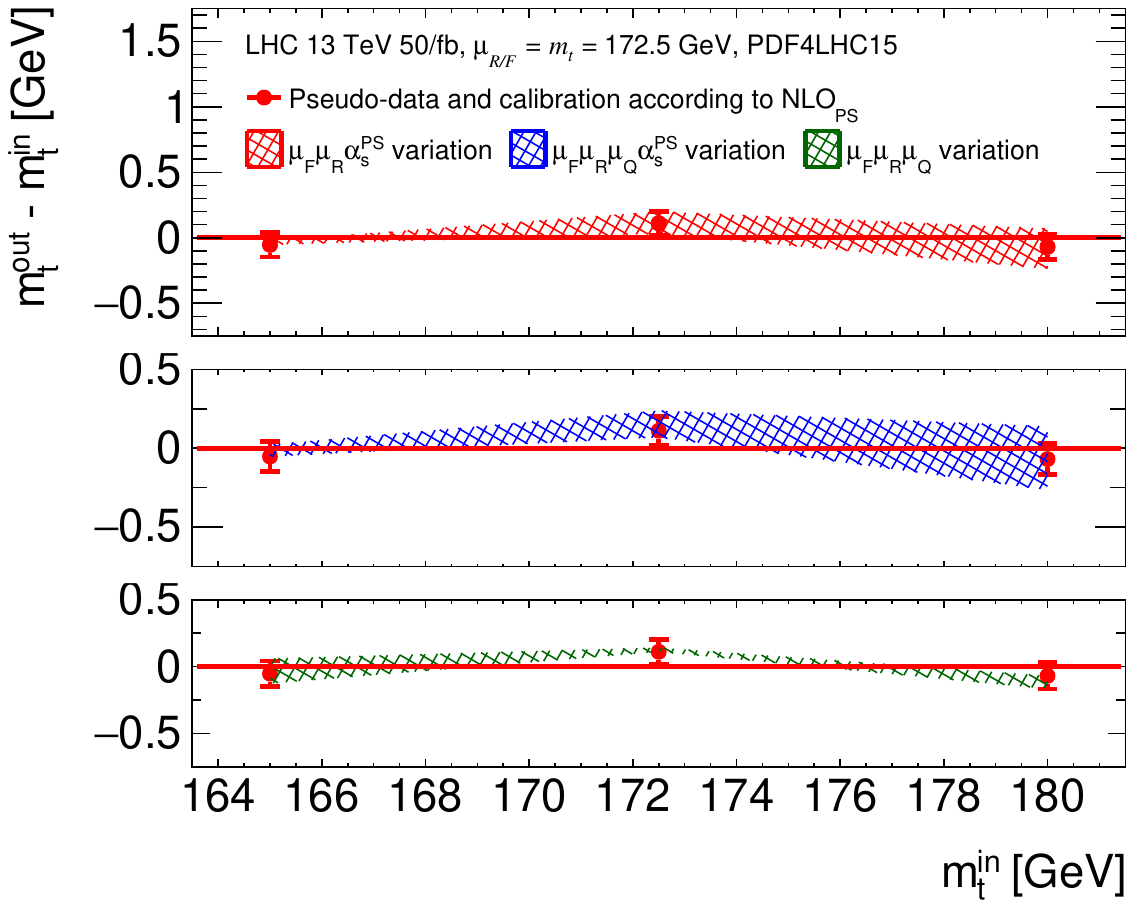}
\vspace{\TwoFigBottom em}
\caption{\label{fig:PS_scale_var_mt2}}
\end{subfigure}
\hfill
\begin{subfigure}{0.495\textwidth}
\includegraphics[width=\textwidth]{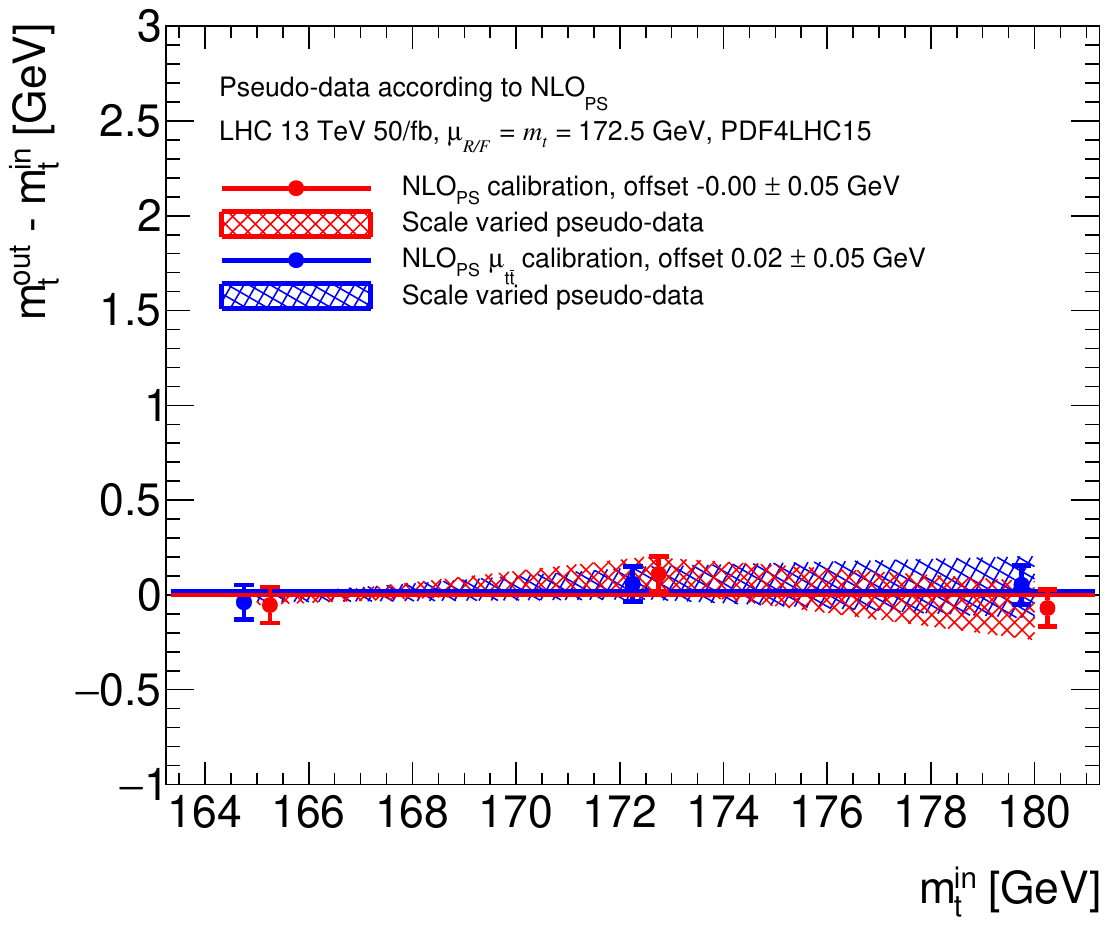}
\vspace{\TwoFigBottom em}
\caption{\label{fig:PS_scale_NLO_mt2}}
\end{subfigure}
\caption{\label{fig:PS_scale_mt2}%
  Same as Fig.~\ref{fig:PS_scale} but for the observable $m_{T2}$.
}
\end{figure}

The investigations performed for the \mlb\ observable are repeated for \mtwo.
 The results corresponding to Figs.~\ref{fig:LOvsNLO_NWA_decayLO}
 to~\ref{fig:PS_scale_NLO} are shown in Figs.~\ref{fig:LOvsNLO_NWA_decayLO_mt2}
 to~\ref{fig:PS_scale_NLO_mt2}.
 Also for \mtwo, the offsets obtained when using the corresponding pair of
 pseudo-data and calibration are consistent with zero, i.e.~the method is
 closed.

 While most observations are consistent for the \mlb\ and \mtwo\ observables,
 there are some remarkable differences.
 For \mtwo, comparing distributions with LO and NLO in production generally results in
 an \mt\-dependent offset. 
This indicates that the NLO prediction has a weaker mass dependence
than the LO one.
 The slope of the \mlb\ distribution in
 Fig.~\ref{fig:NLO_PS_vs} is less steep than the one in Fig.~\ref{fig:NLO_PS_vs_mt2}. This indicates a
 different effect of the parton shower
 on the more inclusive \mtwo, retaining a higher sensitivity to the
 top quark mass.

 The offsets observed for the various pairs of pseudo-data and calibration are
 given in Table~\ref{tab:offset}.
 The comparison of the offsets obtained for \mtwo\ with those for \mlb\
 exhibits a very similar pattern.
 To investigate whether the sensitivity of the observables to differences in the
 theoretical predictions coincides, the differences in their offsets are
 expressed by a \Chiq\ calculated from the offsets, using the fact that the
 offsets are uncorrelated for their statistical uncertainties.\footnote{Given
   $\Oi{o}{i}\pm \Oi{u}{i}$ for the offsets \Oi{o}{i}\ and their uncertainties
   \Oi{u}{i}\ with $i=1,2=\mlb, \mtwo$, the \Chiq\ is defined as:
   $\Chiq=(\Oi{o}{1}-\Oi{o}{2})^2/ (\Oi{u}{1}^2+\Oi{u}{2}^2)$.}
For a number of pairs the differences of the offsets for \mlb\ and \mtwo\ are
 consistent with zero, leading to small values of \Chiq,
for example when comparing $\lodec$ with $\mathrm{LO_{NWA}^{LOdec}}$
 (Figs.~\ref{fig:LOvsNLO_NWA_decayLO} and
 \ref{fig:LOvsNLO_NWA_decayLO_mt2}).
  In contrast, most notably for the pair $\nlops$ and $\nlodec$
(Figs.~\ref{fig:PS_vs_ttbar_NLO_nlodc} and \ref{fig:PS_vs_ttbar_NLO_nlodc_mt2}),
  the difference is significant, leading to a large \Chiq.
 This means, at the expected statistical precision of the $13\tev$ LHC, the two
 estimators exhibit different sensitivities to this difference in the
 theoretical prediction.

\clearpage

\section{Conclusions}
\label{sec:conclusions}

We have studied the impact of various theoretical descriptions for top
quark pair production on measurements of the top quark mass in the
di-lepton channel. In particular, we have compared the NLO QCD results
for $W^+W^- b\bar{b}$ production ($\nlofull$) to results based on the narrow-width
approximation, combining $t\bar{t}$ production at NLO with ($i$) LO top
quark decays ($\lodec$), ($ii$) NLO top quark decays
($\nlodec$) and ($iii$) a parton shower ($\nlops$). 
We have assessed the theoretical uncertainties associated with the
different theory descriptions via the variation of renormalisation,
factorisation and shower scales, and investigated the top quark mass
sensitivity of the observables $\mlb$, $\mtwo$, $\mll$ and $\etdr$.

Based on these results, we then studied the prospects of a top quark
mass extraction from the observables $\mlb$ and $\mtwo$, which we
found to be most sensitive to top quark mass variations.
Using pseudo-data based on our calculations, we employed the template
method to determine the offset in the top quark mass from calibrations
that differ in their underlying theory description. These analyses
show that the behaviour of the observables $\mlb$ and $\mtwo$ is
rather similar in what concerns the observed offsets in the top quark
mass.

More importantly, we found that the NLO corrections to the top quark decay
play a significant role, because they lead to non-uniform scale
uncertainty bands.
As the fits are based on normalised differential cross sections, 
shape differences induced by the scale variations will 
lead to larger theory uncertainties for the top quark mass extraction.
Even though the total scale uncertainties decrease at NLO as to be expected, 
the shape changes on the $\mlb$ distribution induced by scale variations are
particularly pronounced in the cases where the decay is described at NLO.
For both the $\nlofull$ as well as the $\nlodec$ description, the theoretical
uncertainties in determining \mt\ therefore increase by at least a
factor of two compared to the uncertainties emerging when LO decays are
involved.
Furthermore, the direct comparison of theories differing in their treatment
of the top quark decays can lead to offsets of more than $1\gev$ in the
measured \mt value.
This is observed in both cases, i.e.~when confronting $\nlofull$ pseudo-data
with the $\lofull$ calibration and $\nlodec$ pseudo-data with the $\lodec$
calibration.
These findings indicate that the non-resonant and non-factorising contributions
have a smaller effect on the top quark mass extraction than the NLO treatment
of the decay.

Turning to the parton shower ($\nlops$) results of our analysis
approach, we have compared them to the theory models $\nlofull$ and
$\nlodec$, leading to mass shifts of $-0.09\pm0.07\gev$ and
$0.96\pm0.07\gev$, respectively (in the \mlb\ case).
The good agreement between $\nlofull$ and $\nlops$ results can be
attributed to the fact that the two descriptions are rather similar
for an appropriate fit range, but it does not mean that the two descriptions
agree for the entire $\mlb$ range. Resummation effects for low $\mlb$ values
in the $\nlops$ case and off-shell effects affecting the tail in the
$\nlofull$ case are clearly visible in the $\mlb$ distribution.
The differences between $\nlops$ and $\nlodec$ mainly originate from
the regions of small and near-edge $\mlb$ values, where resummation
corrections play an important role.

To better understand these differences, we investigated the parton shower
behaviour in more detail.
We considered results where we limit the number of emissions in both the
production and the decay showers, and indeed observe that the predictions of
such restricted parton showers move closer to the fixed-order $\nlodec$ result.
These investigations also showed that the resummation corrections incorporated
by the unrestricted showers may lead to effects on the top quark mass
determination that can be as large as $1\gev$.
In addition, we have switched off the shower emissions in either production or
decay, and found that both the production and the decay showers impact our
analysis in a significant manner.
Different ways to assess the shower scale uncertainties within the $\nlops$
description were also studied but their effect turned out to be small. 
The choice of a different central scale also had only a minor impact on the mass determination.

We finally investigated how the choice of the fit range impacts our
results and found that the corresponding offsets do not change considerably
if the fit range is altered (in a way that still leads to acceptable closure).

Based on our results, we expect that the non-uniform scale variation bands
in the $\mlb$ distribution, induced by NLO corrections to the decay as present
in the $\nlofull$ calculation, would not level out largely if a parton
shower was matched to $\nlofull$.
It is therefore conceivable that a top quark mass extraction based on LO
(or shower approximated) decays may underestimate the theoretical
uncertainties, even if higher perturbative orders in the top quark pair
production process are taken into account.

In the future, it would be very interesting to see how the pseudo-data
used here compare to real data. In this context, the impact of hadronisation
and colour reconnection effects should be studied.
Owing to the rather strong impact of the resummation, it would also be useful
to perform a dedicated comparison of different parton shower
prescriptions such as different evolution variables and recoil strategies.
Furthermore, it would be worthwhile to investigate how the NLO
results for the full $W^+W^- b\bar{b}$ final state, ideally matched to a
parton shower, compare to NNLO results for top quark pair production in the
narrow-width approximation, combined with different descriptions of the top
quark decay.

\section*{Acknowledgements}

We would like to thank Stefan H\"oche for numerous discussions and
help with \tsc{Sherpa} generator issues.
GH would like to thank the CERN LPCC Theory Institute workshop ``LHC
and the Standard Model: Physics and Tools'' for hospitality while
parts of this project have been carried out.
JW gratefully acknowledges funding by the CERN Theory Department and
the Max Planck Institute for Physics in Munich, and is thankful for
great hospitality while parts of this work were completed.
Furthermore, JW's work is supported by the National Science
Foundation under Grant No.~PHY--141732.
We further acknowledge support of the Research Executive Agency (REA)
of the European Union under the Grant Agreement number
PITN-GA2012316704 (HiggsTools).
This research was supported in part by the European Union through the ERC
Advanced Grant MC@NNLO (340983).
We also acknowledge support and resources provided by the Max
Planck Computing and Data Facility (MPCDF).


\bibliographystyle{JHEP}
\bibliography{refs_mt}

\end{document}